\documentclass[preprints,article,accept,moreauthors,pdftex]{Definitions/mdpi} 

\usepackage[title]{appendix}
\usepackage{multirow}
\usepackage{algorithm,algcompatible}
\usepackage[utf8]{inputenc}
\usepackage[T1]{fontenc}
\usepackage{textcomp}
\usepackage{subfig}
\usepackage{pgf}
\usepackage{makecell}
\definecolor{bluecite}{HTML}{0875b7}
\usepackage[colorlinks,citecolor=bluecite,urlcolor=bluecite,bookmarks=false,hypertexnames=true]{hyperref}

\firstpage{1} 
\pubvolume{xx}
\issuenum{1}
\articlenumber{5}
\pubyear{2021}
\copyrightyear{2021}
\history{\textbf{This is a preprint. For a revised version and its published version refer to \url{https://doi.org/10.3390/rs13050956}}}

\Title{Outlier detection at the parcel-level in wheat and rapeseed crops using multispectral and SAR time series}

\Author{Florian Mouret$^{1, 2, *}$, Mohanad Albughdadi $^{1}$, Sylvie Duthoit $^{1}$, Denis Kouamé $^{3}$, Guillaume Rieu $^{1}$ and Jean-Yves Tourneret $^{2}$}

\AuthorNames{Florian Mouret, Mohanad Albughdadi, Sylvie Duthoit, Denis Kouamé, Guillaume Rieu, and Jean-Yves Tourneret}

\address{%
$^{1}$ \quad TerraNIS, 12 Avenue de l’Europe, 31520 Ramonville-Saint-Agne, France\\
$^{2}$ \quad University of Toulouse / IRIT-INP-ENSEEIHT / TéSA, 2 Rue Charles Camichel, 31000 Toulouse, France \\
$^{3}$ \quad University of Toulouse / IRIT-UPS, 118 Route de Narbonne, 31062 Toulouse Cedex 9, France}

\corres{Correspondence: florian.mouret@terranis.fr; florian.mouret@irit.fr}

\abstract{This paper studies the detection of anomalous crop development at the parcel-level based on an unsupervised outlier detection technique. The experimental validation is conducted on rapeseed and wheat parcels located in Beauce (France). The proposed methodology consists of four sequential steps: 1) preprocessing of synthetic aperture radar (SAR) and multispectral images acquired using Sentinel-1 and Sentinel-2 satellites, 2) extraction of SAR and multispectral pixel-level features, 3) computation of parcel-level features using zonal statistics and 4) outlier detection. The different types of anomalies that can affect the studied crops are analyzed and described. The different factors that can influence the outlier detection results are investigated with a particular attention devoted to the synergy between Sentinel-1 and Sentinel-2 data. Overall, the best performance is obtained when using jointly a selection of Sentinel-1 and Sentinel-2 features with the isolation forest algorithm. The selected features are VV and VH backscattering coefficients for Sentinel-1 and 5 Vegetation Indexes for Sentinel-2 (among us, the Normalized Difference Vegetation Index and two variants of the Normalized Difference Water). When using these features with an outlier ratio of 10\%, the percentage of detected true positives (i.e., crop anomalies) is equal to 94.1\% for rapeseed parcels and 95.5\% for wheat parcels.}


\begin{document}
\section{Introduction}\label{sec:intro}

Monitoring crop growth and status is a major challenge in remote sensing for agriculture \citep{WEISS2020111402}. Multispectral images have been used for this purpose for many years, thanks to their convenient interpretation and exploitation \citep{Bannari1995, GOMEZ201655, Inglada_2017, VERRELST2015review}. Synthetic aperture radar (SAR) images have also been widely studied since they are available regardless of sunlight and cloud coverage conditions \citep{Betbeder2016, Khabbazan2019, Kumar2013, Liu2019SAR, McNairn2016}. The complementarity of these two types of images has been used to address problems including crop type classification \citep{Inglada_2016, Orynbaikyzy2019}, estimation of crop water requirement \citep{Navarro_2016} and change detection \citep{prendes2015icassp, prendes2015tip, prendes2015rfpt}. The joint use of SAR and multispectral images is also encouraged by the large amount of free data provided by the Sentinel-1 (S1) and Sentinel-2 (S2) satellites operated by the European Space Agency (ESA). The fine spatial and temporal resolutions of S1 and S2 images allow working at the parcel-level with a high revisit frequency, which is well suited for precision agriculture \citep{DEFOURNY2019551, Vreugdenhil_2018}. Various studies have introduced methodologies for deriving crop classification maps using S1 and S2 data \citep{Denize2018, Kussul2018, Pouya2018}. A comprehensive analysis of the temporal behavior of S1 and S2 data has also been proposed \citep{Navarro_2016, VELOSO2017415}.

This work addresses an interesting remaining challenge in precision agriculture, namely the automatic detection of crop parcels that have an anomalous vegetation development. Detecting crop parcels whose phenological behaviors significantly differ from the others could help users such as farmers or agricultural cooperatives to optimize agricultural practices, disease detection or fertilization management. It could also be valuable in areas such as subsidy control or crop insurance.

In the literature, the problem of finding objects that are unusual or different from the majority of the data is known as outlier detection (also referred to as anomaly detection). Outlier detection techniques have received a considerable attention \citep{Aggarwal2017, Chandola2009, Pimentel2014} since they are used in a large variety of application domains, \textit{e.g.,} fraud detection or medical diagnosis. In the Earth observation area, various methods have been introduced to detect abnormal vegetation areas at the country-level using time series constructed from the Normalized Difference Vegetation Index (NDVI). These approaches aim at modeling NDVI time series using historical data and detecting potential anomalies by comparing new observations with their corresponding predicted values \citep{Atzberger2011noground, Beck2006, Klisch2016, Meroni2019, Verbesselt2012}. Recent studies have investigated similar techniques with S1 and S2 data, as for instance in \citet{Kanjir_2018} where Breaks for Additive Season and Trend (BFAST) are used to detect land use anomalies. However, these approaches can be difficult to implement for crop monitoring since modeling the normal behavior of the data implies having access to normal representative examples, which can be difficult and time consuming in practice. Crop rotation, lack of historical data and the inconsistency of S2 time series due to the cloud coverage are other factors leading to a harder practical implementation. In previous work conducted in \citet{Albughdadi2017}, agronomic features have been extracted from multispectral images to detect outliers in crop parcels. Nevertheless, this approach was based on clustering (using the mean shift algorithm), whose parameters are not easy to adjust for the detection of abnormal crop parcels. This literature overview motivates the need to investigate new approaches for outlier detection dedicated to crop monitoring.

Rather than detecting inter-annual abnormalities, a method investigated in this paper consists in detecting the most abnormal parcels within a growing season (or a part of a growing season). In the literature, such method is referred to as point outlier detection \citep[Section 2.2]{Chandola2009}. Point outlier detection consists in comparing each instance of the dataset (here, each crop parcel) to the rest of the data to find the most different instances, which are isolated from the majority of the observed data. Point outlier detection approaches are well suited to our problem since 1) they are unsupervised by nature (\textit{i.e.}, a training set using nominal data is not necessary), 2) they can be used with multiple indicators providing information about the crop development and 3) they can be applied to data acquired within the growing season (\textit{i.e.}, no historical data are needed).

This paper provides a methodology to detect anomalous crop development using SAR and multispectral images acquired by S1 and S2 sensors. The novelty of this work for crop monitoring is that it uses unsupervised outlier detection algorithms within a single growing season analysis, without any prior knowledge on the normal behavior of the parcels and using SAR and multispectral data jointly. The two main objectives of this study are: 1) to provide a detailed analysis of the different anomalies detectable with S1 and S2 data that could affect crops such as wheat and rapeseed and 2) to investigate the different factors that influence the detection results such as the feature sets. 

This paper is structured as follows. Section~\ref{sec:studyarea} presents the study area and the data used for the detection of outlier crop parcels. Section~\ref{sec:method} proposes the methodology suggested to detect anomalous crop development at the parcel-level. In particular, a description of the different types of agronomic outliers encountered during the study is provided. Section~\ref{sec:results} validates the detection results for rapeseed and wheat crops. Finally, some conclusions and future work are reported in Section~\ref{sec:conclusion}.

\section{Study area and data}\label{sec:studyarea}

\subsection{Study area}

\begin{figure}[ht!]
\centerline{\includegraphics[width=1\textwidth]{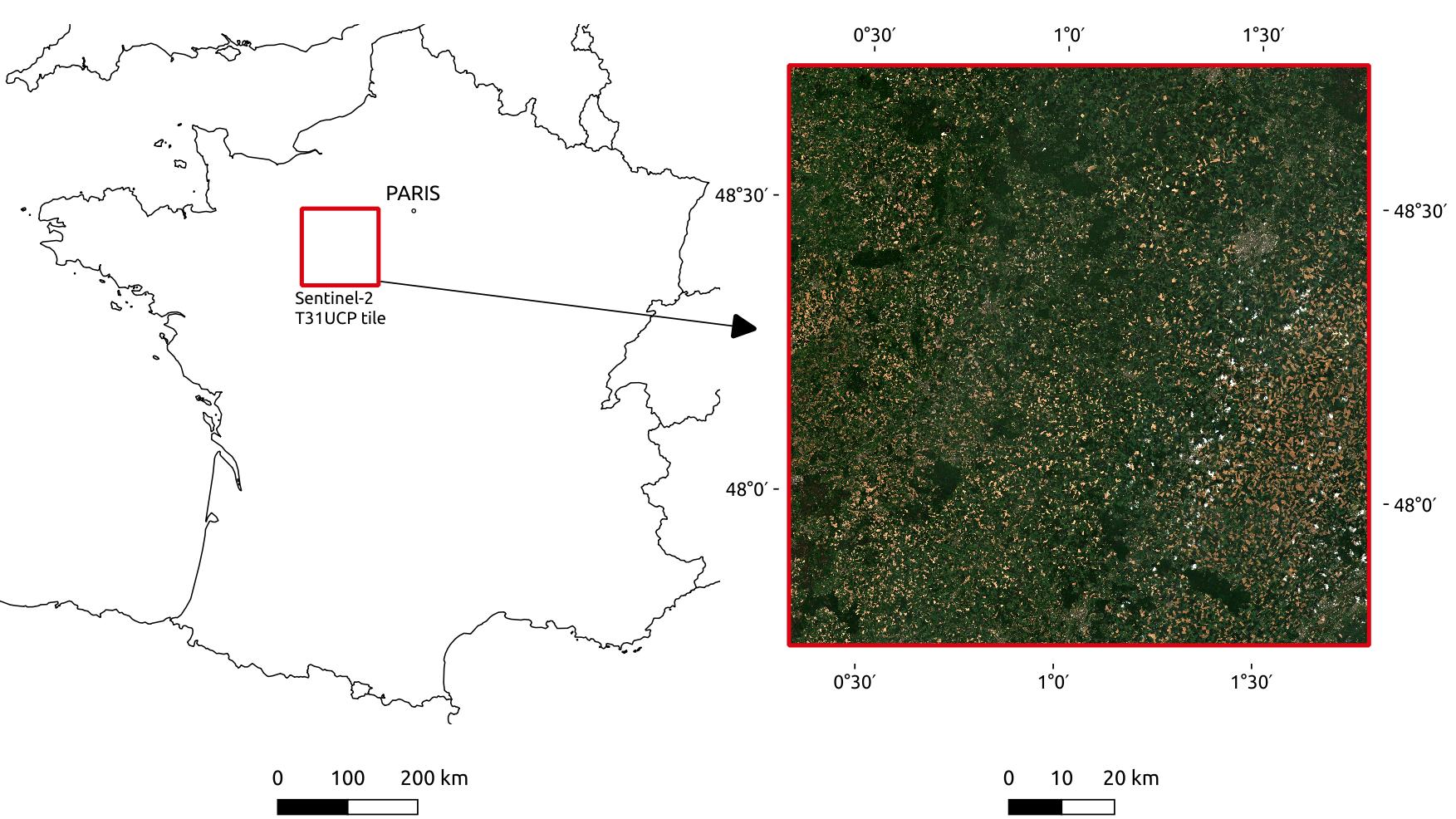}}
\caption{The Sentinel-2 tile T31UCP considered in this work is located in the Beauce area (near Paris) and delimited by the red box. On the right, the S2 image processed in level 2A acquired in May 19 2018 is displayed in natural colors.}
\label{fig:zone_etude}
\end{figure}

The analyzed area is located in the Beauce region in France. The area has an extent of $109.8 \times 109.8$ $\textrm{km}^2$ and is centered approximately at 48°24'N latitude and 1°00'E longitude (corresponding to the T31UCP S2 tile). \autoref{fig:zone_etude} shows the tile location and the studied area, which was chosen due to its richness of large crop fields such as wheat and rapeseed.

\subsection{Parcel data}

The analysis is conducted on a total of 2218 rapeseed parcels (associated with the 2017/2018 growing season) and 3361 wheat parcels (associated with the 2016/2017 growing season). To avoid problems in parcel boundaries, a buffer of $10$~m was applied allowing too small parcels (area less than 0.5 ha) to be discarded from the database. In the Supplementary Materials, the robustness of the proposed method to changes in the parcel boundaries is validated using 2118 rapeseed parcel delineations resulting from the French Land Parcel Identification System (LPIS) \citep{BARBOTTIN2018281}, which is available in open license.

\subsection{Remote sensing data}\label{sec:remote_sensing_data}

The S2 and S1 images used in this study were selected and downloaded from the PEPS platform (Plateforme d'Exploitation des Produits Sentinel) of the French National Center for Space Studies (Centre National d’Études Spatiales, CNES)\footnote{\url{peps.cnes.fr/}, online accessed 8 December 2020}. For multispectral data, both S2-A and S2-B satellites were used, which makes a theoretical revisit time of 5 days. S2 images have $13$ spectral bands covering the visible, the near infra-red (NIR) and the shortwave-infrared (SWIR) spectral region \citep{DRUSCH201225}. Details about the spectral bands used in the analysis to extract agronomic features at the pixel-level are reported in \autoref{tab:S2_bands}.

\begin{table}[ht!]
    \caption{Sentinel-2 multispectral bands used for the analysis.}
    \centering
    \footnotesize
    \begin{tabular}{llll}
         Spectral bands & Central wavelength (\textrm{$\mu$}m) & Bandwith (\textrm{$\mu$}m) & Resolution (m)  \\ \hline
         Band 3: Green & $0.560$ & $0.035$ & $10$ \\
         Band 4: Red & $0.665$ & $0.030$ & $10$ \\
         Band 5: Vegetation Red Edge & $0.705$ & $0.015$ & $20$ \\
         Band 8: Near Infrared (NIR) & $0.842$ & $0.115$ & $10$ \\
         Band 11: Shortwave Infrared (SWIR) & $1.610$ & $0.090$ & $20$ \\ \hline
         
    \end{tabular}
    \label{tab:S2_bands}
\end{table}

Radar S1 images are constructed by analyzing the response signal in flight direction of a C-band synthetic aperture radar (SAR) operating at a center frequency of 5.405 GHz. Both S1-A and S1-B satellites were used in ascending orbit (6 days revisit time). Ground Range Detected (GRD) products were used in the Interferometric Wide (IW) swath mode: phase information is lost but the volume of data is considerably reduced. All SAR images were available in dual polarization (VH+VV) with a $10$~m spatial resolution.

\begin{figure}[ht]
\subfloat[]{\includegraphics[width=0.5\textwidth]{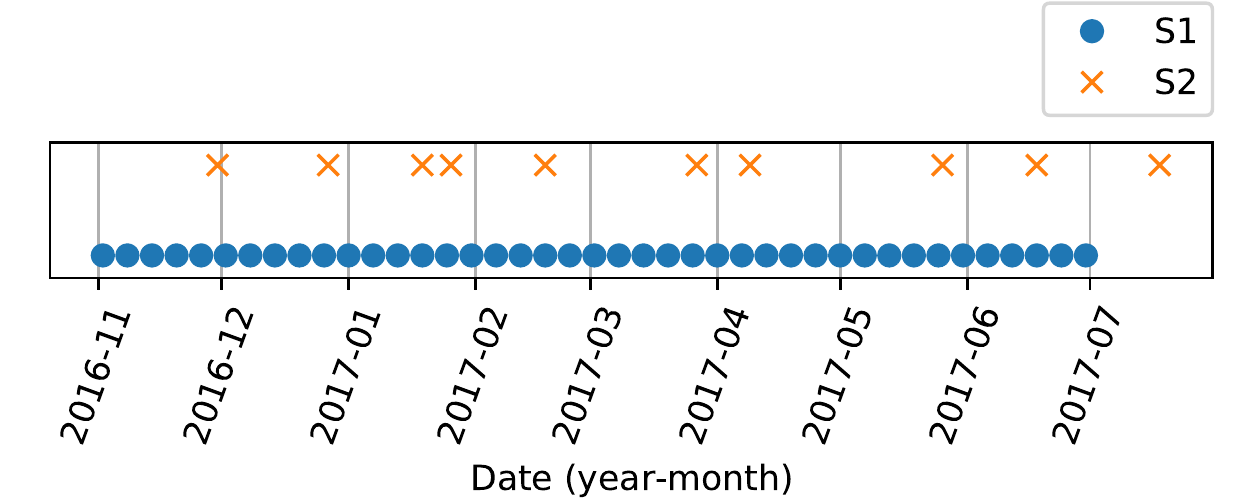}}
        \hfill
\subfloat[]{\includegraphics[width=0.5\textwidth]{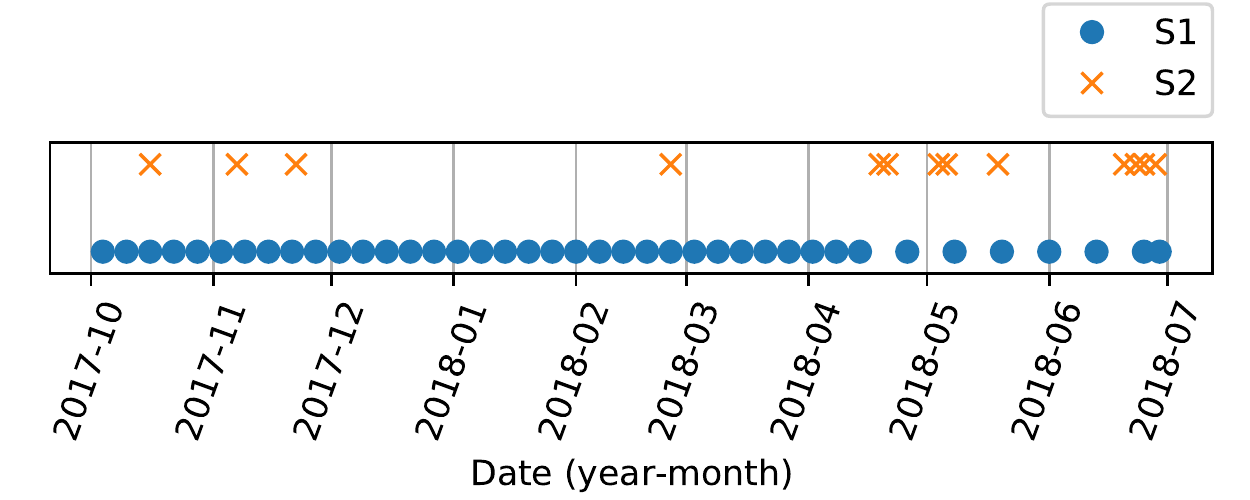}}
\caption{Each marker corresponds to the acquisition date of a used image for the growing season (a) 2016/2017 and (b) 2017/2018.}
\label{fig:calendar}
\end{figure}

The acquisition dates of S1 and S2 images are depicted in~\autoref{fig:calendar} for the 2016/2017 and 2017/2018 growing seasons. It was decided to select S2 images with a low cloud coverage (cloud coverage lower than 20\%). The strategy considered to handle remaining clouds is detailed in Section~\ref{sec:image_preproc}. All S1 images covering the analyzed area in ascending orbit were selected. For the 2016/2017 growing season, 41 S1 images and 10 S2 images were selected whereas 40 S1 images and 13 S2 images were selected in 2017/2018. Due to cloud coverage, the acquisition dates for S2 images are very different for the two growing seasons. Note that a reduced number of S1 images were available between May and July 2018. The absence of data during this period can be observed in all webpages providing S1 data, which confirms problems in data acquisition.

\section{Methods}\label{sec:method}

\begin{figure}[ht!]
\centerline{\includegraphics[scale=0.4]{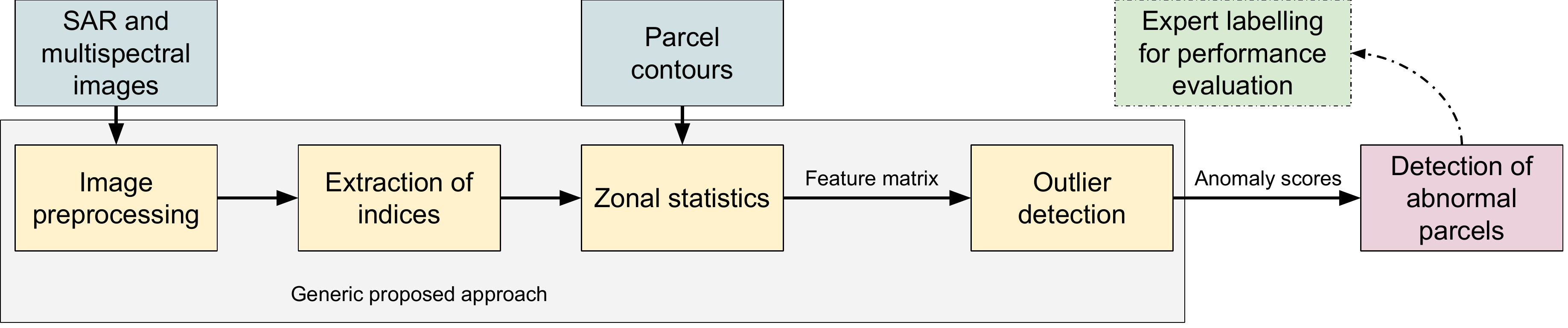}}
\caption{Diagram summarizing the methodological steps for the detection of anomalous crop development.}
\label{fig:workflow}
\end{figure}

The proposed method for detecting abnormal crop development relies on a four-step sequential procedure depicted in~\autoref{fig:workflow} and discussed in detail in what follows. Methods to describe and evaluate the detection results are also provided in this section.

\subsection{Image preprocessing}\label{sec:image_preproc}

S2 images were preprocessed using the online MAJA processing chain \citep{Hagolle2015} available on the PEPS platform of CNES. This preprocessing step provides level-2A ortho-rectified products expressed in surface reflectance. In addition to atmospheric correction, level-2A images are available with a cloud and shadow mask discarding irrelevant pixels in the images. A resampling strategy was adopted to obtain a spatial resolution of $10$~m for the channels with a lower spatial resolution. Parcels fully covered by clouds during at least one time instant were discarded from the database and parcels partially covered by clouds were analyzed using pixels not covered by the cloud mask (the shadow mask was used in a similar way).
 
\begin{figure}[ht!]
    \centering
    \includegraphics[width=0.8\linewidth]{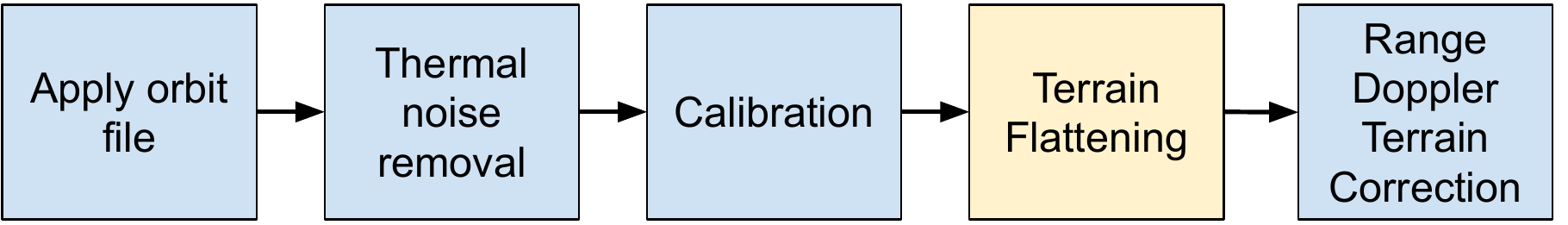}
    \caption{Sentinel-1 processing chain used in the Sentinel Application Platform (SNAP). The yellow box, terrain flattening, was added to the workflow proposed in \citet{Filipponi_2019} to take into account the local incidence angle.}
    \label{fig:SNAP_process}
\end{figure}

To build the database of S1 images, an offline processing (illustrated in \autoref{fig:SNAP_process}) was conducted with the Sentinel Application Platform (SNAP, version 7.0)\footnote{\url{http://step.esa.int/main/toolboxes/snap/}, online accessed 8 December 2020}. This processing is inspired by the workflow proposed in \citet{Filipponi_2019}. A Terrain-Flattening operation was added to take into account the local incidence angles, as the analyzed area is wide and parcel features are compared to each other. This operation uses the Shuttle Radar Topography Mission (SRTM) Digital Elevation Model (DEM). The Range Doppler terrain correction provides orthorectified images. Note that a multi-temporal speckle filtering step was also tested without significant differences on the results (we implemented our own Python version of the filter introduced in Eq. (14) of \citet{Quegan2001}). The best results were obtained with the workflow of \autoref{fig:workflow}.

\subsection{Extraction of SAR and multispectral features at the pixel-level}\label{sec:feat_extract}

The following section describes the pixel-level features derived from multispectral and SAR images considered in this work (reported in \autoref{table:table_optical_indicators}) and their importance for monitoring crop growth. It was observed that choosing irrelevant features can lead to poor detection results, since unsupervised algorithms use all the features available for the analysis. For post-analysis and practical applications, it is also important to choose features whose interpretation is convenient in order to understand why an anomaly has been detected.

\subsubsection{Multispectral vegetation indices}

Many multispectral Vegetation Indices (VIs) have been introduced in the literature \textit{e.g.}, \citep{Bannari1995, Wu2008}. A VI relates the acquired spectral information to the observed vegetation, and thus allows better quantitative and qualitative evaluations of the vegetation covers. The five multispectral VIs considered in this paper are reported in~\autoref{table:table_optical_indicators} and described below. Note that raw S2 bands were also tested without any improvement in the detection precision and a more difficult interpretation of the results when compared to VIs.

The NDVI is a benchmark indicator for agronomic analyses and is mainly related to the plant vigor \citep{rouse1974monitoring, Bannari1995}. The Normal Difference Water Index (NDWI) actually refers to two different widely used indicators. The first version uses Near infrared (NIR) and Short Wave infrared (SWIR) to monitor changes in the water content of leaves \citep{Gao1996}. The second version uses the green band and NIR to monitor changes related to content in water bodies \citep{McFEETERS1996}. Both formulas are similar to NDVI with different bands involved. The SWIR version of NDWI seems to be more appropriate for crop analysis but the GREEN version of NDWI can also provide relevant information, \textit{e.g}., for flooded parcels. The Modified Chlorophyll Absorption Ratio Index (MCARI) was designed to extract information from the chlorophyll content in plants with a resistance to the variation of the Leaf Area Index (LAI). A variant called MCARI/OSAVI uses the Optimized Soil Adjusted Vegetation Index (OSAVI) to minimize the contribution of background reflectance \citep{DAUGHTRY2000, Wu2008}. The Green Red Vegetation Index (GRVI) is similar to NDVI but uses the red and green bands. According to \cite{Motohka2010}, GRVI ``\textit{can be a site-independent single threshold for detection of the early phase of leaf green-up and the middle phase of autumn coloring}'' (referred to as senescence for crops).

\subsubsection{SAR features}

Many investigations have been performed to establish a relationship between SAR images and vegetation and have been reported in two recent reviews \citep{McNairn2016, Liu2019SAR}. The backscattering coefficients (denoted as $\gamma^0_{\text{VH}}$ and $\gamma^0_{\text{VV}}$) have been used intensively in the literature \citep{Whelen2018, Khabbazan2019}. The polarization ratio $\gamma^0_{\text{VH}} / \gamma^0_\text{{VV}}$, also used in various studies \citep{Abdikan2016, Denize2018, VELOSO2017415, Vreugdenhil2018} was tested without showing any clear improvement. The same observation stands for the Radar Vegetation Index (RVI) \citep{Kumar2013}, which has been adapted to S1 with an alternative form $ 4 \sigma_{VH}^0 / (\sigma_{VH}^0 + \sigma_{VV}^0)$ \citep{Nasirzadehdizaji_2019}. Thus, the results presented in this paper have been obtained with the backscattering coefficients reported in \autoref{table:table_optical_indicators}.

\begin{table}[ht]
\caption{Pixel-level features computed from S2 and S1 images used in this paper. For S2, The near infrared (band~8), red edge (band~5), short wave infrared (band~11), green (band~3) and red (band~4) channels are denoted as NIR, RE, SWIR, GREEN and RED, respectively.}
\centering
\begin{tabular}{lll}
\\\hline Sensor type              & Indicator & Formula  \\ \hline & \\
\multirow{10}{*}{Multispectral} & \textnormal{NDVI}      & $\frac{\textnormal{NIR} - \textnormal{RED}}{\textnormal{NIR} + \textnormal{RED}}$ \\ & \\
                         & $\textnormal{NDWI}_{\textnormal{SWIR}}$     & $\frac{\textnormal{NIR} - \textnormal{SWIR}}{\textnormal{NIR} + \textnormal{SWIR}}$     \\ & \\
                         & $\textnormal{NDWI}_{\textnormal{GREEN}}$      & $\frac{\textnormal{GREEN} - \textnormal{NIR}}{\textnormal{GREEN} + \textnormal{NIR}}$     \\ & \\
                         & $\frac{\textnormal{MCARI}}{\textnormal{OSAVI}}$      & $\frac{(\textnormal{RE} - \textnormal{IR}) - 0.2(\textnormal{RE} - \textnormal{RED})}
                        {(1+0.16)\frac{\textnormal{NIR} - \textnormal{RED}}{\textnormal{NIR} + \textnormal{RED} + 0.16}}$        \\ & \\
                         & \textnormal{GRVI}      & $\frac{\textnormal{GREEN} - \textnormal{RED}}{\textnormal{GREEN} + \textnormal{RED}}$      \\ & \\ \hline & \\
\multirow{4}{*}{SAR}     & \makecell[l]{Cross-polarized backscattering \\ coefficient VH}        & $\gamma^0_{\textnormal{VH}}$       \\ & \\
                         & \makecell[l]{Co-polarized backscattering \\ coefficient VV}        & $\gamma^0_{\textnormal{VV}}$\\
                        \\\hline
\end{tabular}
\label{table:table_optical_indicators}
\end{table}

\subsection{Input data for the outlier detection algorithms}
\subsubsection{Extraction of parcel-level features with zonal statistics}

The pixel-level features are averaged through spatial statistics referred to as ``zonal statistics'' in order to provide parcel-level features. Two zonal statistics are considered for the S2 VIs, namely the median and interquartile range~(IQR). The median captures the mean behavior of a given parcel with more robustness than the classical mean as it is not affected by extreme values \citep{Huber2011}. It is used to detect anomalies affecting the entire crop parcel, such as anomalies in crop vigor. IQR is defined as the difference between the 75th and 25th percentiles. It contains information related to the heterogeneity of a given parcel while being robust to the presence of extreme values. These statistics were computed using the Python libraries SciPy version 1.4.1 \citep{2020SciPy-NMeth} and rasterstats version 0.13.0\footnote{\url{https://pythonhosted.org/rasterstats/}, online accessed 8 December 2020}. Since cloud and shadow pixels were discarded, these statistics were computed from the remaining pixels after applying the cloud and shadow masks. Other zonal statistics were also tested, namely the skewness (which is related to the asymmetry of a distribution) and the kurtosis (which can be used to characterize the tail of a distribution) but led to a deterioration of the detection results (Supplementary Figure S14). The SAR feature set was reduced to the median of the backscatter intensities, as IQR of S1 data was found to be directly proportional to the median.

\subsubsection{Feature matrix}\label{sec:feature_matrix}

Each parcel is represented by a vector concatenating the zonal statistics computed for all pixel-level features at each date. The construction of the feature matrix, used as the input of the outlier detection algorithms, is illustrated in \autoref{table:table_feature_matrix} when using the NDVI with 2 statistics. In the general case, the number of columns of this matrix is $N_{col}=  N_{1,im} \times N_{1,f} \times N_{1,s} + N_{2,im} \times N_{2,f} \times N_{2,s} $, where $N_{1,im}$ is the number of S1 images, $N_{1,f}$ is the number of pixel-level features extracted for each S1 image, $N_{1,s}$ is the number of statistics computed for each S1 feature and similar definitions apply to $N_{2,im}$, $N_{2,f}$ and $N_{2,s}$ for S2 images. As each column corresponds to a unique combination statistics/feature/time, it is possible to compare each parcel columnwise. Note that classical preprocessings such as the Principal Component Analysis (PCA) \citep{Jolliffe1986} or the Multidimensional Scaling (MDS) \citep{Borg1997} were applied to this feature matrix without significant improvement regarding the outlier detection results. Thus, these preprocessing were ignored from our analysis.

\begin{table}[ht!]
\caption{Simplified version of the feature matrix using NDVI only and two statistics (median/IQR) for $n$ dates and $M$ parcels. NDVI$_{t_n}$ means NDVI computed for image $\# n$ and median$_{P_M}$ means spatial median of the feature computed inside the parcel $\# M$}.
\centering
\begin{tabular}{|c|c|c|c|c|c|c|}
    \hline
    Parcel \# & Feature 1 & Feature 2 & . & Feature L-1 & Feature L \\
    \hline
    $P_1$ & median$_{P_1}$(NDVI$_{t_0}$) & IQR$_{P_1}$(NDVI$_{t_0}$) & . & median$_{P_1}$(NDVI$_{t_n}$) &  IQR$_{P_1}$(NDVI$_{t_n}$) \\
    \hline
    $P_2$ & median$_{P_2}$(NDVI$_{t_0}$) & IQR$_{P_2}$(NDVI$_{t_0}$) & . & median$_{P_2}$(NDVI$_{t_n}$) &  IQR$_{P_2}$(NDVI$_{t_n}$) \\
    \hline
    ... & ... & ... & . & ... & ...  \\
    \hline
    $P_M$ & median$_{P_M}$(NDVI$_{t_0}$) & IQR$_{P_M}$(NDVI$_{t_0}$) & . & median$_{P_M}$(NDVI$_{t_n}$) &  IQR$_{P_M}$(NDVI$_{t_n}$) \\
    \hline
    
\end{tabular}
\label{table:table_feature_matrix}
\end{table}

\subsection{Outlier detection algorithms}\label{sec:algorithms}

Using the feature matrix whose construction is detailed in the previous section, an outlier detection algorithm attributes to each parcel an \textit{outlier score}. The percentage of the most abnormal parcels to be detected is called the \textit{outlier ratio} (i.e., the parcels that have the highest outlier score).

The experiments presented in this document were conducted using the Isolation Forest (IF) algorithm \citep{Liu2012}, which provided the best results overall. IF is an isolation-based technique that assumes that anomalies can be easily isolated in ``\textit{a tree structure based on random cuts in the values of randomly selected features in the dataset''} \citep{Hariri2018}. IF is considerably fast since it does not need computation of distances or density measures. The number of isolation trees was fixed to $\text{n}_{\text{trees}} = 1000$ and the size of the data sub-sampling was fixed to $\text{n}_{\text{samples}} = 256$, as in the original paper. Changing these two parameters did not have a significant effect on the results, which is a crucial advantage compared to the other algorithms (see Section~\ref{sec:discussion} for more details).

Three other outlier detection algorithms were also tested, namely: One-Class Support Vector Machine (OC-SVM) \citep{Scholkopf1999}, Local Outlier Probabilities (LoOP) \citep{Kriegel2009}, and AutoEncoder (AE) \citep[Section 3.6]{Aggarwal2017}. These algorithms are based on different ideas, making them interesting for comparison purposes. The effect of changing the algorithm used for the detection is discussed in Section~\ref{sec:discussion} and details regarding the choice of the hyperparameters are provided in the Supplementary Materials. To run the experiments conducted in this study, the Python Scikit-learn (version 0.23.0) implementations of OC-SVM and IF were used \citep{scikit-learn}. The Python library PyNomaly (version 0.3.3) was used for the implementation of the LoOP algorithm \citep{Constantinou2018}. Finally, we implemented our own autoencoder with the Python library Keras\footnote{\url{https://keras.io/}, online accessed 8 December 2020} (version 2.3.0).

\subsection{Experiments conducted to evaluate the proposed method}\label{sec:nomemclature}

\begin{figure}[ht!]
    \centering
    \includegraphics[width=1\textwidth]{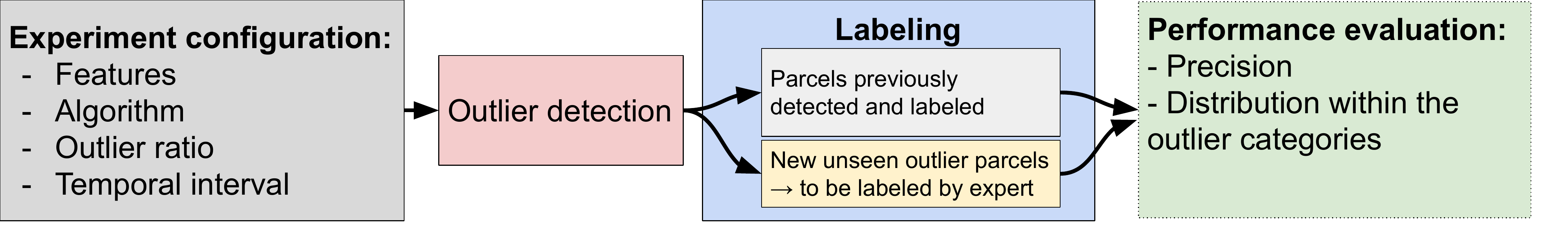}
    \caption{Diagram illustrating the idea behind the different conducted experiments.}
    \label{fig:diag_expe}
\end{figure}

In what follows, what is called ``experiment'' corresponds to an outlier detection conducted with a specific initial configuration (set of features, algorithm, outlier ratio, temporal interval) using one of the two datasets (wheat or rapeseed). Various experiments were conducted to evaluate the proposed approach: each time a new set of features or a new algorithm tuning was tested, the parcels declared as outliers were counterchecked by experts (if not previously detected), confirming the anomaly (true positive) or not (false positive), and determining the type of anomaly (see details later). This iterative procedure is illustrated in \autoref{fig:diag_expe}. 

For the rapeseed dataset, $252$ initial configurations were tested to evaluate the factors that can influence the detection results. Most of these experiments were conducted on a complete growing season to evaluate the capacity of the proposed approach to detect anomalies occurring at different periods of the crop growth, and to determine whether differences between the detected parcels can be observed or not. Some other experiments were also made with a lower amount of data, in particular for a mid season analysis between October and February. Early detection can be of interest for warning purposes at the beginning of the growth cycle and gives more details on the effect of having only few data available for the analysis. The influence of the amount of parcels to be detected (called outlier ratio) is also tested to analyze the relevance of the outlier score given to each parcel. 

For the wheat dataset, 25 experiments were made: the main idea was to determine whether our approach can be applied with minor modifications to other kinds of crops. The different experiments conducted during the study are reported in \autoref{table:experiments}. The main results obtained from a selection of these experiments are presented in this document whereas additional results are discussed in Section~\ref{sec:discussion}.

\begin{table}[ht!]
\caption{Summary of the evaluated factors analyzed throughout the study. In parenthesis, the number of different initial configurations tested (features, algorithm, time interval, outlier ratio).}
\centering
\begin{tabular}{p{35mm}p{25mm}p{60mm}}
\\ 
\hline
Evaluated crop type             & Time interval                     & Evaluated factors  \\\hline & \\
\multirow{8}{*}{\makecell[l]{Rapeseed \\ (252)} } & \multirow{6}{*}{\makecell[l]{Complete season \\ (218)}}   & Outlier detection algorithms                                      \\ & & Feature sets                          \\
                                &                                  & Outlier ratio                                           \\
                                &                                  & Zonal statistics                                                \\
                                &                                  & Missing S2 images                                                                   \\
                                &                                  & Changes in parcel boundaries                                                                   \\
                                & \makecell[l]{Mid season \\(34)}                       & Feature sets, algorithms, outlier ratio                                                      \\ \hline & \\
\multirow{2}{*}{\makecell[l]{Wheat \\ (25)}}                  & \makecell[l]{Complete season \\ (20) }                  & Feature sets, algorithm \\
                                & \makecell[l]{Mid season \\ (5) }       &   Feature sets, algorithm                                                                             \\ \hline & \\                                         
\end{tabular}
\label{table:experiments}
\end{table}

\newpage
\subsection{Description of the outlier parcels}\label{sec:labeling_proc}

The outlier parcels were identified through multiple outlier detection analyses presented in Section~\ref{sec:nomemclature}. With the help of agronomic experts, the labeling of the detected parcels was conducted by visual-interpretation using all the available S1 and S2 images and by using all the time series of the different features/statistics to compare any analyzed parcel to the rest of the dataset. Each labeled parcel was assigned to one of the outlier categories described in what follows.

\subsubsection{Outlier categories}\label{sec:description_anomalies}

The different anomalies analyzed throughout the study can be decomposed into 4 main categories: heterogeneity problems, growth anomalies, database errors and others. The category ``others'' corresponds to non-agronomic outliers that were considered not relevant for crop monitoring (referred to as false positives). A brief description of each category is proposed in \autoref{table:list_cat} and more details and examples are provided below. Additional examples are also available in the Supplementary Materials.

\begin{table}[ht]
\caption{Description of the different categories of anomalies detected during the labeling process. Subcategories were added to have a more precise description. For each category TP means true positive, considered relevant for crop monitoring, and FP means false positive, considered irrelevant for crop monitoring.}
\footnotesize
\begin{tabular}{p{21mm}p{45mm}p{75mm}}
\\ \hline
Category (TP/FP)               & Subcategory                              & Description                      \\ \hline
\multirow{4}{*}{\makecell[l]{Heterogeneity \\ (TP)}} & Heterogeneity                         & Affects the parcel most of the season                                                                                   \\
                               & Heterogeneity (2 different parts)     & The parcel is separated into two homogeneous different parts                                                                      \\
                               & Heterogeneity after senescence        & Occurs during senescence phase                                                                                           \\
                               & Early heterogeneity                   & Occurs during early growing season                                                                                     \\ \hline
\multirow{5}{*}{\makecell[l]{Growth \\ (TP)}}        & Late growth                           & A late development is observed (non-vigorous crop)                                     \\
                               & Vigorous crop                         & A vigorous development is observed                                                               \\
                               & Early flowers                         & Early flowering phase                                                                  \\
                               & Early senescence                      & Early senescence phase                                                                 \\
                               & Late senescence                       & Late senescence phase                                                                   \\ \hline
\multirow{2}{*}{\makecell[l]{Error in database \\ (TP)}}    & Wrong type                            & A wrong crop type is reported in the database                                                                                              \\
                               & Wrong shape                           & The parcel boundaries are not accurately reported                                                                                          \\ \hline
\multirow{5}{*}{\makecell[l]{Others \\ \textbf{(FP)}}} & Normal (counterchecked)             & The parcel was declared normal by the agronomic expert                                                           \\
                                & Too small                             & The parcel is too small, causing abnormal features                                                                \\
                               & SAR anomaly                           & Soil surface conditions causes abnormal SAR features                                       \\
                               & \begin{tabular}[c]{@{}l@{}}Shadow perturbation \\ (cloud or forest)\end{tabular} & Shadows cause abnormality in the features.          \\ \hline                                    
\end{tabular}
\label{table:list_cat}
\end{table}

\newpage
\begin{itemize}\setlength{\itemsep}{0pt}
    \item \textit{Heterogeneity} corresponds to parcels presenting a clear heterogeneous development (i.e., spatially heterogeneous development). The most common cases of heterogeneity can be observed all along the growing season and are for instance related to soil heterogeneity, presence of weed or diseases. An example of heterogeneous parcel is shown in \autoref{fig:ex_heter}. More transient cases of heterogeneity can affect the beginning (\textit{early heterogeneity}) or the end of the growing season (\textit{heterogeneity after senescence}) and can be for instance related to differences in soil characteristics or parcel exposure (Supplementary Figures S2). \textit{Heterogeneity (2 different parts)} parcels have two areas of the same crop separated by a clear frontier (e.g., strong difference in the phenological stages) (Supplementary Figure S1). 
    
    \begin{figure}[ht!]
        \centering
        \subfloat[]{\includegraphics[width=0.49\textwidth]{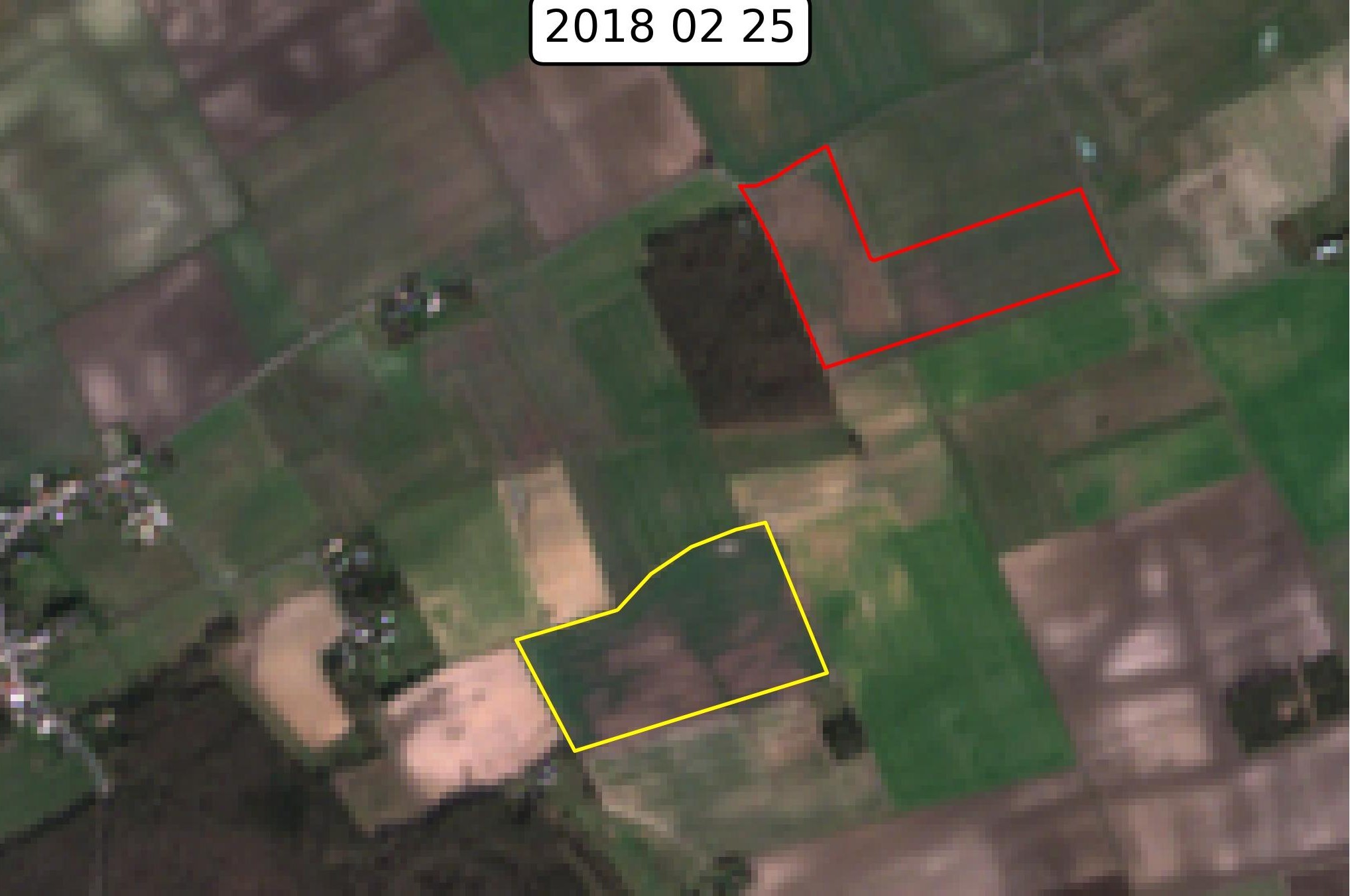}}
        \subfloat[]{\includegraphics[width=0.47\textwidth]{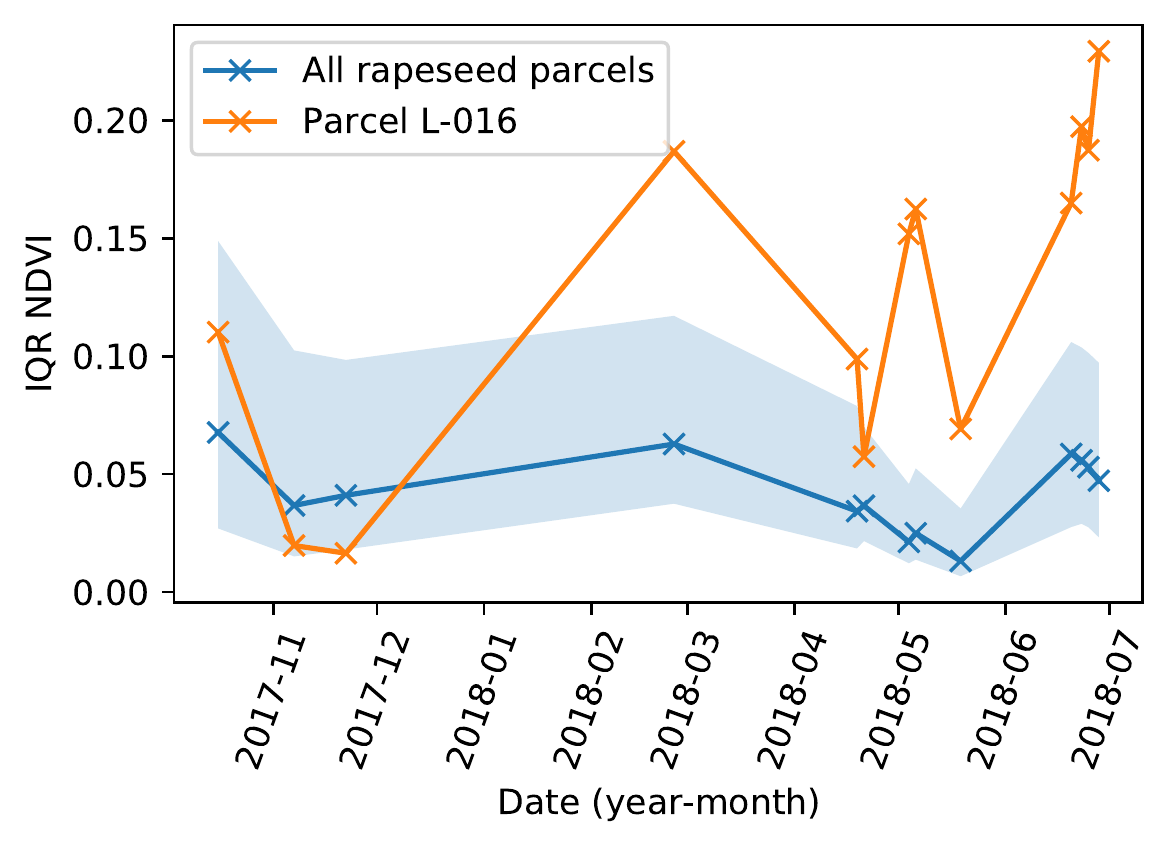}}
        \caption{Example of heterogeneity affecting a parcel. (a): true color S2 image in February. (b): Interquartile range (IQR) time series of the parcel NDVI. The blue line is the median value of the whole dataset. The blue area is filled between the 10th and 90th percentiles. The orange line is the IQR NDVI time series for the analyzed parcel.}
        \label{fig:ex_heter}
    \end{figure}
    
    \newpage
    \item \textit{Growth anomalies} are related to an abnormal development of the crop. The two main categories of growth anomalies are parcels with a low vigor (\textit{late growth}) or, on the contrary, with a high vigor (\textit{vigorous crop}). \autoref{fig:summary_late_anomalies} illustrates how the different growth anomalies can affect the median NDVI of the parcels within a growing season. \autoref{fig:SAR_late_growth} provides an example of growth anomaly where the S1 VH time series is affected by a late growth issue. As for heterogeneity, more transient growth anomalies, such as a delay in the flowering or senescence phase, can affect a crop parcel (Supplementary Figure S5).
    
    \begin{figure}[ht!]
        \centering
        \includegraphics[width=0.49\textwidth]{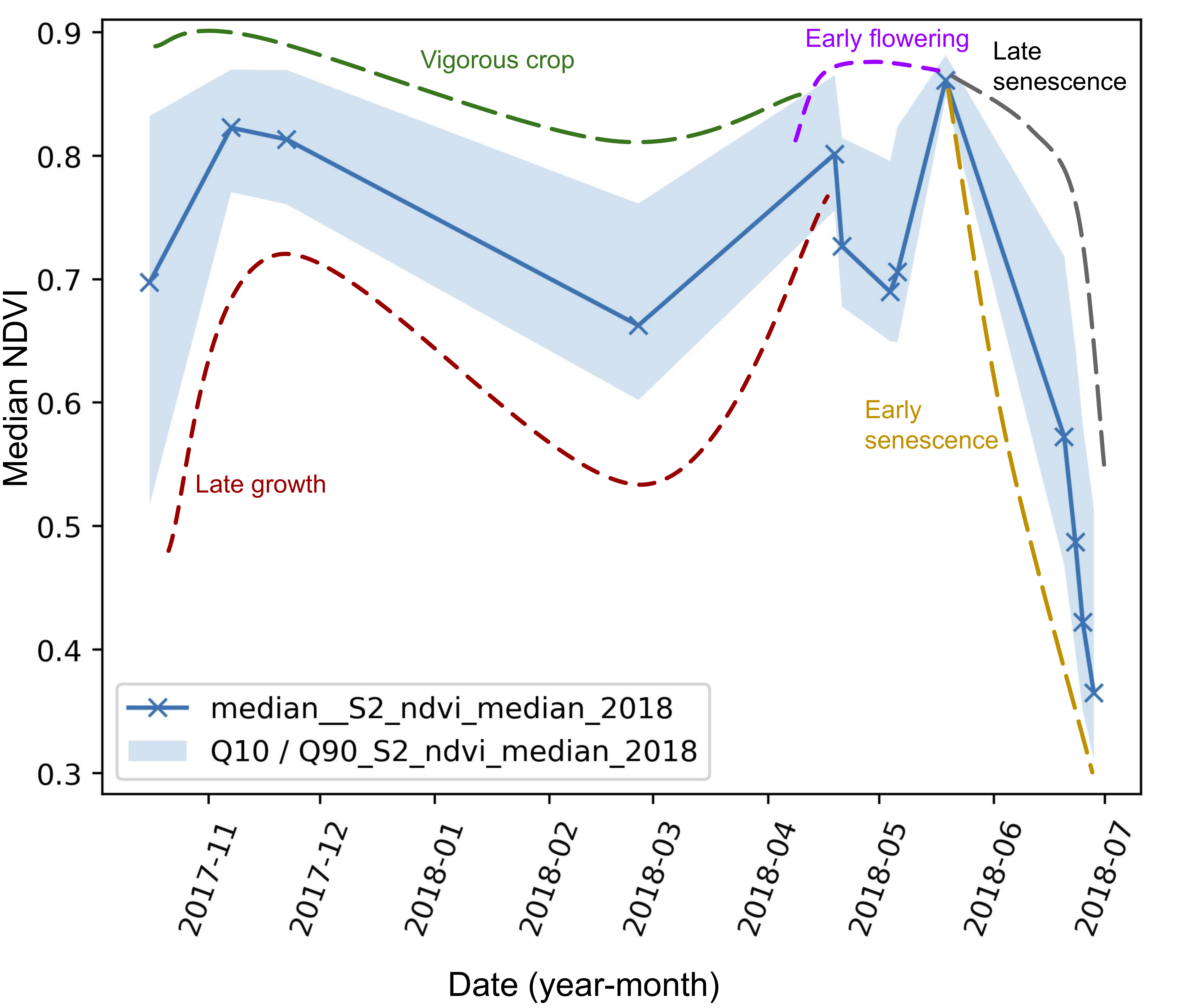}
        \caption{Illustration of the different growth anomalies that were detected and their potential influence on the median NDVI of the parcels (rapeseed crop). The blue line is the median value of the whole dataset. The blue area is filled between the 10th and 90th percentiles. Note that the labeling was conducted using all the S1 and S2 features (not only median NDVI).}
        \label{fig:summary_late_anomalies}
    \end{figure}

    \begin{figure}[ht!]
        \subfloat[]{\includegraphics[width=0.49\textwidth]{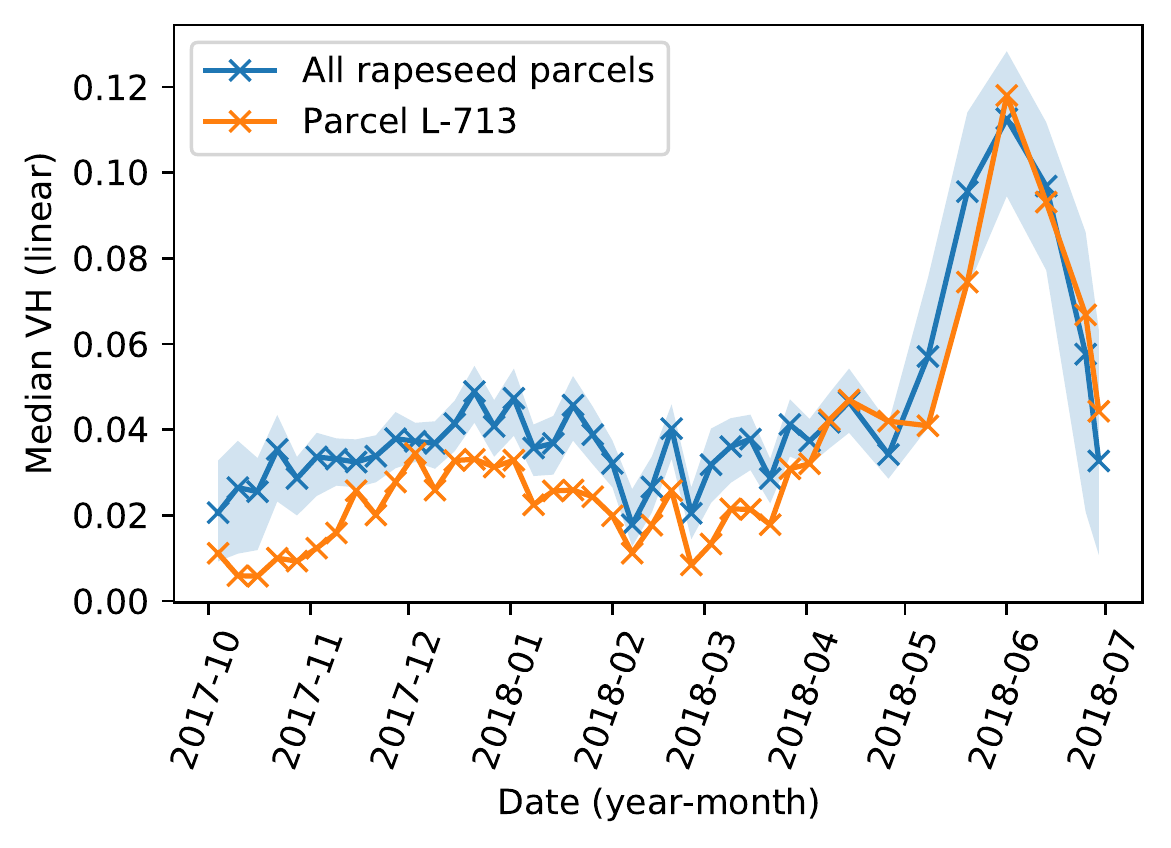}}
        \hfill
        \subfloat[]{\includegraphics[width=0.49\textwidth]{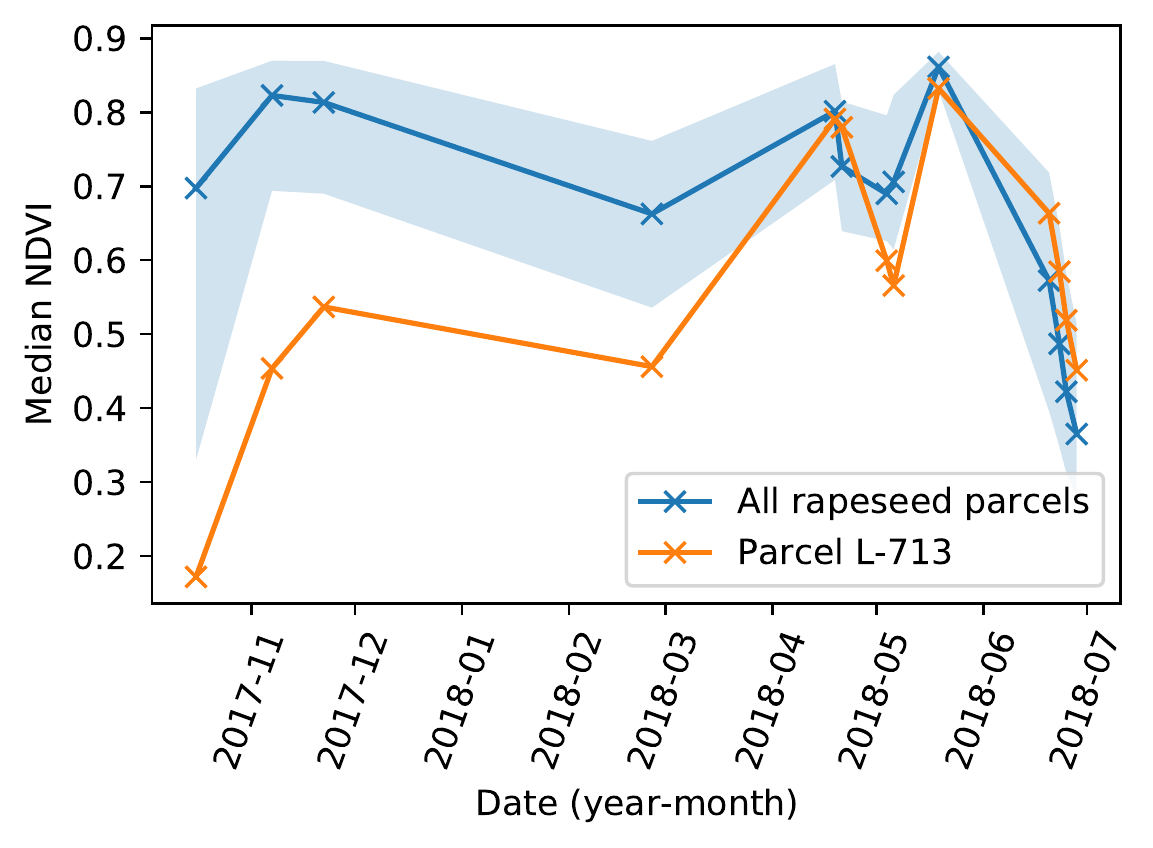}}
        \caption{Example of time series subjected to late growth for a rapeseed parcel: (a) median VH and (b) median NDVI for a rapeseed parcel. The blue line is the median value of the whole dataset. The blue area is filled between the 10th and 90th percentiles. The orange line corresponds to a specific parcel subjected to late growth.}
        \label{fig:SAR_late_growth}
    \end{figure}
    
    \item \textit{Database errors} are considered as relevant anomalies to be detected. This type of error is a common problem in large databases and can be challenging and time consuming to be detected manually. Examples of ``\textit{wrong shape}'' and ``\textit{wrong type}'' reported in the database are provided in \autoref{fig:img_wrong_shape_type}. This category of anomalies presents in general a strong sign of abnormality.
    
    \begin{figure}[ht!]
        \centering
        \subfloat[]{\includegraphics[width=0.49\textwidth]{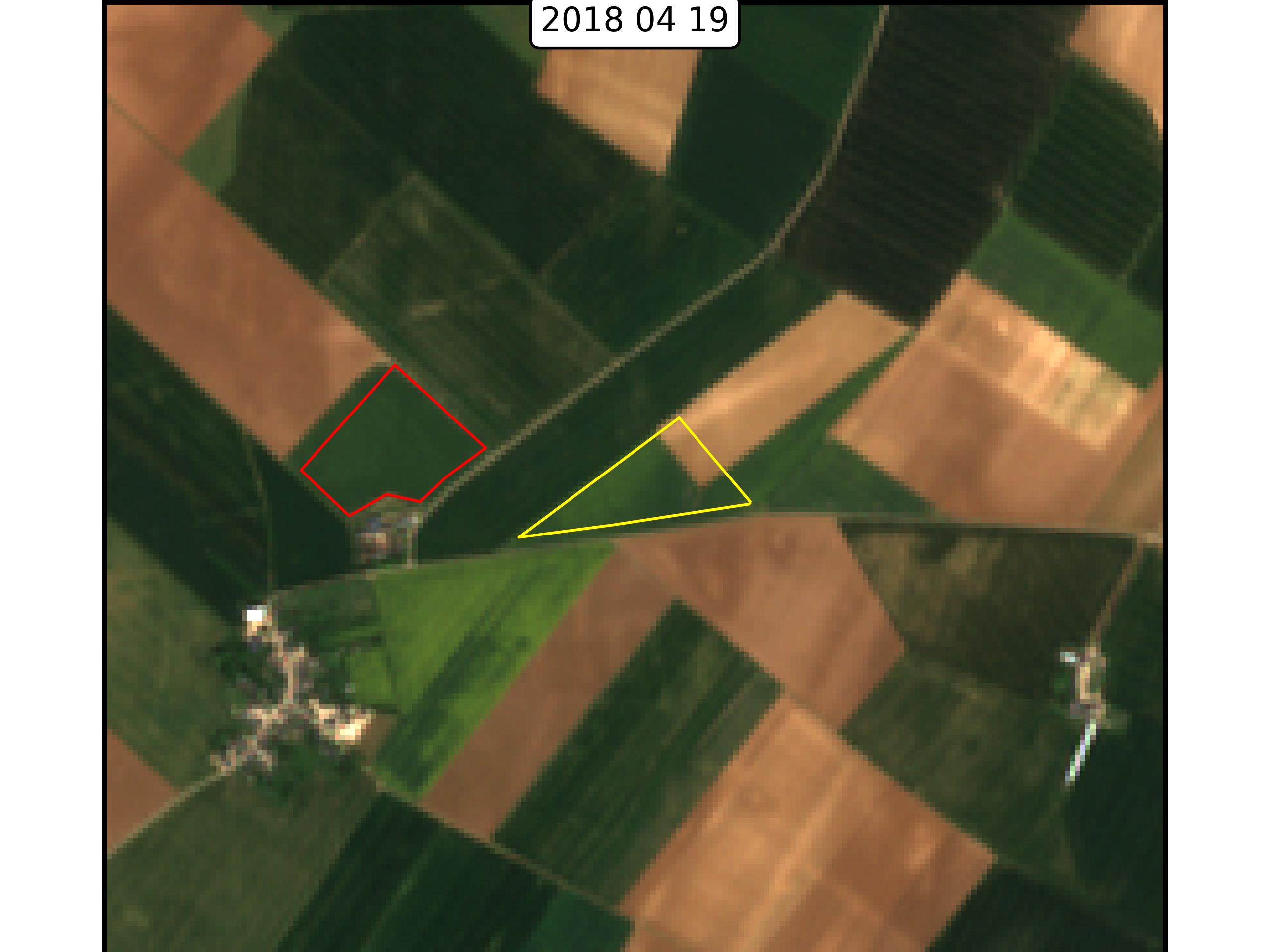}}
        \subfloat[]{\includegraphics[width=0.49\textwidth]{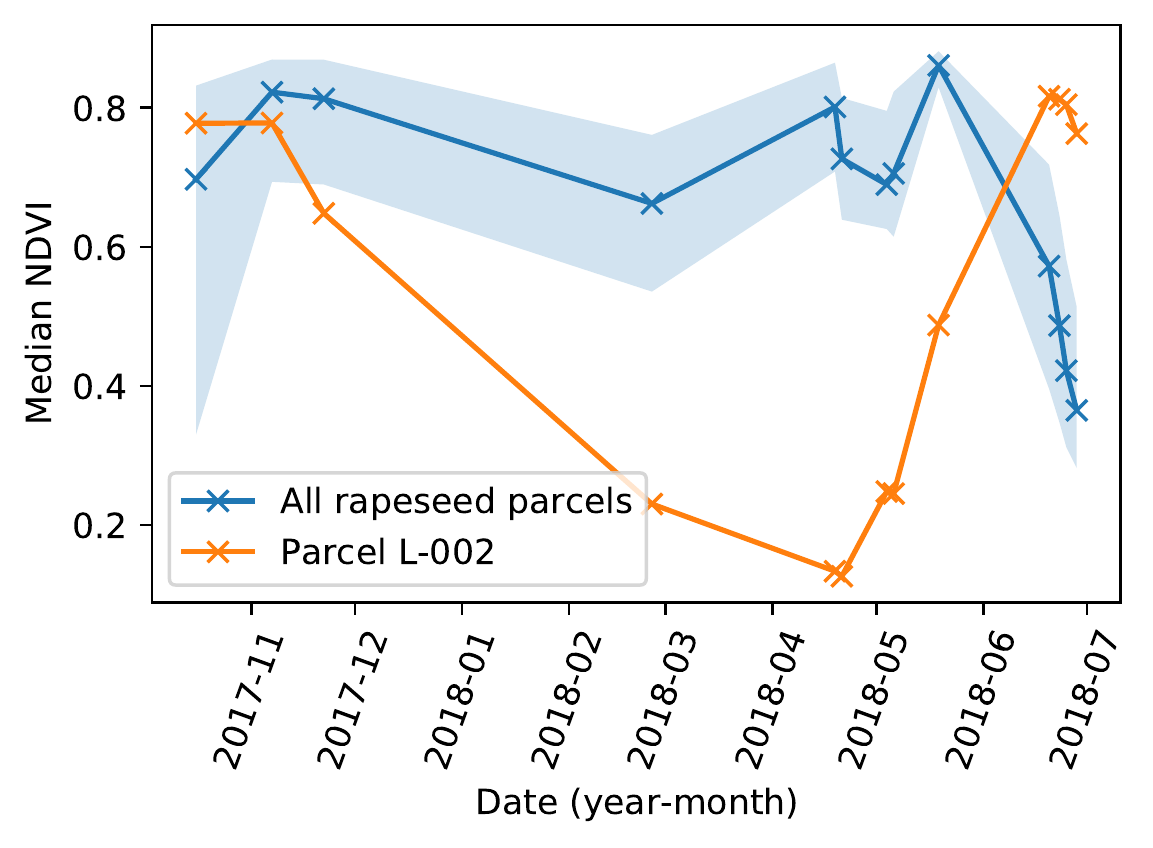}}
        \caption{Two examples of error in the parcel contour database (a): an error in the parcel delineation is visible (true color S2 image). (b): median NDVI time series for a parcel having a wrong crop type declared.}
        \label{fig:img_wrong_shape_type}
    \end{figure}
    
    \newpage
    \item The \textit{``Normal (checked)''} label was given to parcels that were labeled as normal after inspecting the features and images. In some cases, some few extreme values were observed explaining why the parcel was detected as abnormal by the outlier detection algorithms. In any case, all these parcels should have an outlier score (\textit{i.e.}, the score given by an outlier detection algorithm) lower than the parcels affected by agronomic anomalies (\textit{e.g.}, heterogeneity or growth anomaly).
    
    \item Other non-agronomic anomalies considered as false positives concern a few percentage of the analyzed parcels. Some very small parcels were still present in the dataset and are labeled as \textit{``too small''} (it is sometimes difficult to clean efficiently too small parcels that are long and narrow). Analyzing this type of parcels is not possible due to the spatial resolution of Sentinel data. These parcels were kept in the database to illustrate problems that can occur in practical applications. \textit{``Shadow''} is another kind of non-agronomic anomaly that can be caused by forests near the parcel (Supplementary Figure S6) or clouds that are not detected using the cloud mask.
    \item A subcategory of non-agronomic anomalies are \textit{``SAR anomalies''}. These anomalies correspond to parcels where SAR features have an abnormal time evolution in early growing season (\textit{i.e.,} the SAR indicators are abnormal compared to the rest of the data), whereas multispectral images and their features were counterchecked as normal. It is a known issue in crop monitoring with SAR data that was studied in \citet{Wegmuller2006, Wegmuller2011, Marzahn2012}, which is reported as a ``Flashing field'' phenomenon. These anomalies are considered as non-agronomic since SAR data are affected by other factors than the vegetation status such as soil moisture, soil structure, row orientation or soil roughness. This kind of anomalies was observed more frequently for wheat crops and in early growing season when there is a low vegetation cover. The ``flashing field'' terminology can easily be understood looking at the example displayed in \autoref{fig:SAR_anomaly_example}.
\end{itemize}

\begin{figure}[ht!]
\subfloat[]{\includegraphics[width=0.49\textwidth]{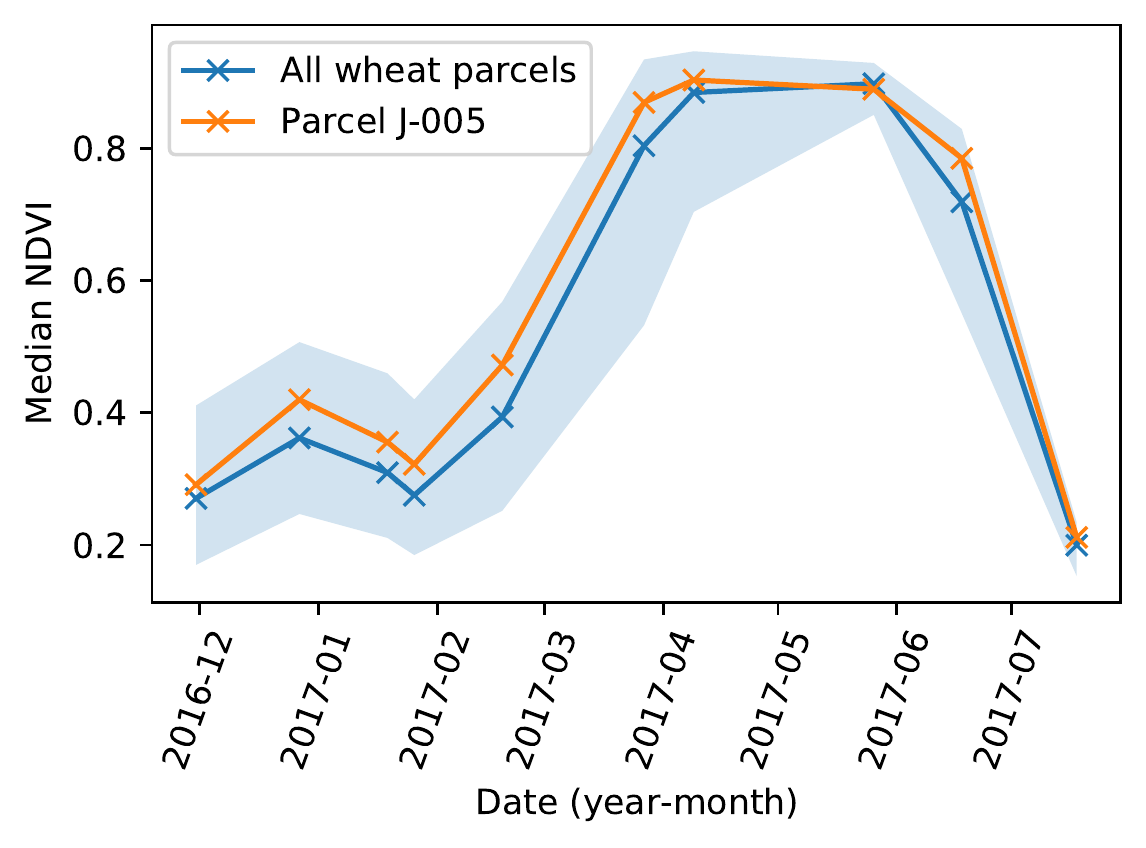}}
\hfill
\subfloat[]{\includegraphics[width=0.49\textwidth]{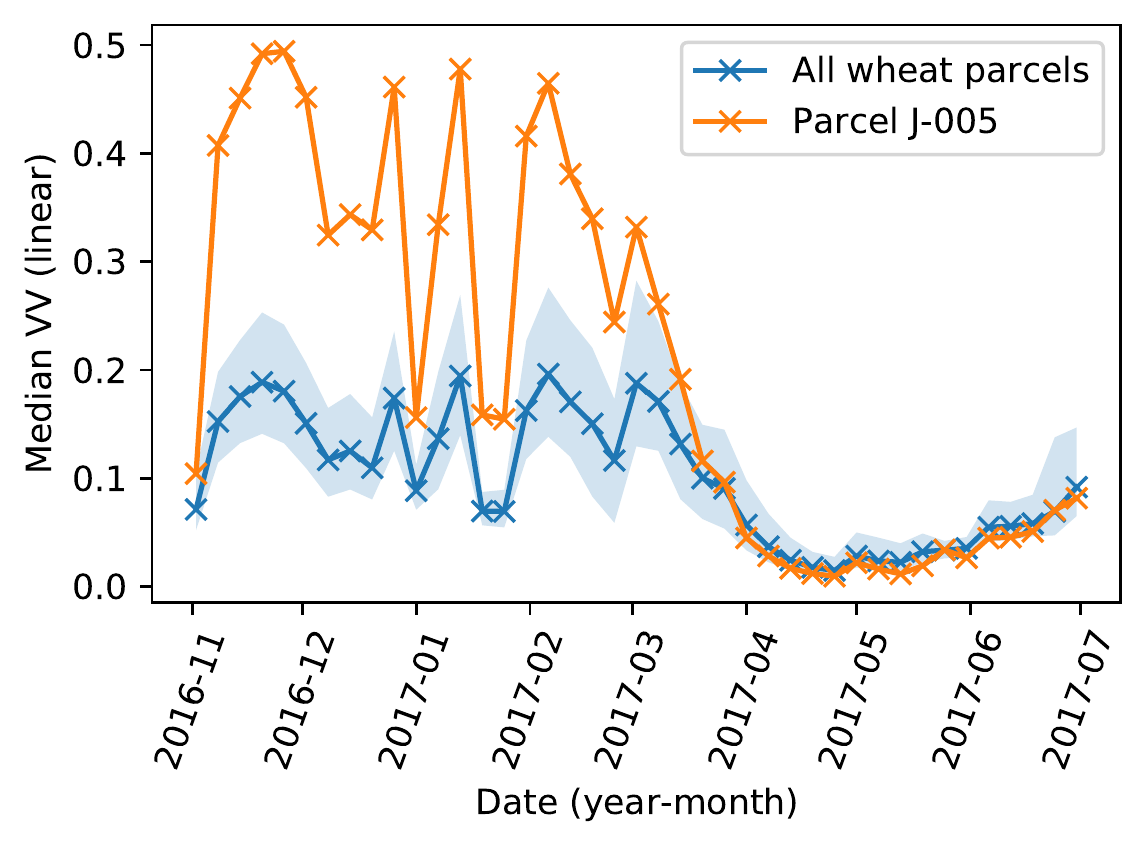}}
\\
\subfloat[]{\includegraphics[width=0.49\textwidth]{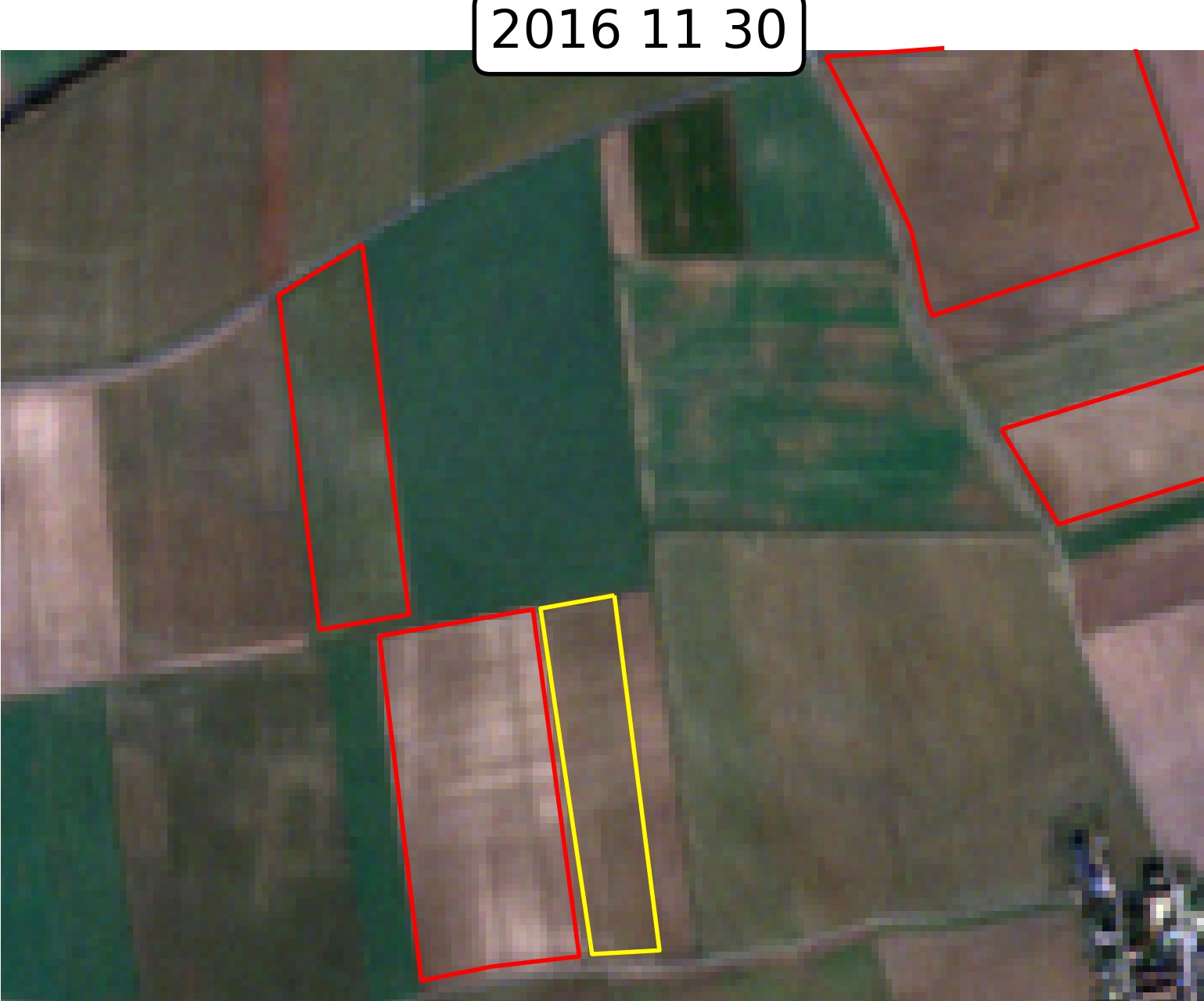}}
\hfill
\subfloat[]{\includegraphics[width=0.49\textwidth]{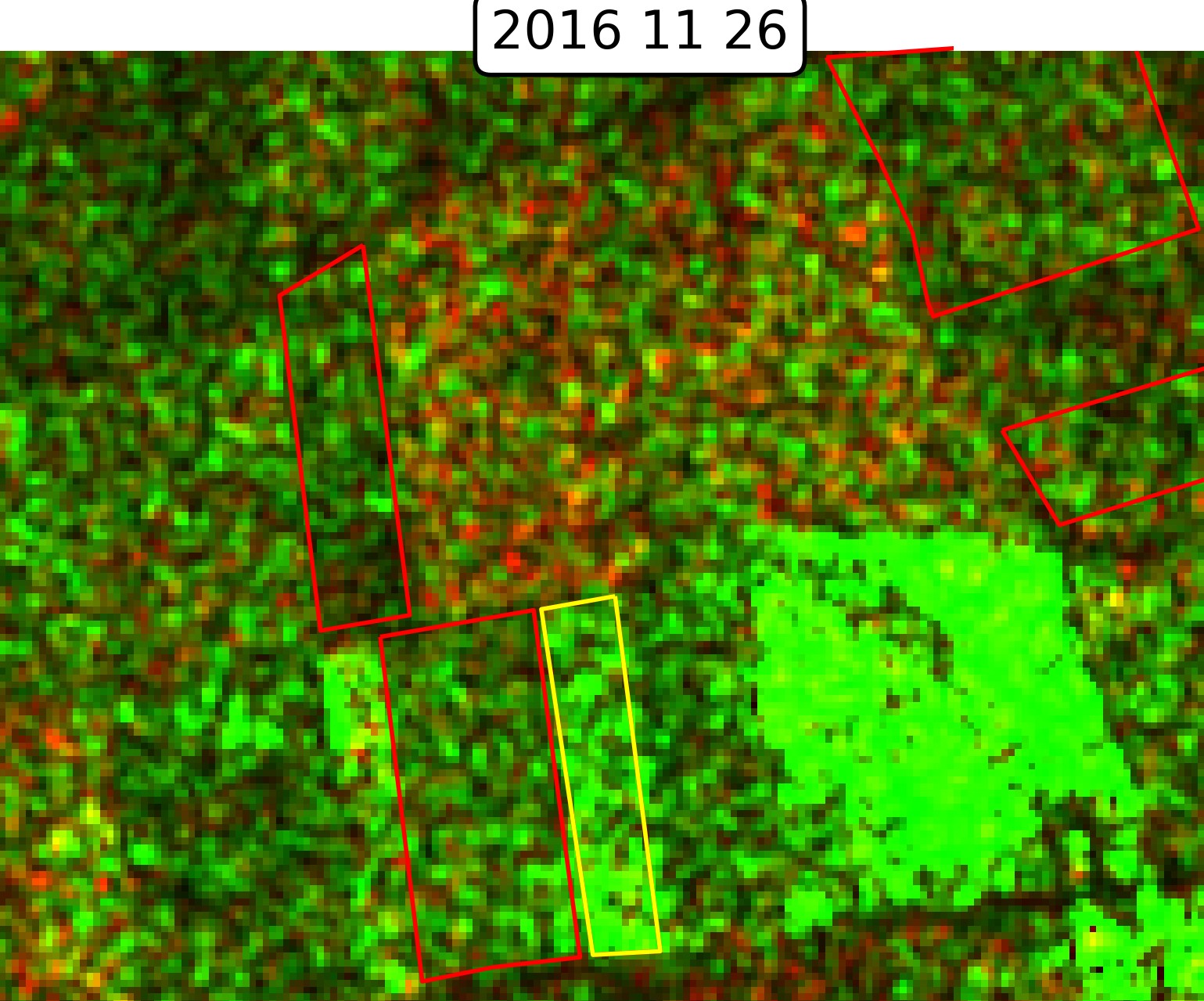}}
\caption{Time series of (a) median NDVI and (b) median VV polarization for a wheat parcel. The blue line is the median value of the whole dataset. The blue area is filled between the 10th and 90th percentiles. The orange line corresponds to a specific flashing-field parcel. Images acquired at the end of November: (c) true color S2 image and (d) S1 composite image (Green=VV, Red=VH).}
\label{fig:SAR_anomaly_example}
\end{figure}

\newpage
\subsubsection{Distribution of the outlier parcels in the two datasets}\label{sec:distribution_dataset}

\begin{figure}[ht!]
\subfloat[]{{\includegraphics[width=0.42\textwidth]{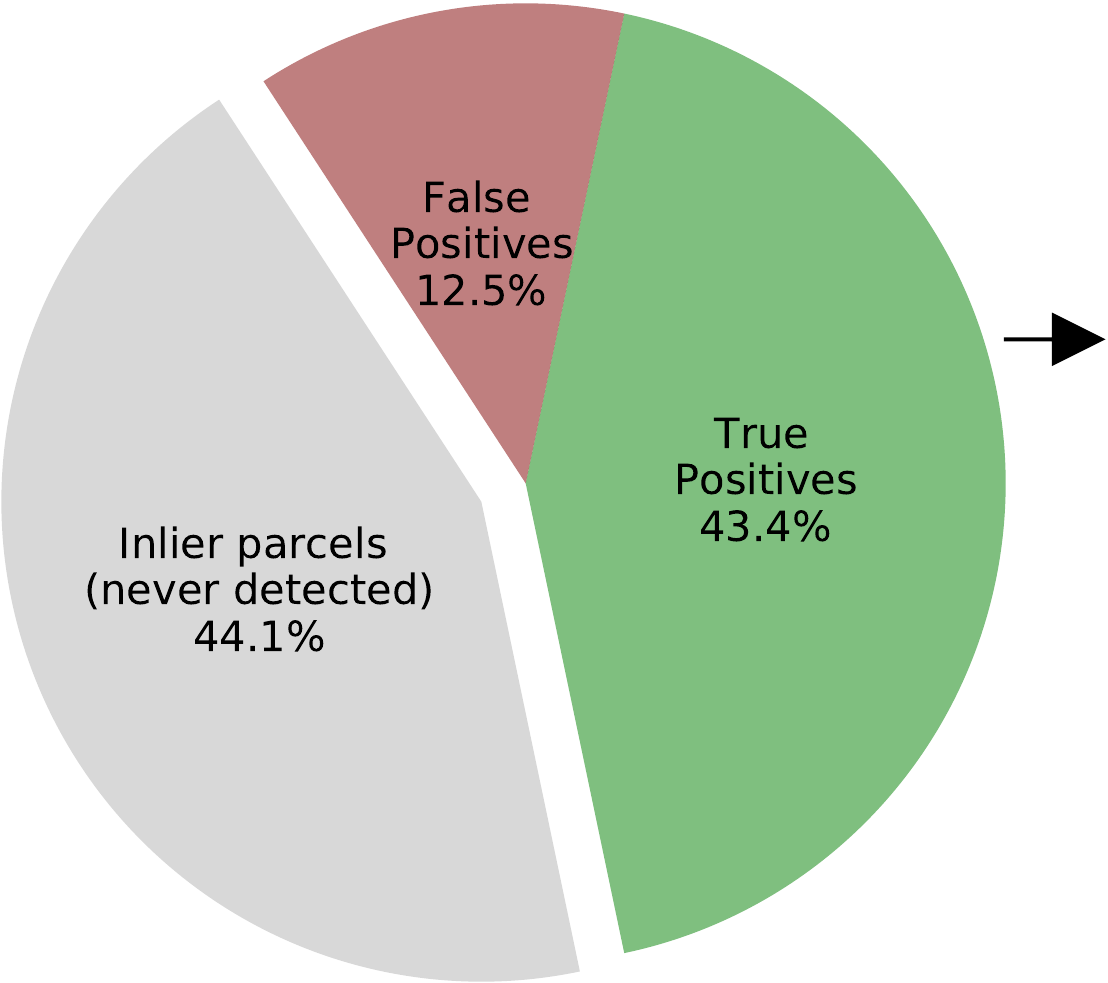}}}
\hfill
\subfloat[]{{\includegraphics[width=0.54\textwidth]{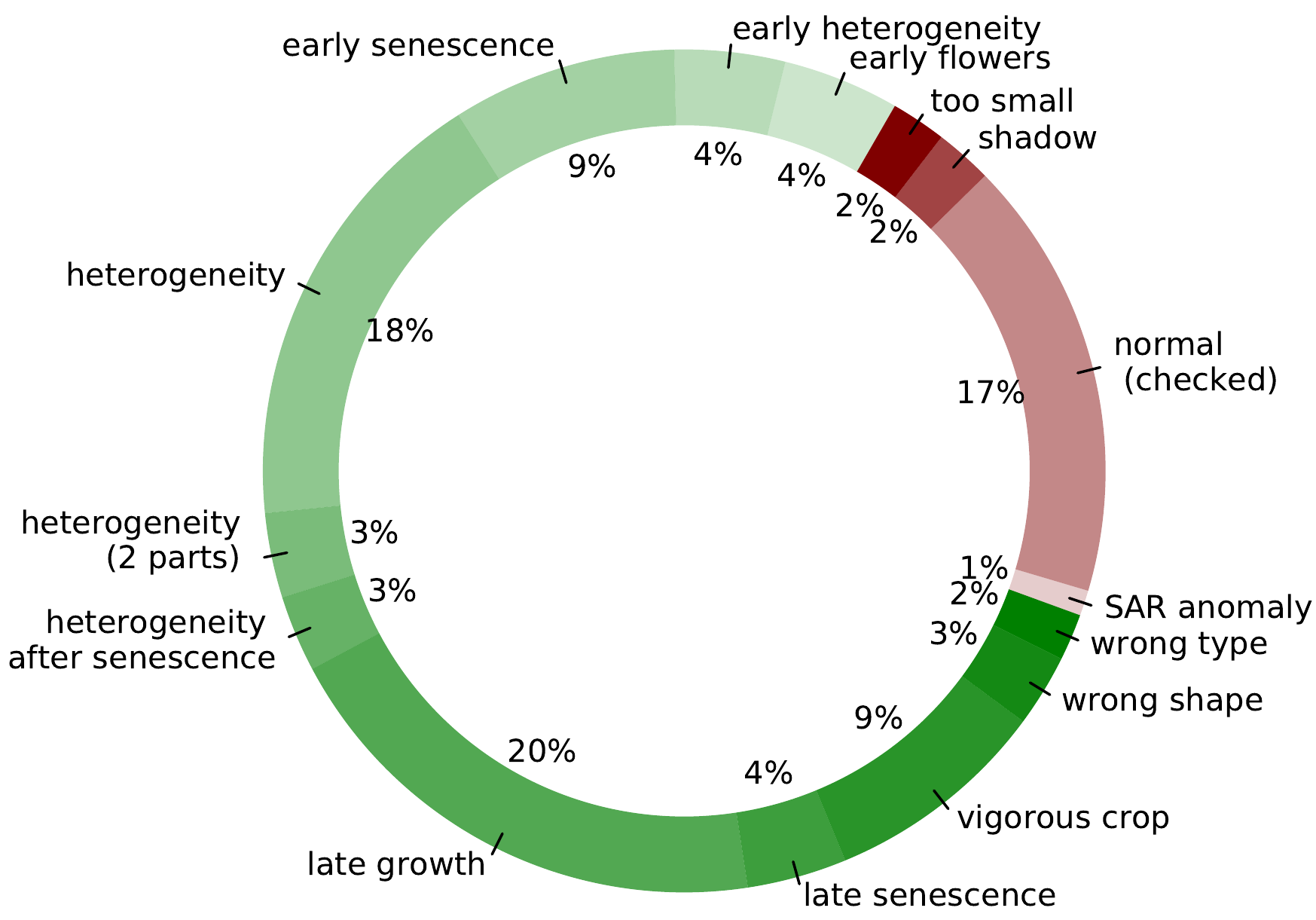}}}
\\
\subfloat[]{{\includegraphics[width=0.42\textwidth]{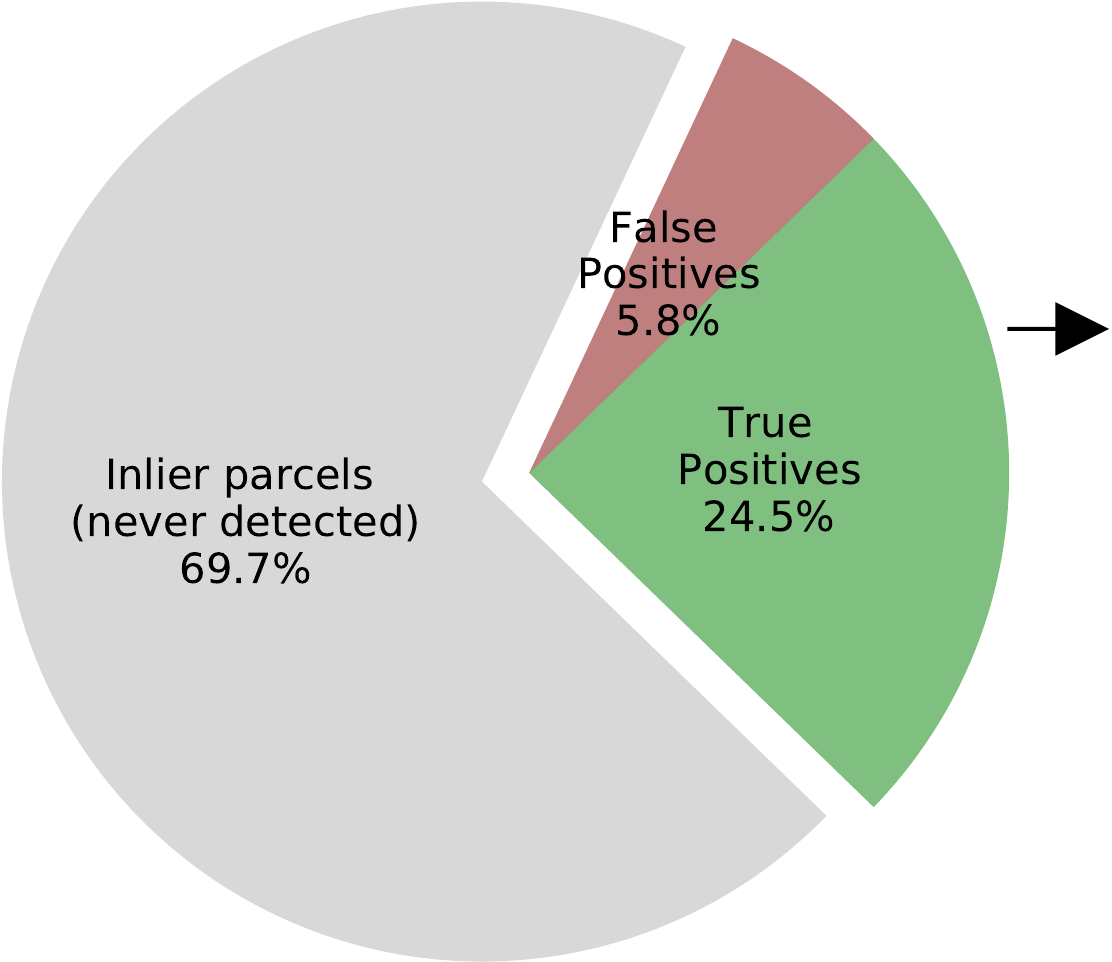}}}
\hfill
\subfloat[]{{\includegraphics[width=0.54\textwidth]{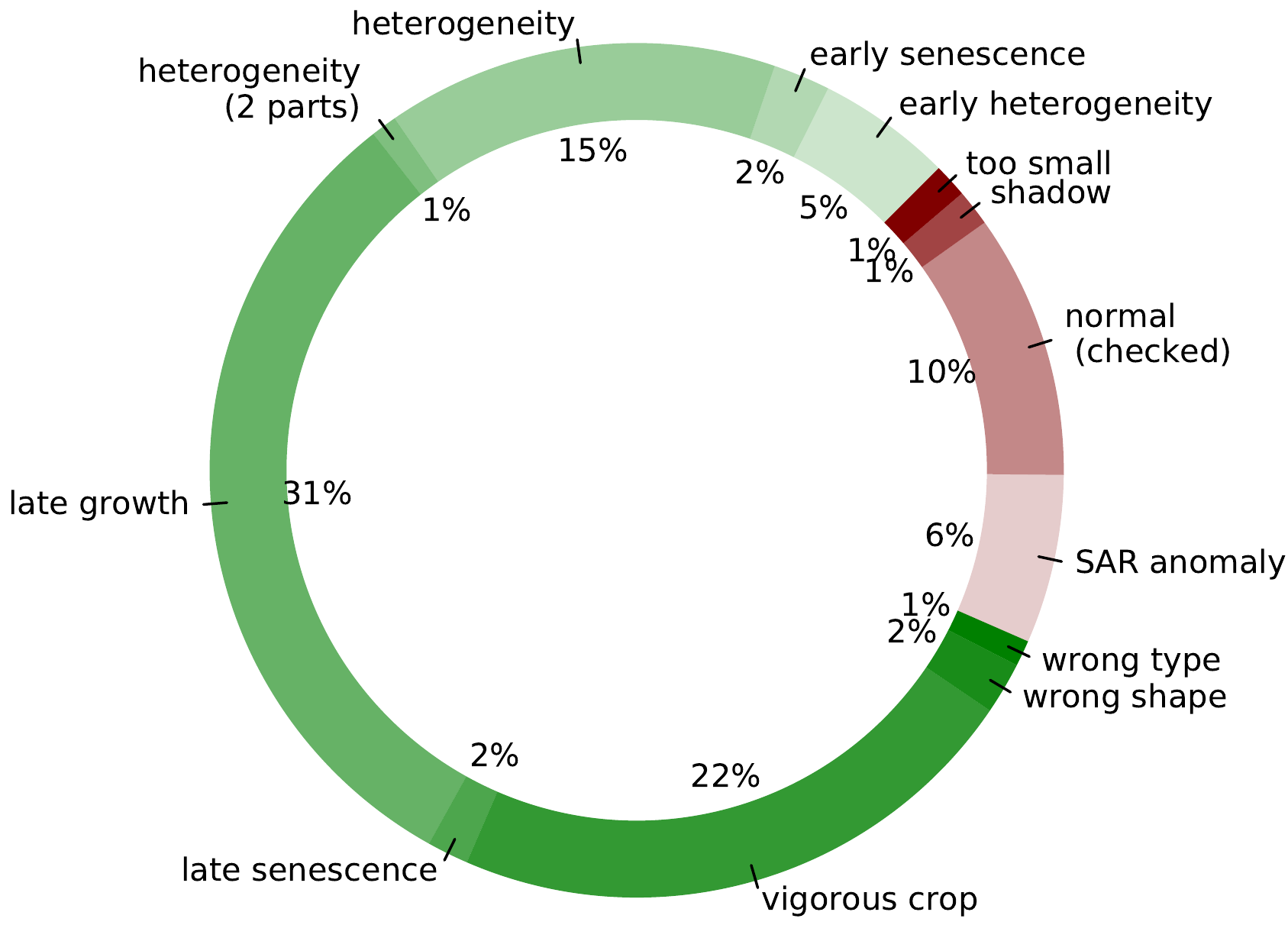}}}

\caption{Distribution of the outlier parcels (labeled by experts) and the inlier parcels (never detected) for (a, b) rapeseed crops and (c,d) wheat crops. Green categories correspond to relevant anomalies considered as true positives and red categories to non-agronomic anomalies considered as not relevant.}
\label{fig:pie_charts}
\end{figure}

\autoref{fig:pie_charts} summarizes the distribution of the anomaly categories for both wheat and rapeseed crops. Approximately 55\% of the rapeseed dataset was checked by the agronomic experts, ensuring that the outlier parcels analyzed in the study are representative. Similarly, 30\% of the most abnormal wheat parcels were checked to validate the relevance of our method when applied to another crop type. \autoref{fig:pie_charts} shows that heterogeneity and growth problems are the most detected anomalies for both types of crops. 

\newpage
\subsection{Performance evaluation}

The precision is used to evaluate the quality of a detection and is defined as
\begin{linenomath}\begin{equation}
    \text{precision} = \frac{\text{TP}}{\text{TP} + \text{FP}}
    \label{eq:precision_formula}
\end{equation}\end{linenomath}
where TP and FP are the numbers of true positives and false positives, respectively. The precision expresses the percentage of detected parcels that are true positives (here, agronomic anomalies checked by the experts).

The precision can be computed for different values of the outlier ratio. Plotting precision vs. outlier ratio curves is a good way to compare various detection results: for a given outlier ratio, a good algorithm or feature choice has generally detection results with a higher precision. These curves are similar to the Receiver Operating Characteristics (ROC) but have the advantage to be more adapted to outlier detection since the outlier ratio can be fixed. It was observed that both types of curves lead to the same conclusions, with an easier interpretation for the precision vs. outlier ratio curves. The area under the precision vs. outlier ratio curve (AUC) can be used to provide a quantitative measure of detection performance summarizing the information contained in the whole curve. In the analysis, we computed the AUC for outlier ratios in the range [0, 0.5] in order to focus on realistic values of the percentage of outliers. The AUC was then divided by 0.5 to normalize the obtained value: the resulting score can be seen as the average precision for outlier ratios in the range [0, 0.5]. Note that this representation does not give information regarding the distribution of the different detected categories since two algorithms can have the same precision without detecting the same parcels (\textit{e.g}, one algorithm can detect more heterogeneous parcels whereas another one detects more late growth anomalies). Using the distribution of the different types of anomalies detected for a given outlier ratio is a complementary way to address this limitation.

\section{Results and discussion}\label{sec:results}

The different feature combinations tested in this section are identified in the figures using abbreviations that are defined in \autoref{table:table_abbrev_names}. 

\begin{table}[ht!]
\caption{Abbreviations used with their corresponding sets of features used for outlier detection. Each abbreviation can be read as follows: ``S2: pixel-level features (parcel-level statistics), S1: pixel-level features (parcel-level statistics)''.}
\footnotesize
\begin{tabular}{p{60mm}p{70mm}}
    \\
    \hline
    Abbreviated name & Features used \\
    \hline & \\
    S1: VV, VH (median) & Median of S1 features listed in Section~\ref{sec:feat_extract} \\ & \\
    S2: all (median / IQR) & Median and IQR of all S2 features listed in Section~\ref{sec:feat_extract} \\ & \\ 
    S2: all (median / IQR), S1: VV, VH (median) & Median and IQR of all the S2 features and median of the 2 S1 features VV and VH. \\ & \\
    \hline
\end{tabular}
\label{table:table_abbrev_names}
\end{table}

\subsection{Anomaly detection results for rapeseed crops}\label{sec:results_rapeseed}

The results presented in this section were conducted by analyzing the complete rapeseed dataset with the IF algorithm. First, the outlier detection is conducted using S1 features only, since SAR data are available permanently through all the crop cycle, which is important for crop monitoring applications. Then, the effect of using S2 features only is investigated. Finally, S1 and S2 features are used jointly to study the effect of combining the contribution of both types of sensors.

\subsubsection{Outlier detection with S1 features}

\begin{figure}[ht!]
    \centering
    \includegraphics[width=0.8\textwidth]{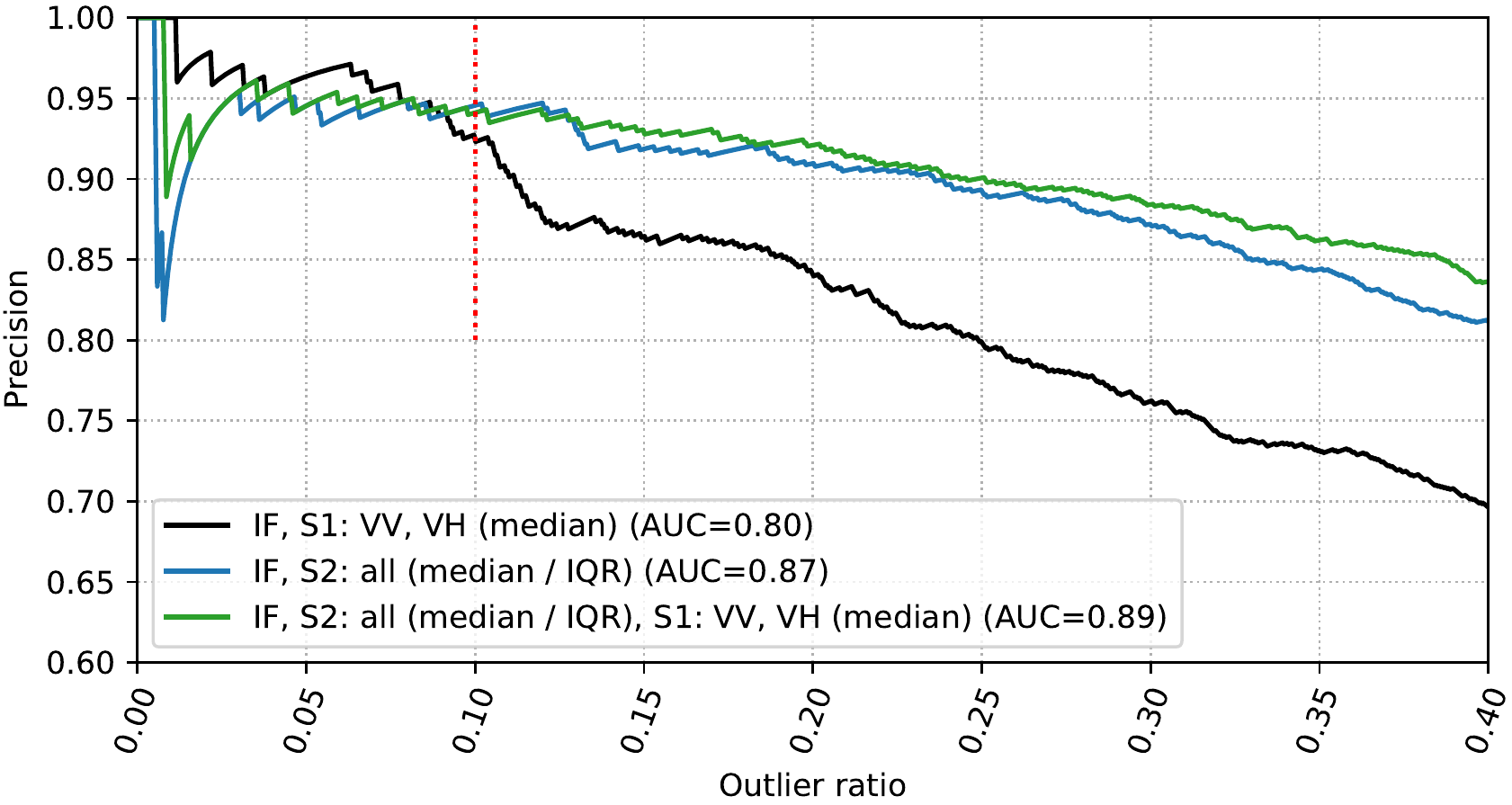}
    \caption{Precision vs. outlier ratio using the IF algorithm on the rapeseed parcels. Black: S1 features only, blue: S2 features only, green: S1 and S2 features jointly. The red line corresponds to the outlier ratio used in \autoref{fig:histo_main_rapeseed}}
    \label{fig:POR_main_rapeseed}.
\end{figure}

\begin{figure}[ht!]
    \subfloat[]{\includegraphics[width=1\textwidth]{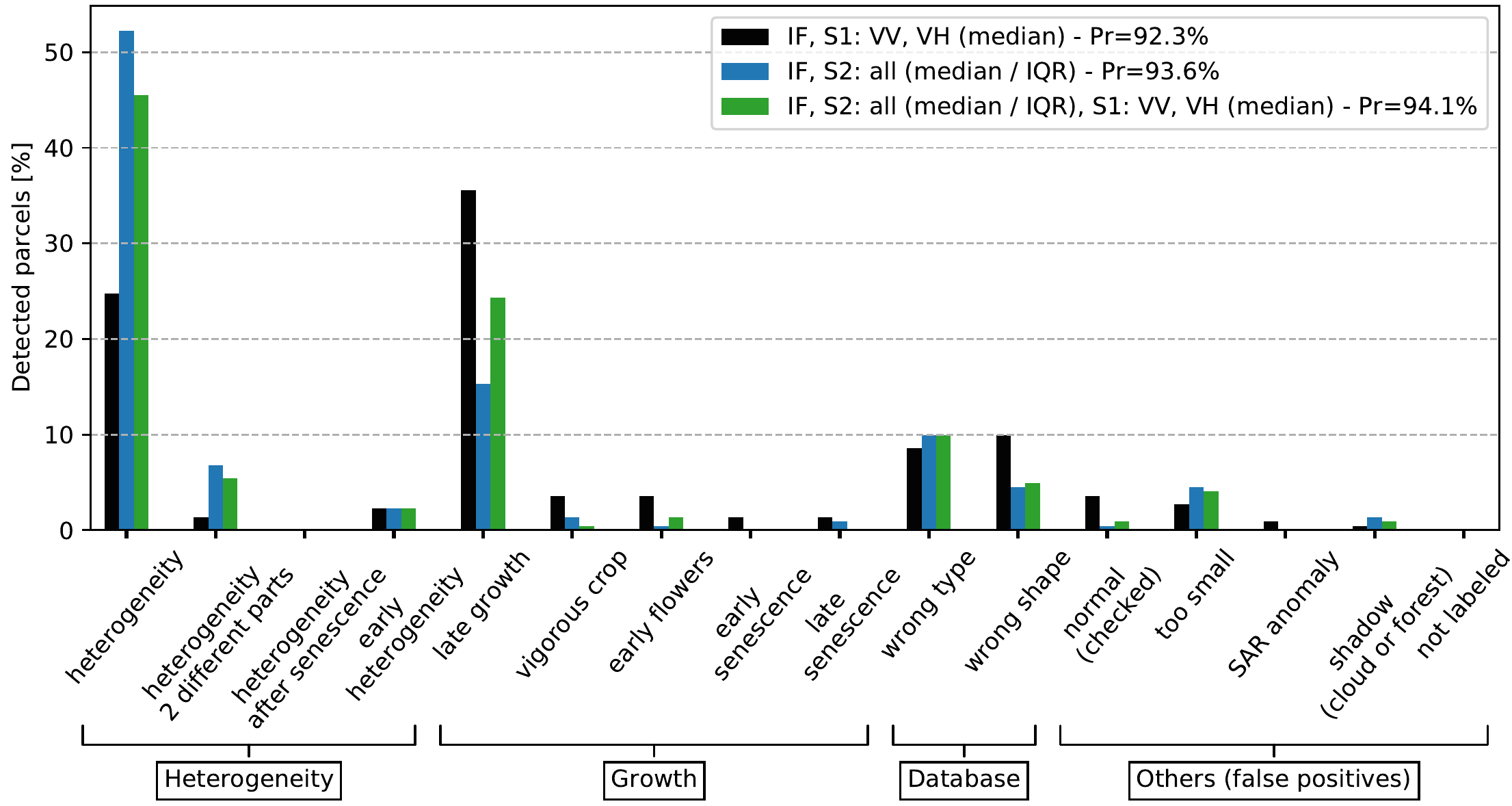}} \\
    \subfloat[]{\includegraphics[width=1\textwidth]{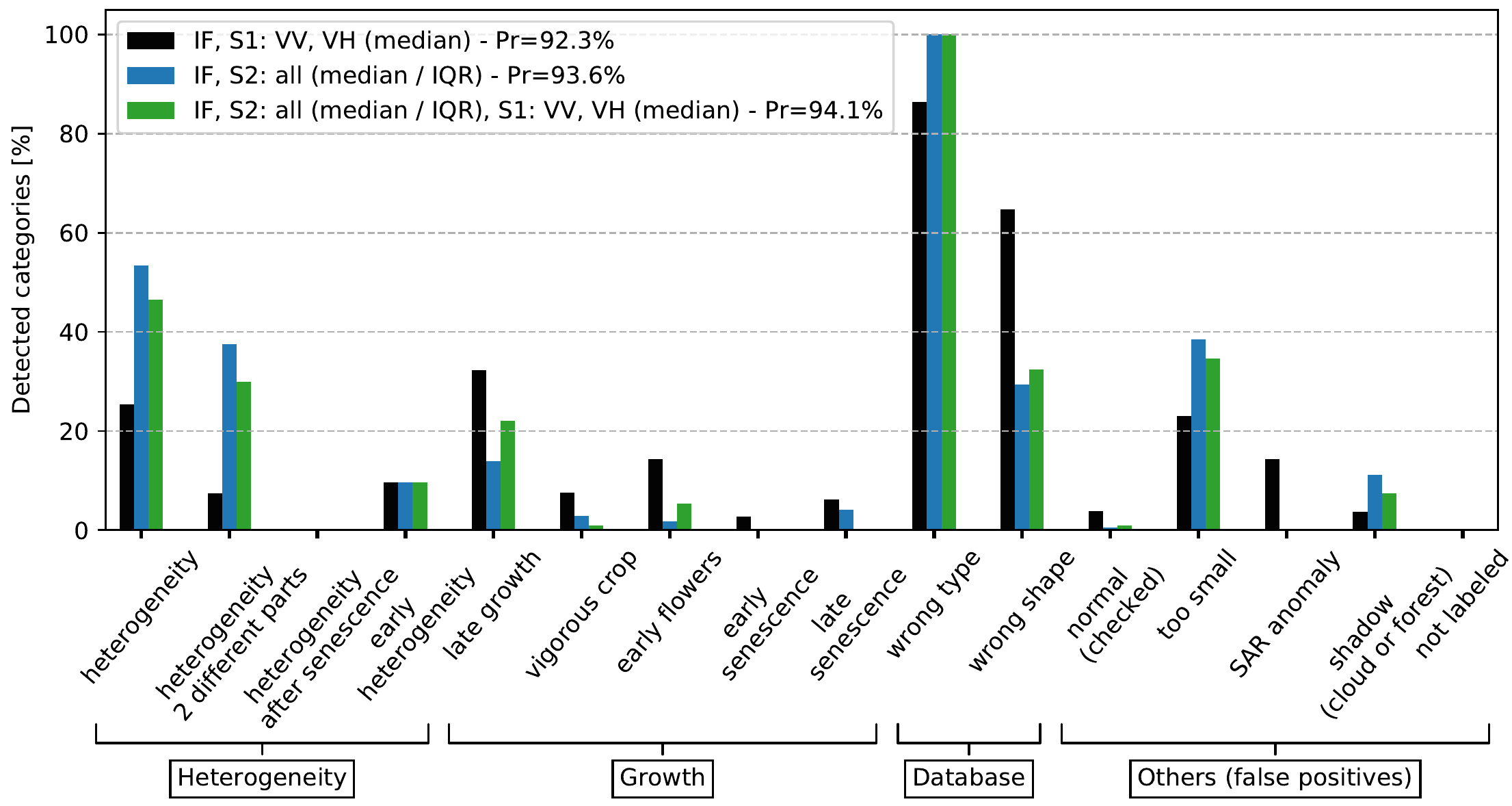}}
    \caption{(a) 100$\times$(Number of detected parcels in each category / Number of detected parcels) (b) 100$\times$(Number of detected parcels in each category / Number of parcels in each category). The analysis is conducted using rapeseed parcels with an outlier ratio equal to 10\% and the IF algorithm. Black: S1 features only, blue: S2 features only, green: S1 and S2 features jointly. The precision (Pr) of the results for each feature set is added in the legend.}
    \label{fig:histo_main_rapeseed}
\end{figure}

The strength of S1 data for crop anomaly detection is confirmed when analyzing \autoref{fig:POR_main_rapeseed} (black curve): the precision is equal to 92.3\% for an outlier ratio fixed to 10\%. For lower outlier ratios, the precision obtained when using S1 features is slightly higher than the precision obtained when using S2 features (which will be discussed later). For higher outlier ratios, the precision decreases (more false positives are detected) but remains close to 85\% for an outlier ratio equal to 20\%. These results highlight the ability of the IF algorithm to provide relevant outlier scores: the parcels with the highest outlier scores are more likely to be true positives. \autoref{fig:histo_main_rapeseed}(a) shows the distribution of the detected parcels in the different anomaly categories. The majority of the detected parcels are affected by \textit{late growth} (35\%) and \textit{heterogeneity} (25\%). Anomalies coming from an error in the database (\textit{wrong shape} and \textit{wrong crop type} reported) are also largely detected (18.5\%). To further investigate these results, \autoref{fig:histo_main_rapeseed}(b) depicts for each category the percentage of detected parcels. All parcels of the category \textit{wrong type} are detected, which can be understood since this anomaly strongly affects the features at a parcel-level. Using S1 features leads to detect more parcels of the category \textit{wrong shape} when compared to using S2 features. A similar observation can be done for \textit{vigorous crops} and \textit{early flowering} to a lesser extent (for this outlier ratio, only few of these transient anomalies are detected).

\newpage
\subsubsection{Outlier detection with S2 features}

Although S2 time series have lower temporal resolution when compared to S1 time series, they are useful for outlier analysis as shown in \autoref{fig:POR_main_rapeseed} (blue curve). For an outlier ratio fixed to 20\%, the precision of the detection obtained using S2 features only is still above 90\%. Moreover, the average precision for outlier ratios in the range [0, 0.5] is equal to 87\% whereas it is 80\% when using S1 features only. For a complete growing season, having 13 S2 images is sufficient to detect a majority of relevant anomalies. However, it appears that S1 and S2 features tend to detect different types of anomalies as highlighted in \autoref{fig:histo_main_rapeseed}(a). When using S2 features, the IF algorithm detects a majority of \textit{heterogeneous} parcels (52\%) and less \textit{late growth} parcels (15\%). This observation justifies the joint use of S1 and S2 features, which is investigated below. \autoref{fig:histo_main_rapeseed}(b) shows that 40\% of the parcels affected by two parts heterogeneity are detected when using S2 features (only 10\% are detected when using S1 features). Moreover, a larger amount of \textit{too small} parcels are detected when using S2 features (around 40\% whereas it is close to 20\% when using S1 features). This last observation should be put in perspective with the small amount of parcels belonging to this category (less than 5\% of the detected parcels).

\subsubsection{Outlier detection with S1 and S2 features}

One of the main objectives of this study is to investigate the joint use of S1 and S2 for outlier detection in agricultural crops. \autoref{fig:POR_main_rapeseed} (green curve) shows that the average precision obtained when using S1 and S2 features jointly is close to 89\%, which is the best performance obtained for a complete growing season analysis of the rapeseed parcels. This result means that a larger amount of relevant anomalies are detected for a given outlier ratio when compared to using S1 or S2 features separately. Moreover, it also means that the IF algorithm is able to use efficiently the characteristics of each sensor. \autoref{fig:histo_main_rapeseed}(a) shows that using S1 and S2 features jointly allows the contribution of each sensor to be accounted. In particular, late growth anomalies are more detected when compared to using S2 features only (24\% vs. 15\% of the detected parcels) and heterogeneous parcels are more detected when compared to using S1 features only (45\% vs. 25\% of the detected parcels). These observations are confirmed by \autoref{fig:histo_main_rapeseed}(b).

Overall, the best combination of features obtained throughout the study consists in using S1 and S2 features jointly. This combination exploits the strength of each sensor for crop monitoring. To be more specific, on the one hand some heterogeneous parcels are not impacting the features extracted from S1 images since this sensor is not sensitive to the color of the crop parcels. On the other hand, some anomalies affecting the crop growth are impacting more clearly the S1 time series that are more sensitive to the vegetation structure. Moreover, since S1 time series are dense, it is in some cases easier to detect late growth or senescence problems (\textit{e.g.}, as mentioned for the wheat crop analysis where only few S2 images were available during the senescence phase). These results are confirmed in what follows when analyzing a different crop type.

\subsection{Extension to wheat crops}\label{sec:results_wheat}

\begin{figure}[ht!]
\centerline{\includegraphics[width=1\textwidth]{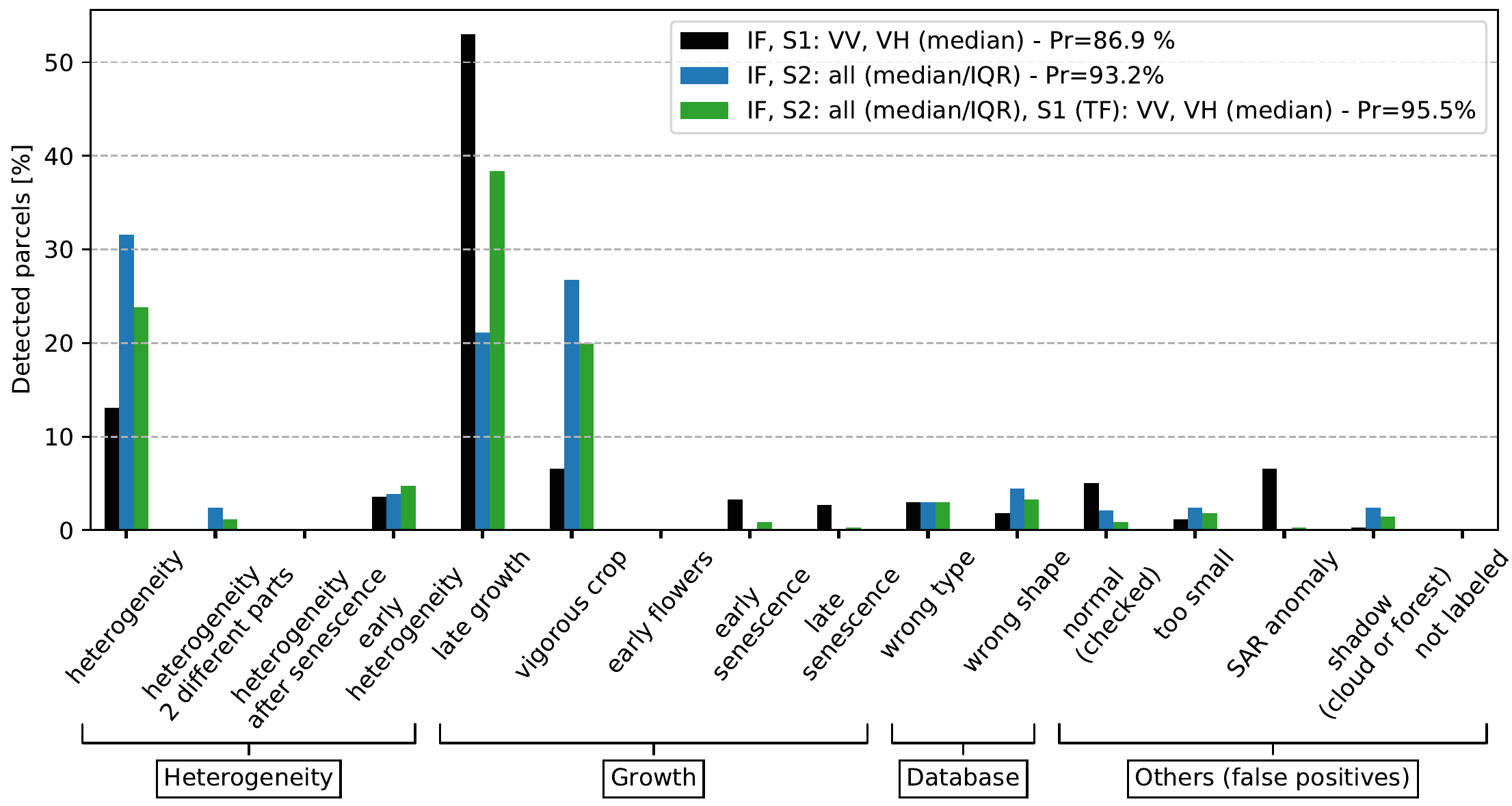}}
\caption{100$\times$(Number of detected parcels in each category / Number of detected parcels). The analysis is conducted using wheat parcels with an outlier ratio equal to 10\% and the IF algorithm. Black: S1 features only, blue: S2 features only, green: S1 and S2 features jointly. The precision (Pr) of the results for each feature set is added in the legend.}
\label{fig:histo_all_season_all_wheat}
\end{figure}

A complementary analysis was conducted to measure the robustness of the proposed method to a change in the crop type. An experiment is presented with the selection of the best features used for rapeseed crops analysis, \textit{i.e.}, all the features listed in \autoref{table:table_abbrev_names}. The IF algorithm was used to detect abnormal wheat parcels for a complete growing season with an outlier ratio of 10\%. The distribution of the detected anomalies in the different categories is depicted in \autoref{fig:histo_all_season_all_wheat}, which also indicates the precision obtained for each detection. Again, combining S1 and S2 data leads to the best precision (95.5\%). Similar to rapeseed crops, using S1 data allows more growth anomalies to be detected when compared to S2 data only. The precision obtained using S1 features only is lower due to a higher number of SAR anomalies (i.e., $22$ SAR anomalies) but the results are still accurate (precision=86.9\%). As for the rapeseed analysis, no SAR anomaly is detected when using S1 and S2 data jointly. Finally, since less S2 images were available during the senescence phase, using S1 features logically leads to better detect problems affecting this growing phase and confirms the interest of using both types of features. These results confirm the interest of the proposed approach and its robustness to changes in the crop type. 

Some differences were observed after analyzing the results obtained for rapeseed and wheat crops. These differences are interesting to analyze since they provide specific information for the monitoring of each crop type. For the wheat crops, the percentage of detected heterogeneous parcels is lower: when using S2 features, 31\% of the detected wheat parcels belong to this category whereas 52\% of heterogeneous parcels are detected for the rapeseed crops. On the other hand, the amount of detected vigorous parcels is higher (28\% when using S2 features only) whereas only few vigorous parcels were detected during the rapeseed analysis. It is also interesting to note that these parcels are more easily detected using S2 features only whereas late growth anomalies are still detected in higher proportion (52\%) when using S1 features only.

\begin{figure}[ht!]
\subfloat[]{\includegraphics[width=0.49\textwidth]{figs/example_anomalies/late_growth_ndvi_median_rapeseed_L-713.pdf}}
\hfill
\subfloat[]{\includegraphics[width=0.49\textwidth]{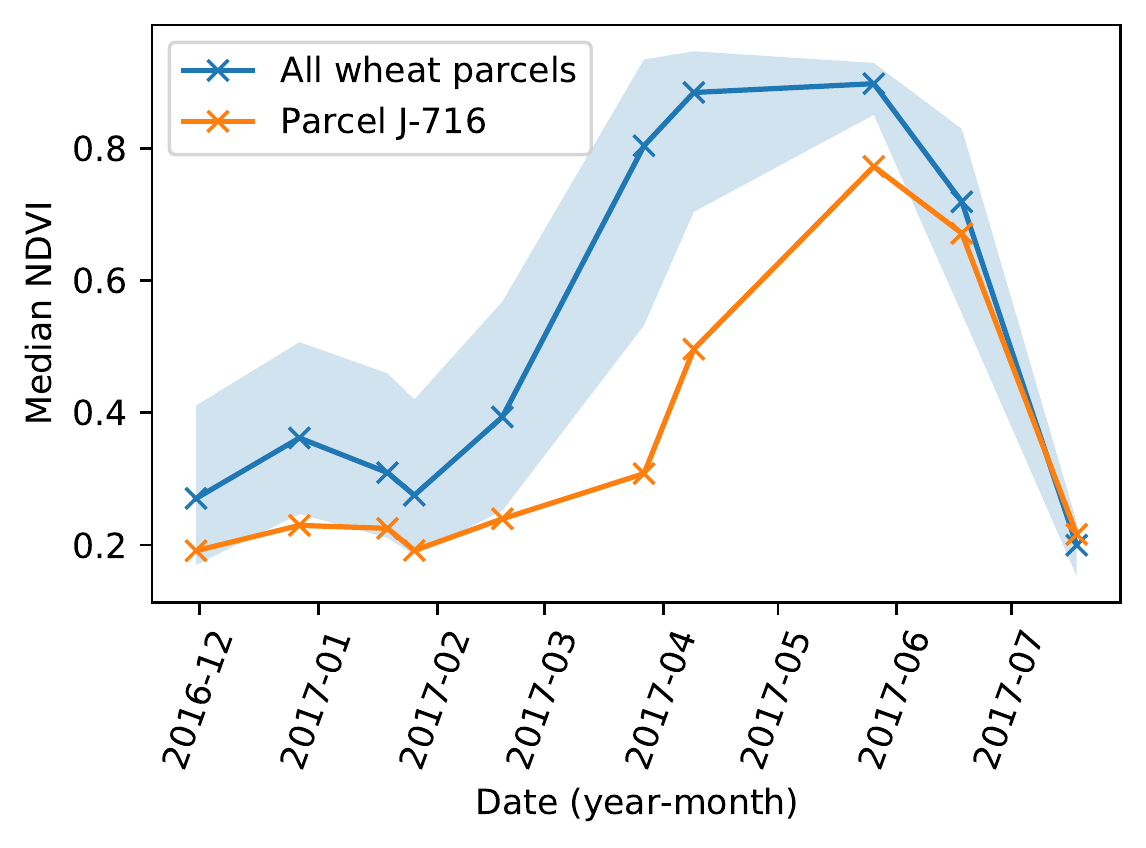}}
\caption{Median NDVI for late growth parcels (a) rapeseed parcel and (b) wheat parcel. The blue line is the median value of the whole dataset. The blue area is filled between the 10th and 90th percentiles. The orange line is a specific parcel analyzed.}
\label{fig:time_series_late_growth_rapeseed}
\end{figure}

The fact that more late growth anomalies have been detected for wheat parcels is coherent with the observations made during the labeling, where it was noticed that late growth problems frequently have a bigger impact on the wheat parcels. A representative example is provided in \autoref{fig:time_series_late_growth_rapeseed}: the rapeseed parcel affected by a late growth anomaly has a normal vigor after the flowering phase, whereas the wheat parcel has a low vigor for the complete growing season. It was also observed that few abnormally vigorous parcels have been detected among the rapeseed dataset: this could be related to an early sowing date and a high vigor shortly after the plant emergence as pointed out in \citet{VELOSO2017415}. Finally, the fact that few abnormally vigorous wheat parcels have been detected when using S1 features only is also coherent with the observations made in \citet{VELOSO2017415}, where it was highlighted that the SAR signal remains stable during early growing season whereas the NDVI starts increasing after the emergence of the plant.

\subsection{Influence of other factors on the detection results}\label{sec:discussion}

Various other factors that can influence the detection results were analyzed in complement of the experiments presented in this document. A summary of the influence of each factor is available at the end of this section in \autoref{table:table_disc_summary}.

Regarding the outlier detection algorithms, different methods for detecting anomalies lead to similar results for a complete growing season analysis (Supplementary Figures S9 and S10). However, we think that the IF algorithm is better suited for two main reasons: 1) it provides a higher precision overall during the various experiments, showing its robustness to different changes in the initial configuration, 2) the fact that no tuning is necessary is of crucial importance since the detection results are relevant without a preliminary analysis of the data. This does not hold true for autoencoders, whose hyperparameters are challenging to tune and strongly impact the detection results. For OC-SCM and LoOP, poor results were observed in the case of a mid growing season analysis, showing a lack of robustness of these methods when changing the amount of data used for the analysis.

Experiments were conducted by changing the outlier ratio and analyzing their distribution among the different categories of outliers (Supplementary Figure S13). For a low outlier ratio (\textit{e.g.}, 10\%), the detected parcels are affected by strong agronomic anomalies (\textit{e.g.}, global heterogeneity, globally low vigor). It is of crucial importance because it means that the IF algorithm attributes to these parcels the highest anomaly scores, which is relevant from an agronomic point of view. Then, parcels with lower outlier scores are affected by more transient anomalies, such as senescence problems. Using an outlier ratio equal to 20\% ensures the detection of the most important anomalies among the crop parcels, with a low amount of false positives when using both S1 and S2 features.

The robustness of our method was also tested regarding the impact of missing S2 images. A good precision was obtained even with a low amount of S2 images: by using half of the S2 images, a similar precision is obtained thanks to the complementary of S1 data, which is permanently available (Supplementary Figures S15 and S16). Moreover, an outlier analysis conducted on a mid growing season (all images acquired before February) was investigated in more detail. The main reasons were to 1) measure the impact of a reduced temporal interval for the analysis and 2) investigate the interest of such analysis for early warning purposes. The results (Supplementary Figures S17 and S18) show that a large amount of abnormal parcels are detected with high precision and that the presented method can be used for an early growing season analysis.

Finally parcel delineations coming from the French Land Parcel Identification System (LPIS) were investigated to confirm the robustness of the proposed method to small changes in the parcel boundaries for the rapeseed parcels (Supplementary Figures S19 and S20). Our results confirm that the proposed method provides consistent results even when using parcel boundaries of lower precision.

\begin{table}[ht!]
\caption{Summary of the influence of the different analyzed factors for the detection of anomalous crop development.}
\begin{tabular}{p{50mm}p{95mm}}
    \\
    \hline
    Evaluated factor & Effect and recommendation \\
    \hline & \\
    Outlier detection algorithm & Similar results obtained with various algorithm. IF is recommended for its robustness and easy tuning. \\
    Outlier score & Strongest anomalies have a higher outlier score than transient anomalies, which is interesting for crop monitoring. \\
    Missing S2 data & The proposed method is robust to missing S2 data. Using S1 dense time series improves the results. \\
    Mid growing season analysis & Results with high precision are obtained, early analysis is possible. \\
    Changes in parcel delineation & Small changes in the parcel delineation do not affect the detection results \\
    \hline
\end{tabular}
\label{table:table_disc_summary}
\end{table}

\section{Conclusion}\label{sec:conclusion}

This paper studied a new anomaly detection method for crop monitoring based on outlier analysis at the parcel-level using Sentinel-1 and Sentinel-2 features. This method is decomposed into 4 main steps: 1) preprocessing of multispectral and synthetic aperture radar (SAR) images, 2) computation of pixel-level features, 3) computation of zonal statistics at the parcel-level for all pixel-level features at each date, 4) detection of abnormal crop parcels using the isolation forest algorithm with the multi-temporal zonal statistics. The proposed method is fully unsupervised and can be used without historical data. It can be applied to different kinds of crops (such as rapeseed or wheat considered in this paper) and is able to detect a majority of parcels that are abnormal in an agronomic sense. Moreover, a relevant anomaly score is attributed to each parcel: agronomic anomalies affecting most of the growing season have a higher score than transient anomalies.

This study showed that S1 and S2 features are complementary for the detection of abnormal parcels in agricultural crops. Regarding S1 features, it is recommended to use median statistics computed at the parcel-level from VV and VH backscattering coefficients. For S2 features, median and IQR statistics computed at the parcel-level from the Normalized Difference Vegetation Index (NDVI), the Green-Red Vegetation Index (GRVI), two variants of the Normalized Difference Water Index (NDWI) and a variant of the Modified Chlorophyll Absorption Ratio Index (MCARI/OSAVI) provided the best results, especially when combined with S1 features. Finally, the Isolation Forest algorithm is the outlier detection algorithm that provided the best results for identifying abnormal agricultural parcels with a simple parameter tuning.

Further investigation should be conducted to determine whether other multispectral features, \textit{e.g.}, biophysical parameters such as the Leaf Area Index (LAI) or the fraction of green vegetation cover (fCover) \citep{DJAMAI2019416, VERRELST2015260}, can improve crop monitoring. Regarding SAR features, the use of SLC images could also be investigated to extract polarimetric parameters such as entropy or volume scattering and in particular the new Dual polarimetric radar vegetation index (DpRVI) considered in \citet{MANDAL2020111954}. Another line of research is to take into account the temporal structure of vegetation indices to potentially improve detection of agronomic anomalies and estimate the dates where the detected anomalies have appeared. For instance, contextual outlier detection might be interesting for disturbance or inter-annual anomaly detection. Including a contextual outlier detection in the strategy might provide complementary information to detect both inter-annual and intra-annual anomalies. Coupling the detection method with a supervised or unsupervised classification algorithm is another prospect, in order to assign to each detection an anomaly type that could be helpful for example to identify heterogeneity or growth problems automatically. It could also be interesting to investigate other types of crops such as soybean (a low biomass crop contrary to wheat and rapeseed that are both considered as high biomass crops). Finally, an anomaly map could possibly be generated for large areas, by taking for instance large rectangular windows of vegetation status.

\supplementary{Supplementary data to this article can be found in the document provided in complement of the manuscript. Figure S1: A rapeseed crop parcel (yellow boundaries) affected by a two-part heterogeneity. Figure S2: (a) A rapeseed crop parcel (yellow boundaries) affected by an heterogeneity after senescence. (b): Associated Interquartile Range(IQR) of the parcel NDVI time series. Figure S3: Example of time series subjected to a red channel problem in March 27 for a wheat parcel. Figure S4: A rapeseed crop parcel affected by a red channel problem (yellow boundaries), which is also highly vigorous. Figure S5: Time series of median NDVI for a rapeseed parcel presenting sign of (a) early senescence and (b) early flowering. Figure S6: Rapeseed parcels: the parcel with yellow boundaries is affected by shadow caused by the trees located next to the parcel. Also, at the bottom a too small parcel is visible. Figure S7: Time series of median SAR features (VV, VH, VH/VV) and median NDVI for a rapeseed parcel. Figure S8: Example of a parcel of rapeseed crop (yellow boundaries) where heterogeneity occurs almost during the complete season. Table S1: hyperparameters used in the different algorithms. Figure S9: Precision vs. outlier ratio for a complete growing season analysis of the rapeseed parcels. Various algorithms are compared using all S1 and S2 features. Figure S10: 100×(Number  of  detected  parcels  in  each  category  /  Number  of  detected  parcels). Rapeseed parcels are analyzed with various outlier detection algorithm and with an outlier ratio equal to 20\%. Figure S11: Precision vs. outlier ratio for a complete growing season analysis of the rapeseed parcels. Various sets of features using the IF algorithm are compared. Figure S12: 100×(Number of detected parcels in each category / Number of detected parcels). Various sets of features are compared with the IF algorithm and an outlier ratio equal to 20\% for a complete growing season analysis (rapeseed crops). Figure S13: 100×(Number of detected parcels in each category / Number of detected parcels). Various outlier ratio are tested with the same set of features and the IF algorithm for a complete growing season analysis (rapeseed crops). Figure S14: Precision vs. outlier ratio for a complete growing season analysis of the rapeseed parcels. Various statistics of the NDVI are compared using the IF algorithm. Figure S15: Precision vs. outlier ratio for a complete season analysis of the rapeseed dataset. Missing dates means that only 1 S2 image out of 2 was taken (6 S2 images instead of 13). Figure S16: Precision vs. outlier ratio for complete season analysis of the rapeseed dataset. Missing dates means that only the S2 images acquired after April were used (7 images). Figure S17: Precision vs. outlier ratio for a mid-season analysis of rapeseed parcels (all images available before February). Various sets of features are compared using the IF algorithm. Figure S18: 100×(Number of detected parcels in each category / Number of detected parcels). Results obtained for a mid season analysis (before February) and a complete growing season analysis are compared for a outlier ratio equal to 10\% in the rapeseed dataset. Figure S19: Example of parcel boundaries (rapeseed crop, growing season 2017/2018). In orange:customer database, in green: LPIS database. Figure S20: 100×(Number of detected parcels in each category / Number of detected parcels). LPIS and proprietary parcellation databases are compared with the IF algorithm and an outlier ratio equal to 20\%.}

\authorcontributions{Conceptualization, F.M., M.A., S.D., D.K., G.R., and J-Y.T.; methodology, F.M., M.A., S.D., D.K., G.R., and J-Y.T.; software, F.M. and M.A.; validation, F.M., M.A., S.D., D.K., G.R., and J-Y.T.; formal analysis, F.M., M.A., S.D., D.K., G.R., and J-Y.T.; investigation, F.M., M.A., and J-Y.T.; resources, S.D., G.R., and J-Y.T.; data curation, F.M.; writing--original draft preparation, F.M. and M.A.; writing--review and editing, F.M., M.A., S.D., D.K., G.R., and J-Y.T.; visualization, F.M.; supervision, J-Y.T.; project administration, S.D, G.R. and J-Y.T.;}

\funding{This document is the results of the research project funded by TerraNIS SAS. and ANRT (convention CIFRE no. 2018/1349).}

\acknowledgments{The authors would like to thanks Alexandre Bouvet, Milena Planells and their colleagues from CESBIO (Centre d'Etudes Spatiales de la BIOsphère) for the help regarding the processing of S1 images.}

\conflictsofinterest{The authors declare no conflict of interest.}

\reftitle{References}
\externalbibliography{yes}
\bibliography{cas-refs}
\end{document}


\section{Complementary examples of outlier parcels}

This section shows some anomalies identified by agronomic experts, provided to illustrate the labeling process of this study in complement of the examples already provided in the main document (Figs. 3 to 7). In what follows, NDVI time series are mostly represented because their interpretations are generally easier and straightforward but the agronomic expert had at his disposal all the features (S1 and S2) and all the multispectral images acquired within a growing season.

\subsection{Heterogeneity}

\begin{figure}[ht!]
\includegraphics[width=0.45\textwidth]{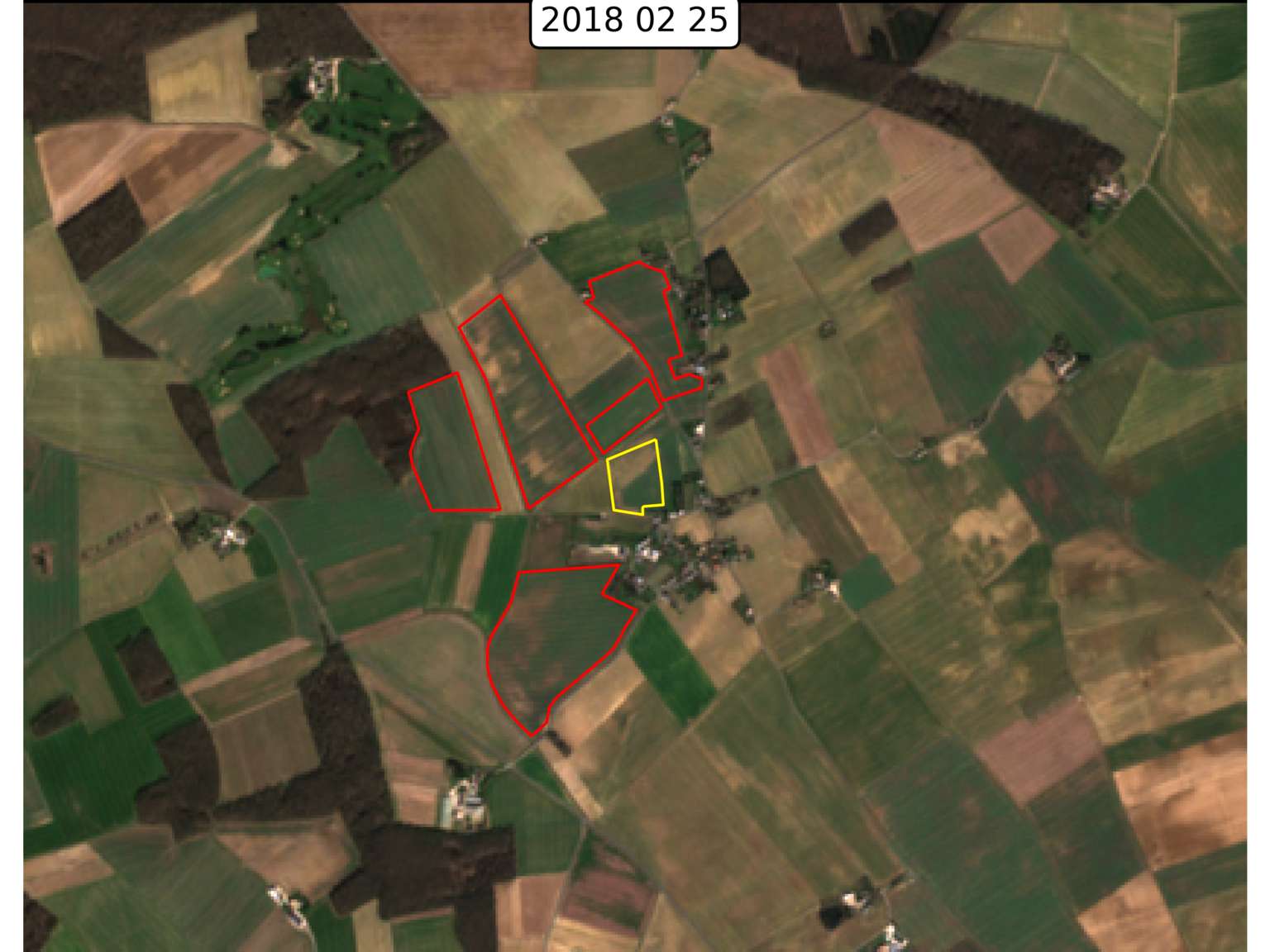}
\hfill
\includegraphics[width=0.45\textwidth]{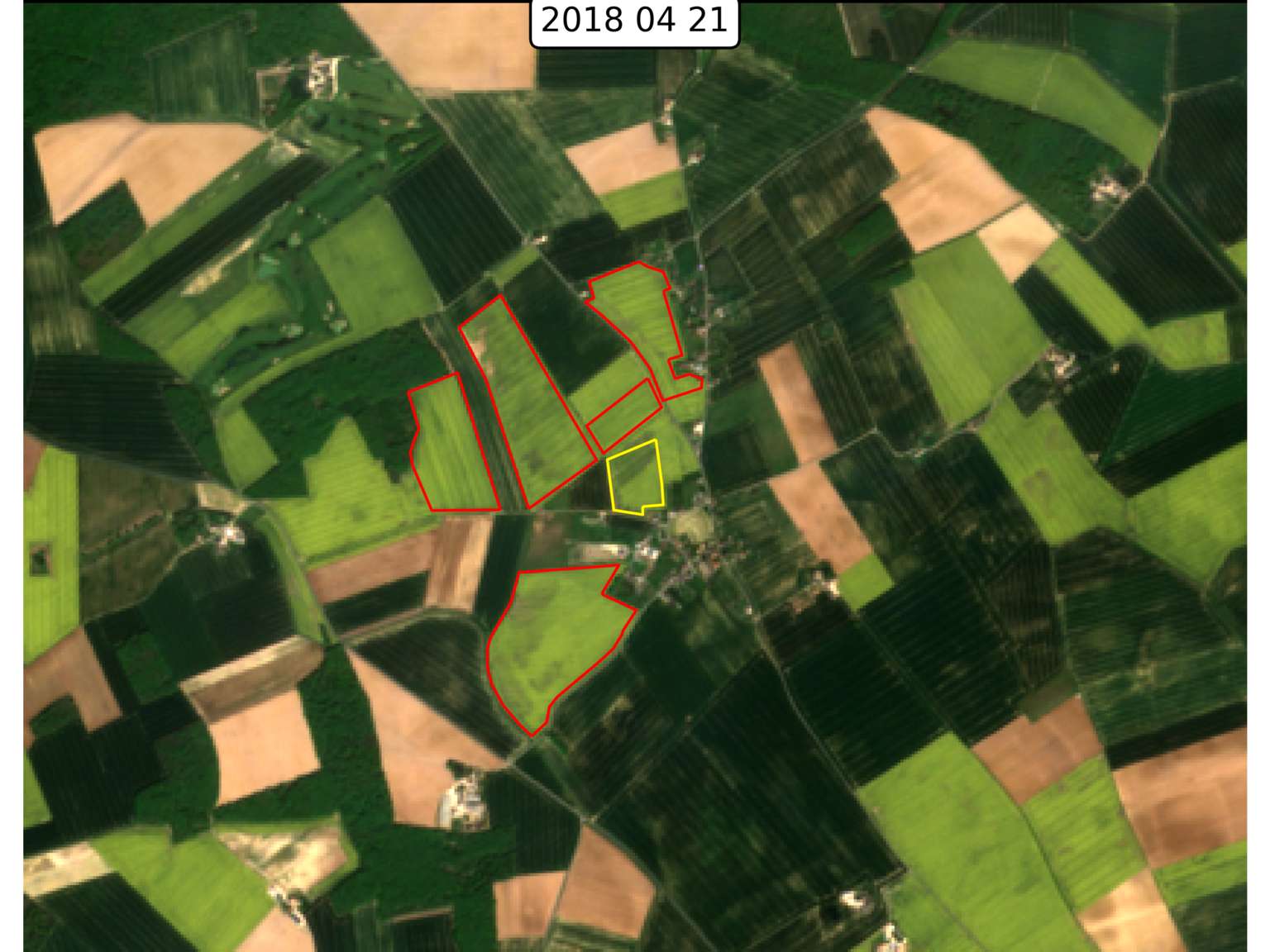}
\caption{A rapeseed crop parcel (yellow boundaries) affected by a two-part heterogeneity. The left image was acquired in February 2018 and the right image in April 2018.}
\label{fig:het_2parts_example}
\end{figure}

\begin{figure}[ht!]

\subfloat[]{
\includegraphics[width=0.45\textwidth]{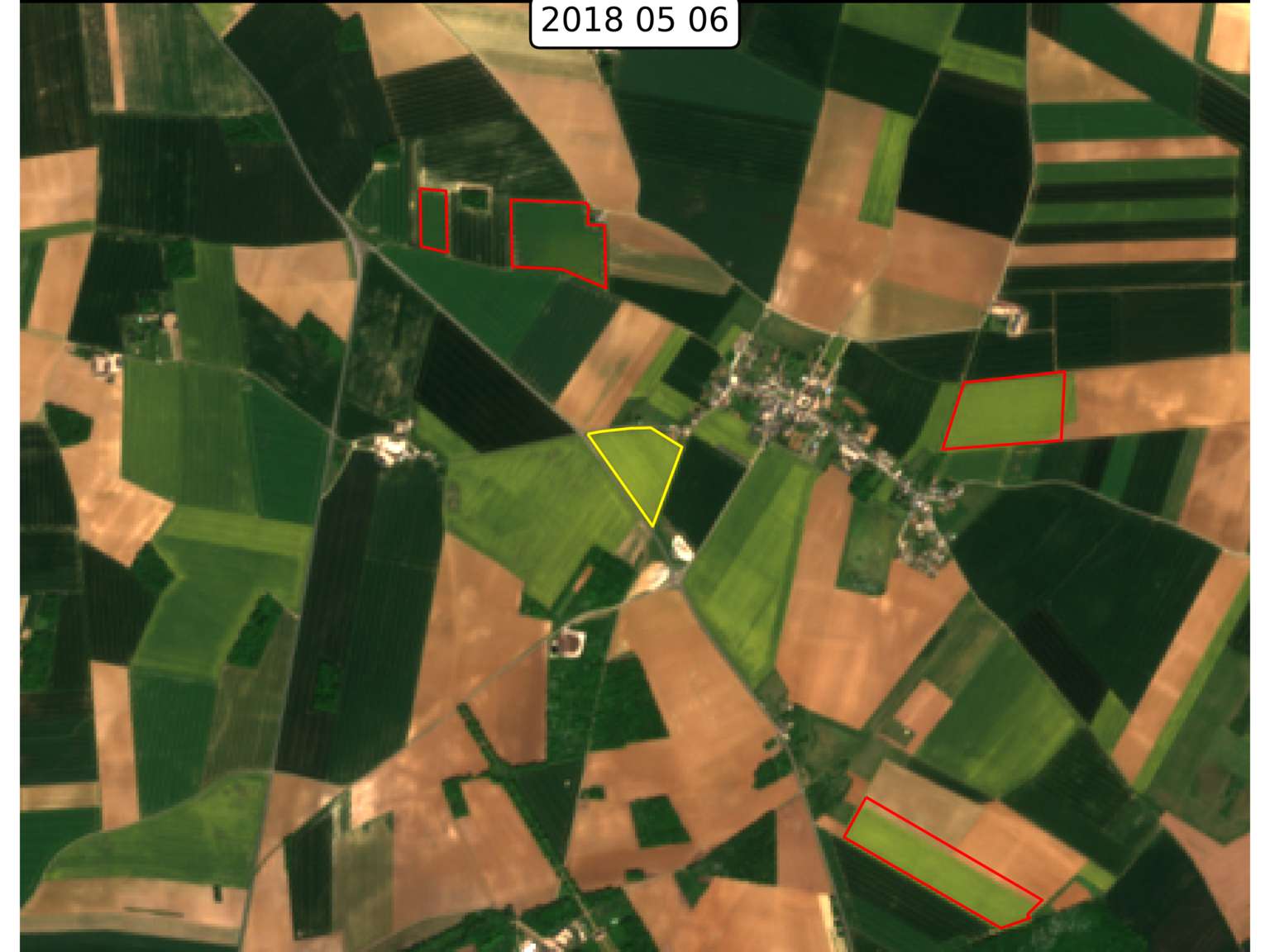}
\hfill
\includegraphics[width=0.45\textwidth]{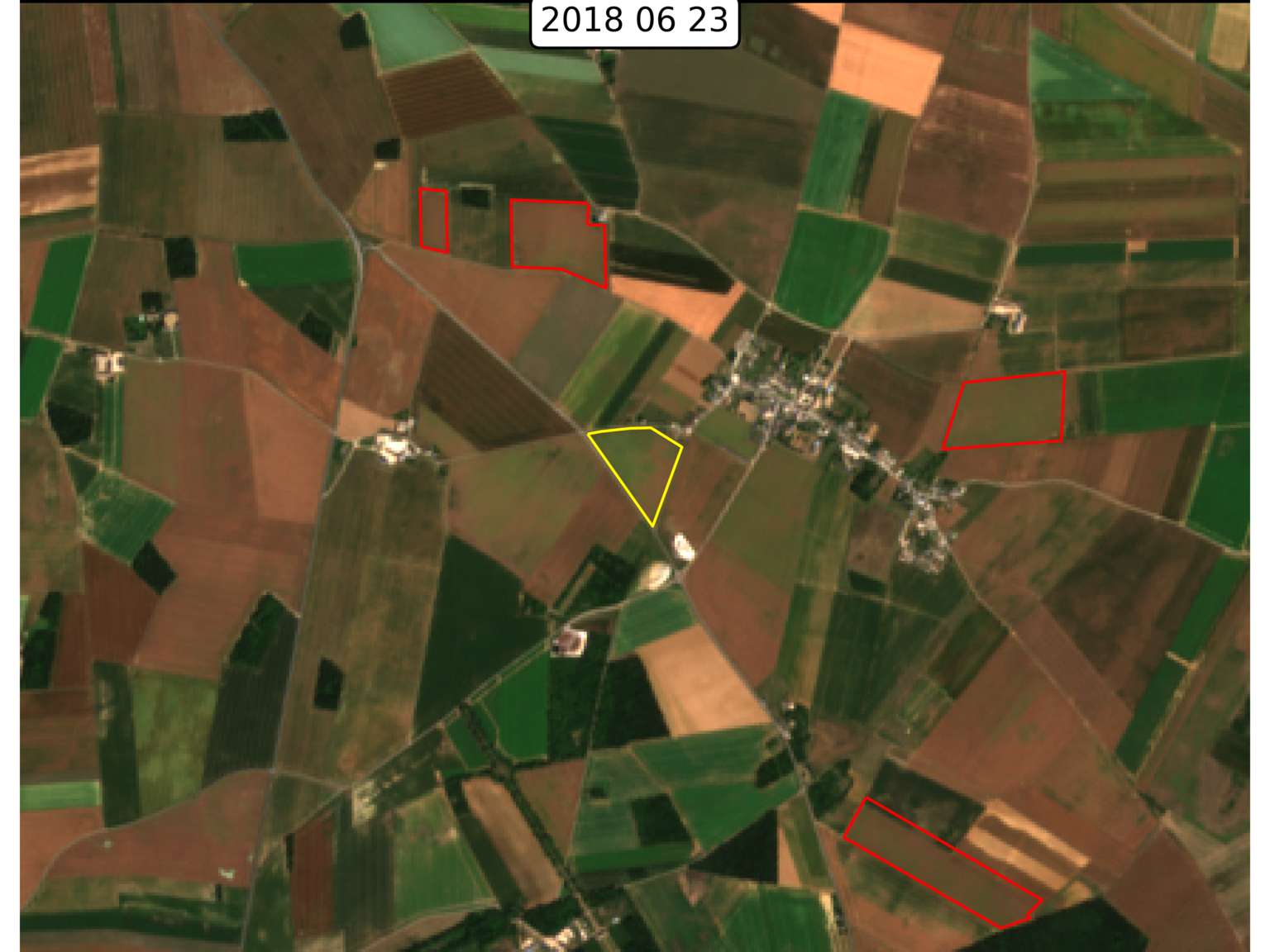}
}

\subfloat[]{\centerline{\includegraphics[width=0.6\textwidth]{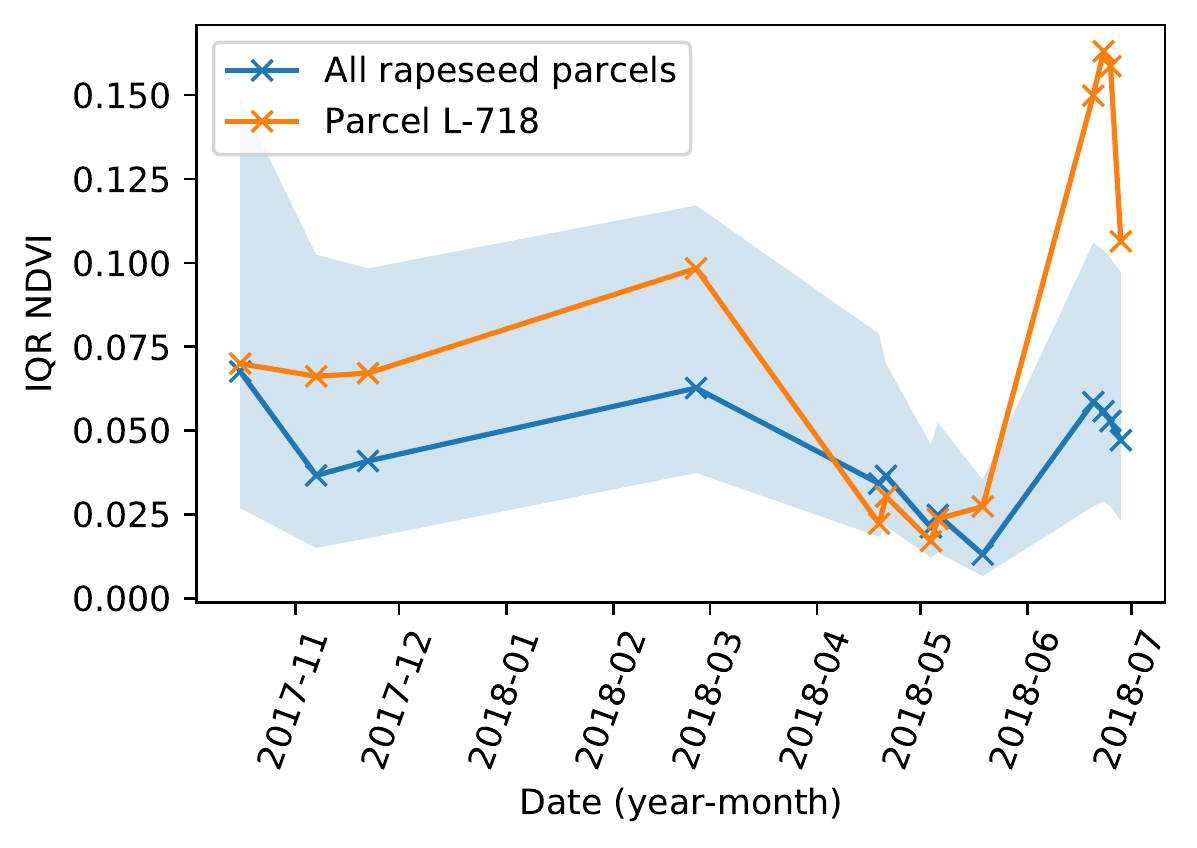}}}
\caption{(a) A rapeseed crop parcel (yellow boundaries) affected by an heterogeneity after senescence. The left image was acquired in May and the right image in June. (b) Associated Interquartile Range (IQR) of the parcel NDVI time series. The blue line is the median value of the whole dataset. The blue area is filled between the 10th and 90th percentiles.}
\label{fig:het_aftersen_example}
\end{figure}

In Fig. 3 of the main document, a case of global heterogeneity was presented. \autoref{fig:het_2parts_example} depicts an example of heterogeneity where two different parts are clearly visible. This example shows that it is difficult to decide if wrong boundaries were provided, if two different varieties of rapeseed were sowed or if soil differences led to heterogeneity. \autoref{fig:het_aftersen_example} shows an example of heterogeneity after senescence, where the parcel is perfectly homogeneous in May whereas it presents some heterogeneity in June. As explained in the paper, this phenomenon is likely to be related to differences in the soil water content and could be of interest for inter season analysis. \autoref{fig:het_aftersen_example}(b) shows the IQR NDVI time series of the same parcel (in yellow) and shows that late heterogeneity can be visible and that its IQR value is greater than 90\% of the rest of the data. The same observations can be made for early heterogeneity. In order to compare the analyzed parcel to the rest of the data, the median, the 10th and the 90th percentiles of the whole dataset were displayed (similarly to a boxplot visualization). This representation allows the agronomic expert to know if the observed parcel has indicator values higher (or lower) than 90\% of the data.


\newpage
\subsection{Growth anomalies}

Growth anomalies are a bit more difficult to analyze when looking at a single composite multispectral image and visualizing the corresponding time series is in general more relevant. Examples of late growth affecting rapeseed and wheat parcels are shown in Figure 8 and 15 of the main document. 

\begin{figure}[ht!]
\subfloat[]{\includegraphics[width=0.49\textwidth]{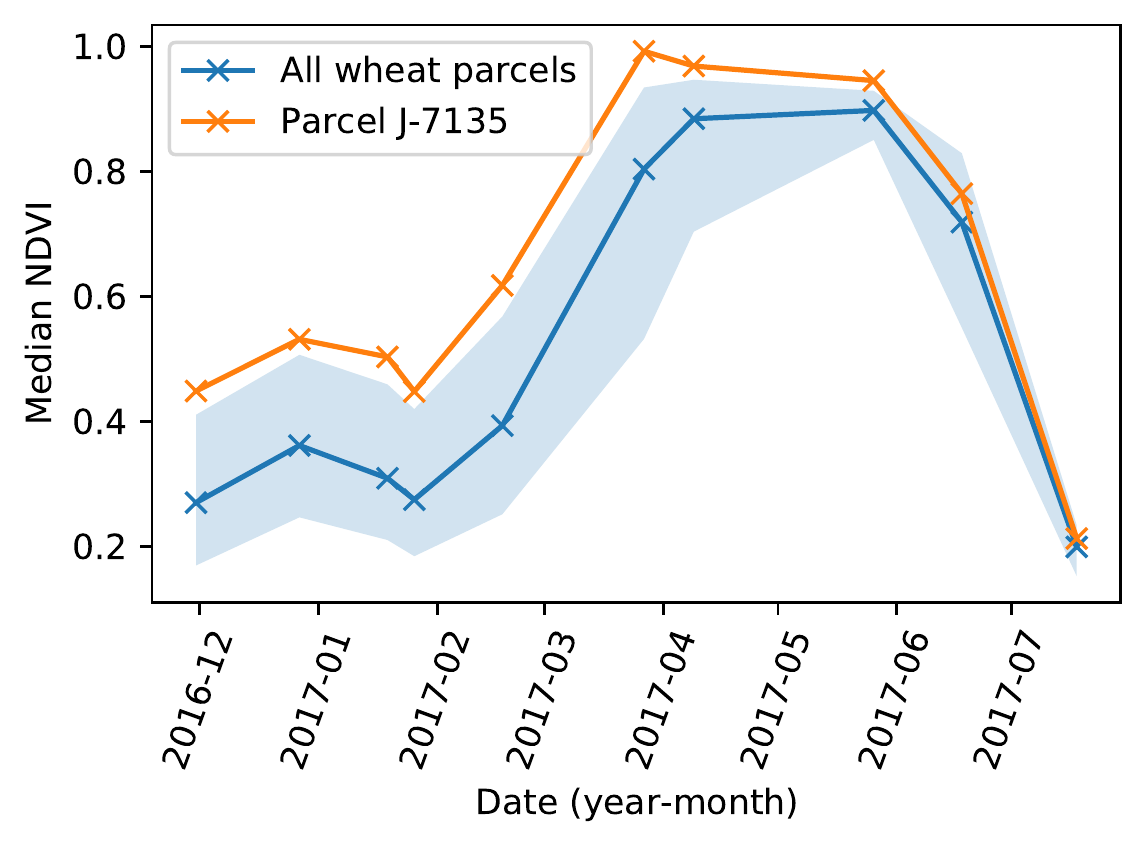}}
\hfill
\subfloat[]{\includegraphics[width=0.49\textwidth]{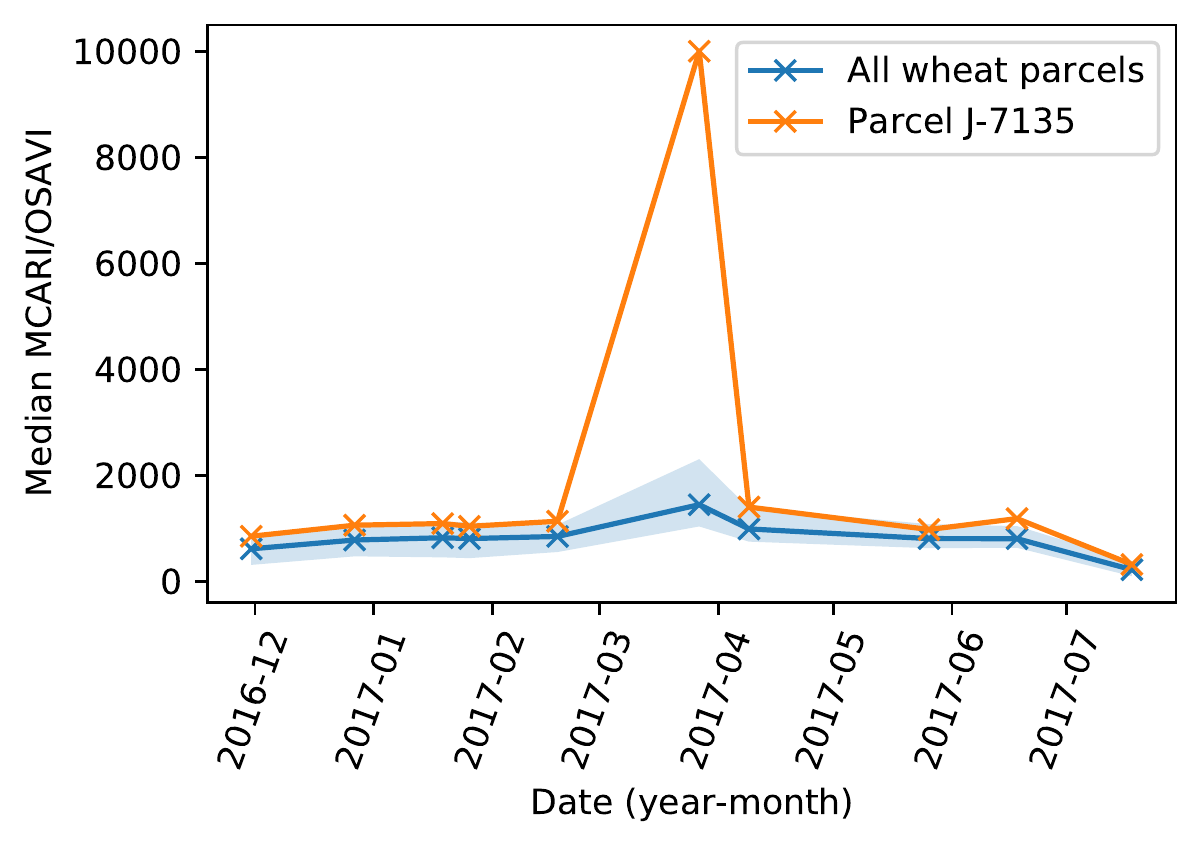}}
\caption{Example of time series subjected to a red channel problem in March 27 for a wheat parcel: (a) median NDVI and (b) median MCARI/OSAVI. The blue line is the median value of the whole dataset. The blue area is filled between the 10th and 90th percentiles. The orange line is a specific parcel analyzed.}
\label{fig:time_srie_red_channel_example}
\end{figure}

\begin{figure}[ht!]
\centerline{\includegraphics[scale=0.63]{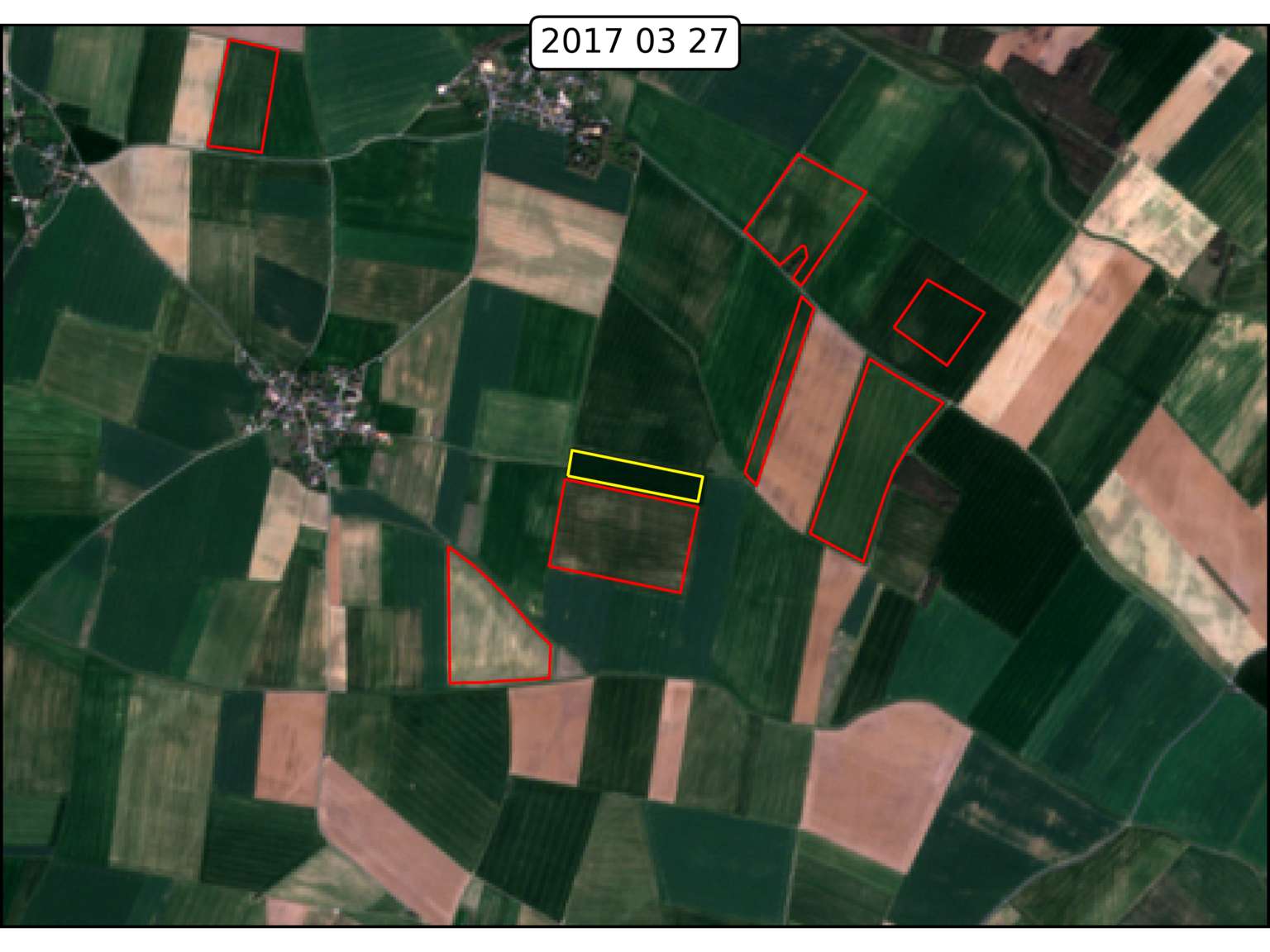}}
\caption{A rapeseed crop parcel affected by a red channel problem (yellow boundaries), which is also highly vigorous. A parcel with a late growth can be observed at the bottom of the image (triangle red boundaries).}
\label{fig:img_red_channel}
\end{figure}

We noticed that in the S2 images acquired in March 27 2017, a small amount of vigorous wheat parcels have the majority of their red pixels equal to zero, causing extreme values of the S2 features as illustrated in \autoref{fig:time_srie_red_channel_example}. This issue was caused by the MAJA processing\footnote{\url{https://labo.obs-mip.fr/multitemp/using-ndvi-with-atmospherically-corrected-data/}, online accessed 10 March 2020} and could be easily fixed using another processing chain like Sen2core or with a threshold for the red band. Since this phenomenon affected only a small amount of vigorous parcels, it did not change the quality of the detection results. \autoref{fig:img_red_channel} shows the S2 true color composite of the parcel analyzed in \autoref{fig:time_srie_red_channel_example}, which is darker than its neighbors. 

\begin{figure}[ht!]
\subfloat[]{\includegraphics[width=0.49\textwidth]{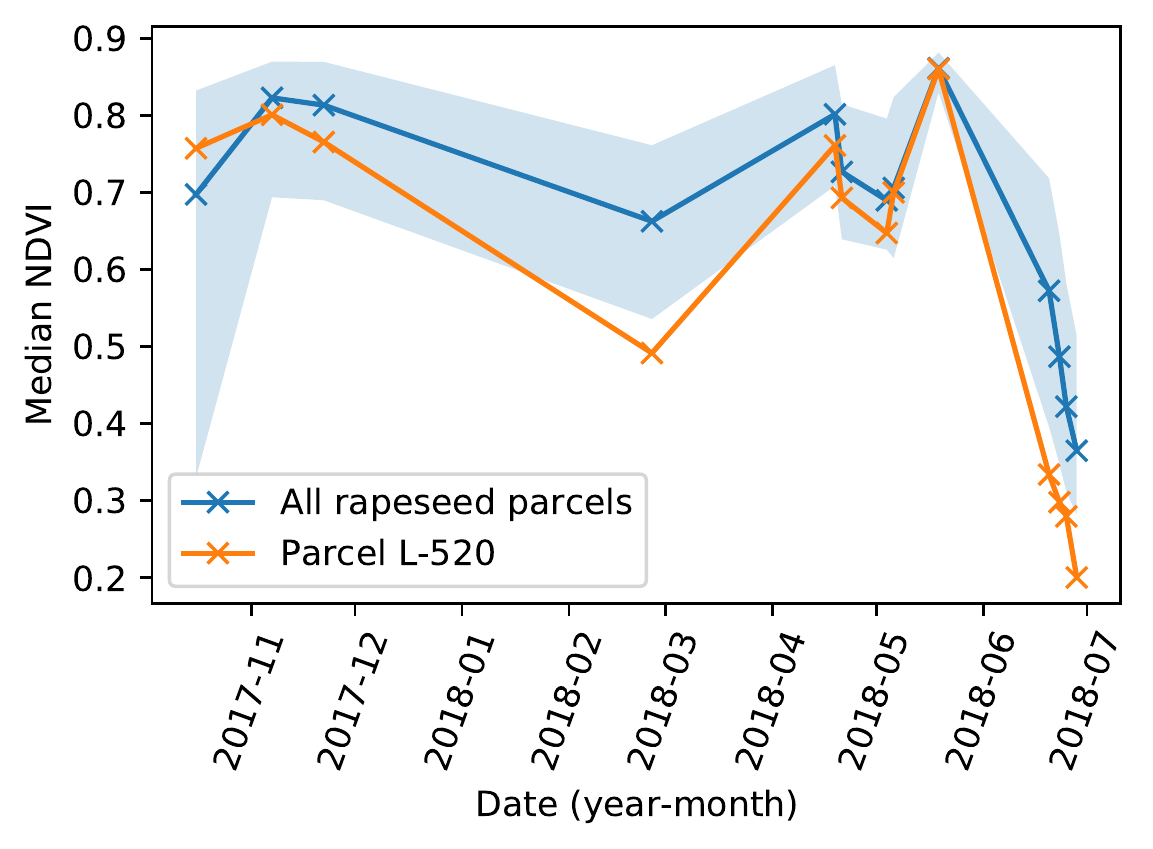}}
\hfill
\subfloat[]{\includegraphics[width=0.49\textwidth]{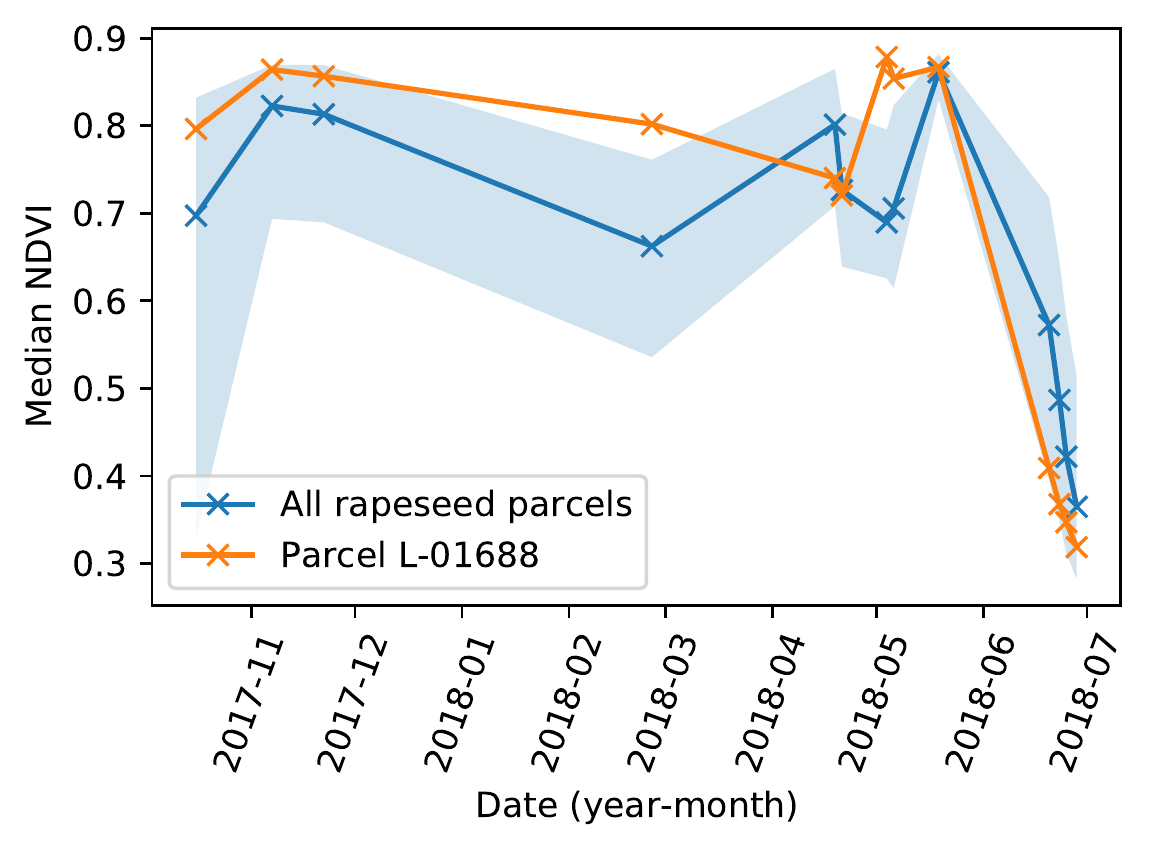}}
\caption{Time series of median NDVI for a rapeseed parcel presenting sign of (a) early senescence and (b) early flowering. The blue line is the median value of the whole dataset. The blue area is filled between the 10th and 90th percentiles. The orange line is a specific parcel analyzed.}
\label{fig:time_early_sen_ex}
\end{figure}

\newpage
\autoref{fig:time_early_sen_ex}.a displays an example of early senescence. As mentioned in the paper this type of anomaly is more subtle than a global heterogeneity or a late growth problem because it occurs only at the end of growing season. In this particular example, a low median NDVI is also observed in February. It could be a problem during winter but it is challenging to confirm abnormality with only one value of NDVI (S1 time series are valuable in that case). \autoref{fig:time_early_sen_ex}.b is an example of early flowering. This case is also subtle and was not observed frequently. In general, the early flowering category is also affected by early senescence.

\newpage
\subsection{Non-agronomic anomalies}

\begin{figure}[ht!]
\centerline{\includegraphics[scale=0.25]{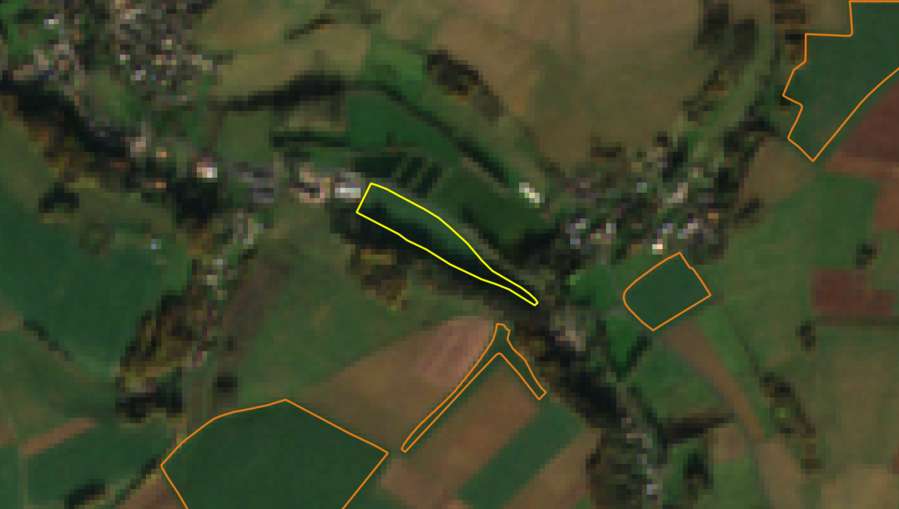}}
\caption{Rapeseed parcels: the parcel with yellow boundaries is affected by shadow caused by the trees located next to the parcel. Also, at the bottom a too small parcel is visible.}
\label{fig:img_shadow_too_small}
\end{figure}

\autoref{fig:img_shadow_too_small} displays examples of two different types of anomalies. The yellow parcel is an example of a parcel affected by shadows. A case of a too small parcel is also visible in the image, the particular form of the boundaries causes difficulties to filter the parcel only using its area. 

\newpage
\section{Complementary information about SAR images and their anomalies}

\begin{figure}[ht!]
\subfloat[]{\includegraphics[width=0.49\textwidth]{figs/example_anomalies/late_growth_plot_S1_VH_median.pdf}}
\hfill
\subfloat[]{\includegraphics[width=0.49\textwidth]{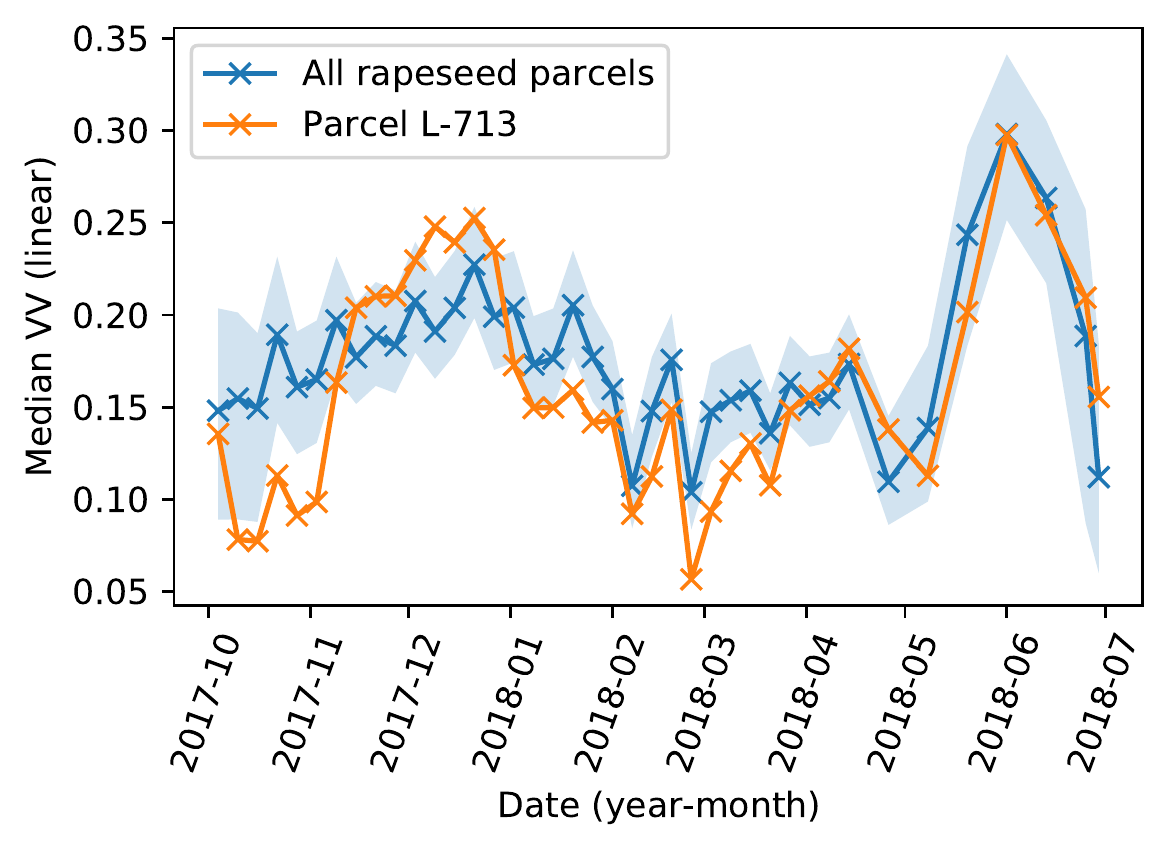}}
\hfill
\subfloat[]{\includegraphics[width=0.49\textwidth]{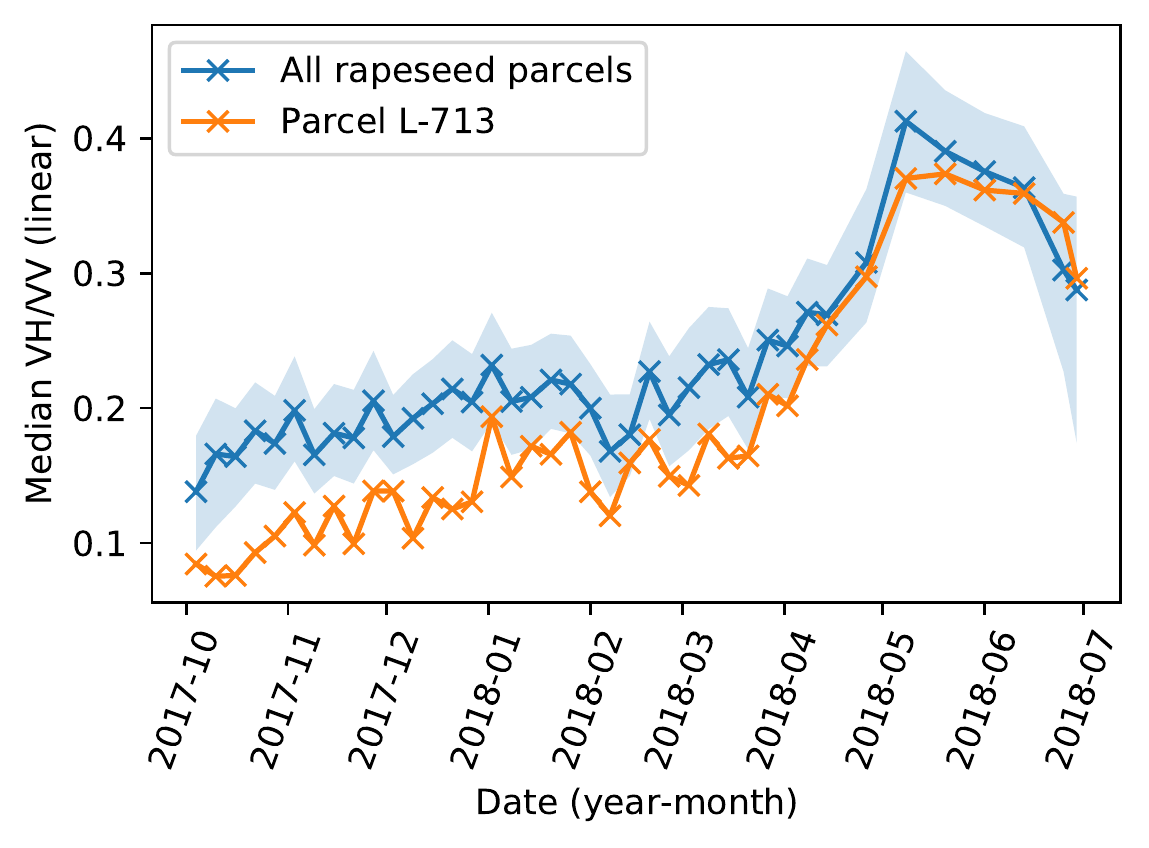}}
\hfill
\subfloat[]{\includegraphics[width=0.49\textwidth]{figs/example_anomalies/late_growth_ndvi_median_rapeseed_L-713.pdf}}
\caption{Time series of median SAR features (VV, VH, VH/VV) and median NDVI for a rapeseed parcel. The blue line is the median value of the whole dataset. The blue area is filled between the 10th and 90th percentiles. The orange line is a specific parcel analyzed.(a): median VH, (b): median VV, (c): median ratio VH/VV, (d): median NDVI}
\label{fig:SAR_late_growth}
\end{figure}

\begin{figure}[ht!]
\includegraphics[width=0.49\textwidth]{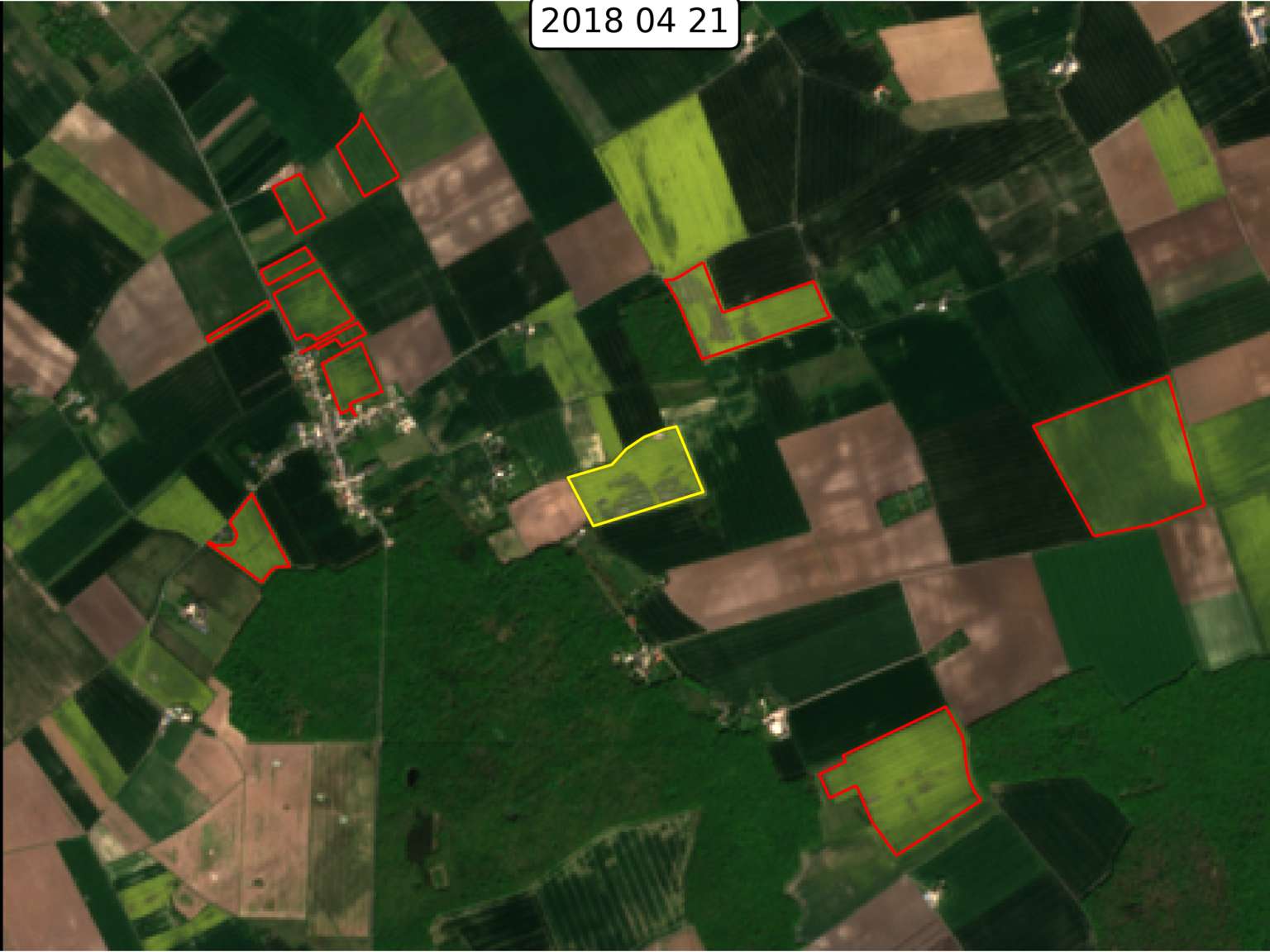}
\hfill
\includegraphics[width=0.49\textwidth]{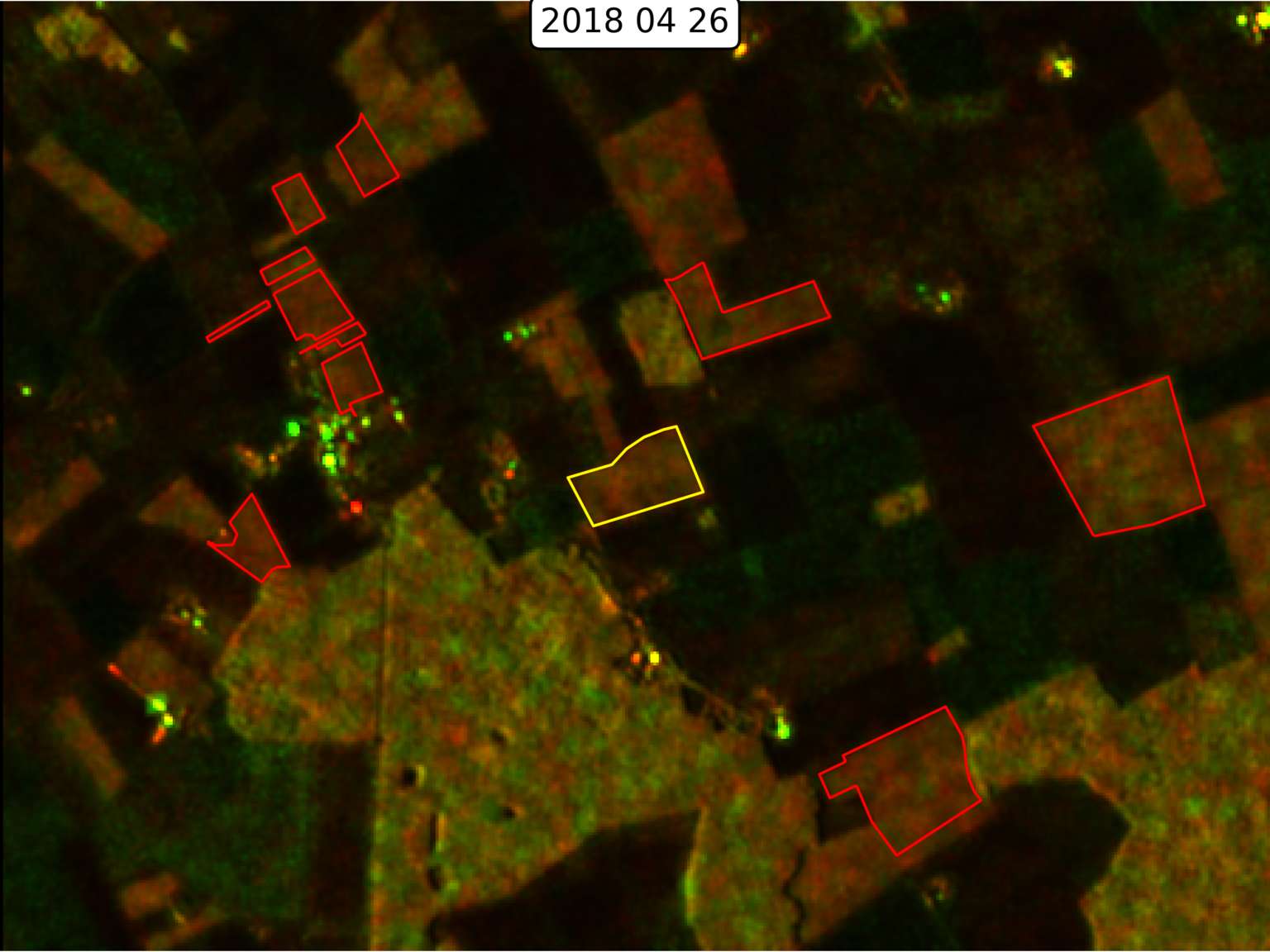}
\caption{Example of a parcel of rapeseed crop (yellow boundaries) where heterogeneity occurs almost during the complete season. Some other parcels show some signs of heterogeneity too. Left: true color S2 image acquired in May, right: composite SAR image (green channel is VV polarization and red channel is VH polarization) acquired in May with multi-temporal speckle filtering.}
\label{fig:heterogeneity_example_SAR}
\end{figure}

A strong correlation between SAR and plant vigor (late growth / vigorous crop) was observed in this study. \autoref{fig:SAR_late_growth} illustrates the effect of late growth on S1 features. \autoref{fig:heterogeneity_example_SAR} shows that SAR images are not always affected by heterogeneity within the crop parcel, as highlighted in the main document. Heterogeneity is detectable with SAR features when the crop structure is affected, which is undestrandable considering the nature of the sensor.

\newpage
\section{Hyperparameter tuning of the outlier detection algorithms}\label{sec:algos}

This section summarizes the different hyperparameter tunings that have been considered for each algorithm leading to the values reported in \autoref{table:table_hyperparameters}.

For the LoOP algorithm, the number of nearest neighbors $knn$ was fixed by grid search to $k=701$. The extent parameter of LoOP was fixed to $\lambda=2$ as recommended in \citet{Constantinou2018}. 

For the OC-SVM algorithm, an efficient heuristic \citep{Jaakkola1999, Aggarwal2017_kernel} consists of estimating the parameter $\sigma$ as the median of the pairwise Euclidean distances between vectors from the learning set $\mathcal{X}$, denoted as $\text{median}(\text{dist}(\mathcal{X}))$.  This estimator of $\sigma$ provided good results without a need to a manual tuning for each new dataset. 

The IF algorithm was used with a number of iTrees equal to $n_{trees}=1000$ and a subsampling fixed to $\text{n}_{\text{samples}}=256 $ as in the original paper \citep{Liu2008}. Changing these two parameters did not have a significant effect on the results, which is a crucial advantage compared to the other algorithms. 

The parameters of the AE were tuned by grid search. We considered a classical structure similar to the one proposed in the Python library for outlier detection PyOD \citep{zhao2019pyod}: 4 hidden layers with respectively 64, 32, 32 and 64 neurons. A Relu activation function was used for all layers except for the output using a sigmoidal function. The regularization parameter of the layers (referred to as ``activity regularizer'' in Keras) was set to $10^{-3}$. Note that this specific regularization significantly improved the detection results, contrary to changes in the network structure (e.g., number of neurons).

Because they are using distances, the OC-SVM, LoOP and AE algorithms also require a normalization in order to have input features in the interval $[0, 1]$, while this step is not mandatory when using the IF algorithm.

\begin{table}[h]
\caption{Hyperparameters used in the different algorithms}
\centering
\begin{tabular}{lll}
    \\
    \hline
    Algorithm & Hyperparameter & Value \\
    \hline 
    \multirow{2}{*}{IF} & $\text{n}_{\text{trees}}$  &  $1000$ \\
    & $\text{n}_{\text{samples}}$  & $256$ \\ \hline
    \multirow{2}{*}{LoOP} & $\text{k}$ &  $701$ \\ 
     & $\lambda$ &  $2$ \\ \hline
    \multirow{2}{*}{AE} & hidden neurons &  $64$, $32$, $32$, $64$ \\ 
    & output regularization &  $10^{-3}$ \\ \hline
    OC-SVM & $\sigma$ &  $ \text{median}(\text{dist}(\mathcal{X}))$\\ \hline
\end{tabular}
\label{table:table_hyperparameters}
\end{table}

\newpage
\section{Complementary results}\label{sec:results}

\subsection{Effect of the algorithm used for crop anomaly detection}\label{sec:effect_algo}

\begin{figure}[htbp!]
\centerline{\includegraphics[width=0.7\textwidth]{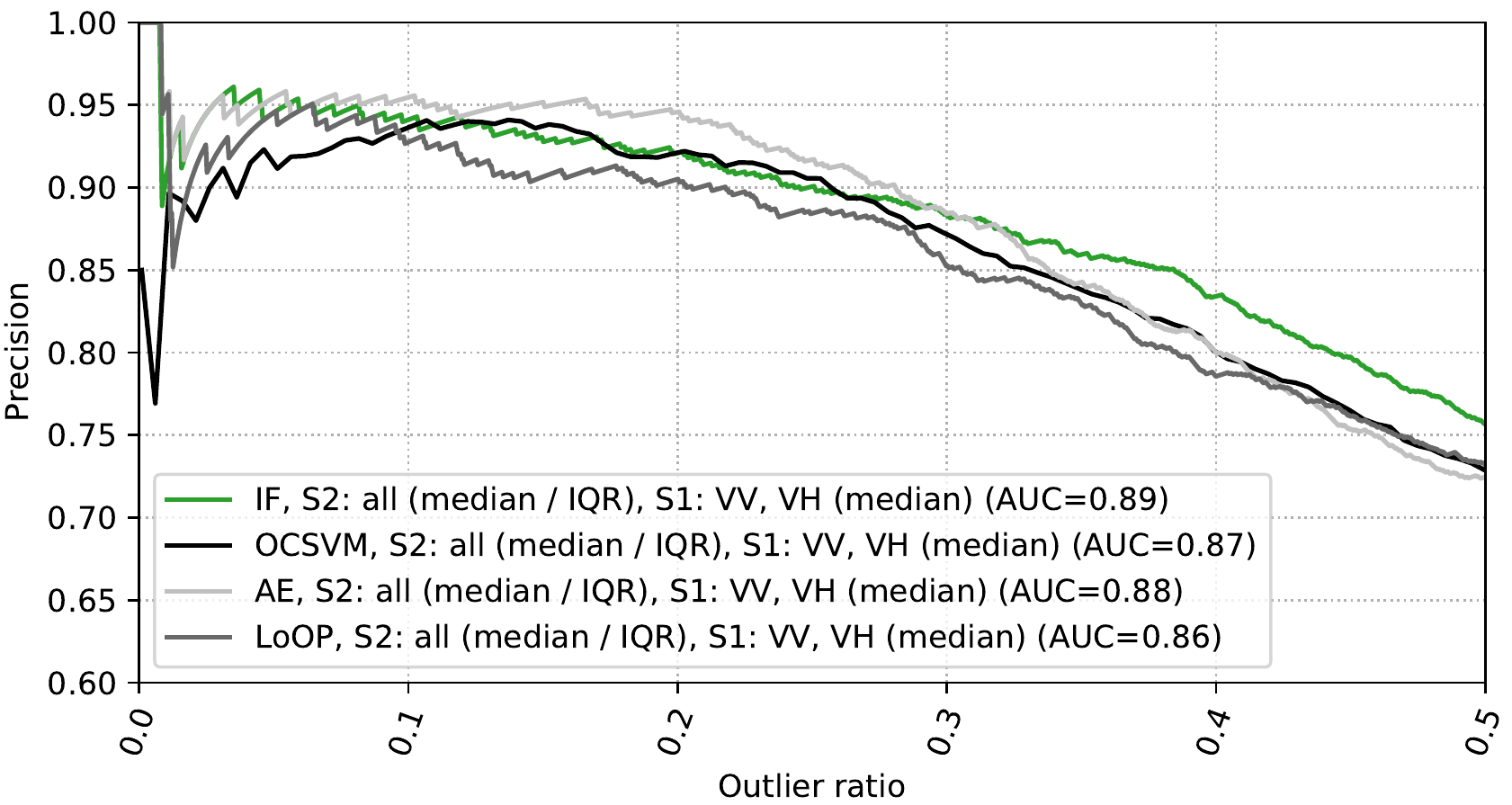}}
\caption{Precision vs. outlier ratio for a complete growing season analysis of the rapeseed parcels. Various algorithms are compared using all S1 and S2 features.}
\label{fig:PCR_all_season_3algos}
\end{figure}

The performance of the different outlier detection algorithms (AE, LoOP, OC-SVM and IF) was first tested for a complete season analysis. A first experiment was made by computing precision vs. outlier ratio curves for the different algorithms. The OC-SVM algorithm does not provide a unique outlier score associated with each parcel as it depends on the maximum amount of outliers defined by the parameter $\nu$. For the other algorithms, an anomaly score is attributed to each parcel and it is then possible to choose the outlier ratio by sorting the parcels according to their anomaly score (in order to select an appropriate percentage of parcels to be detected). \autoref{fig:PCR_all_season_3algos} shows that all the outlier detection algorithms provide similar precision for outlier ratios lower than 30\% with a majority of relevant anomalies detected (more than 90\%), confirming that multiple methods can lead to similar accuracy. The outlier ratio is plotted up to 0.5 to highlight the differences when detecting a large amount of outliers, which is possible for a complete growing season as various transient anomalies can occur. IF and AE perform slightly better overall with a higher AUC and the IF outlier score is more relevant for outlier ratio greater than 30\%.

\begin{figure}[htbp!]
\centerline{\includegraphics[width=1\textwidth]{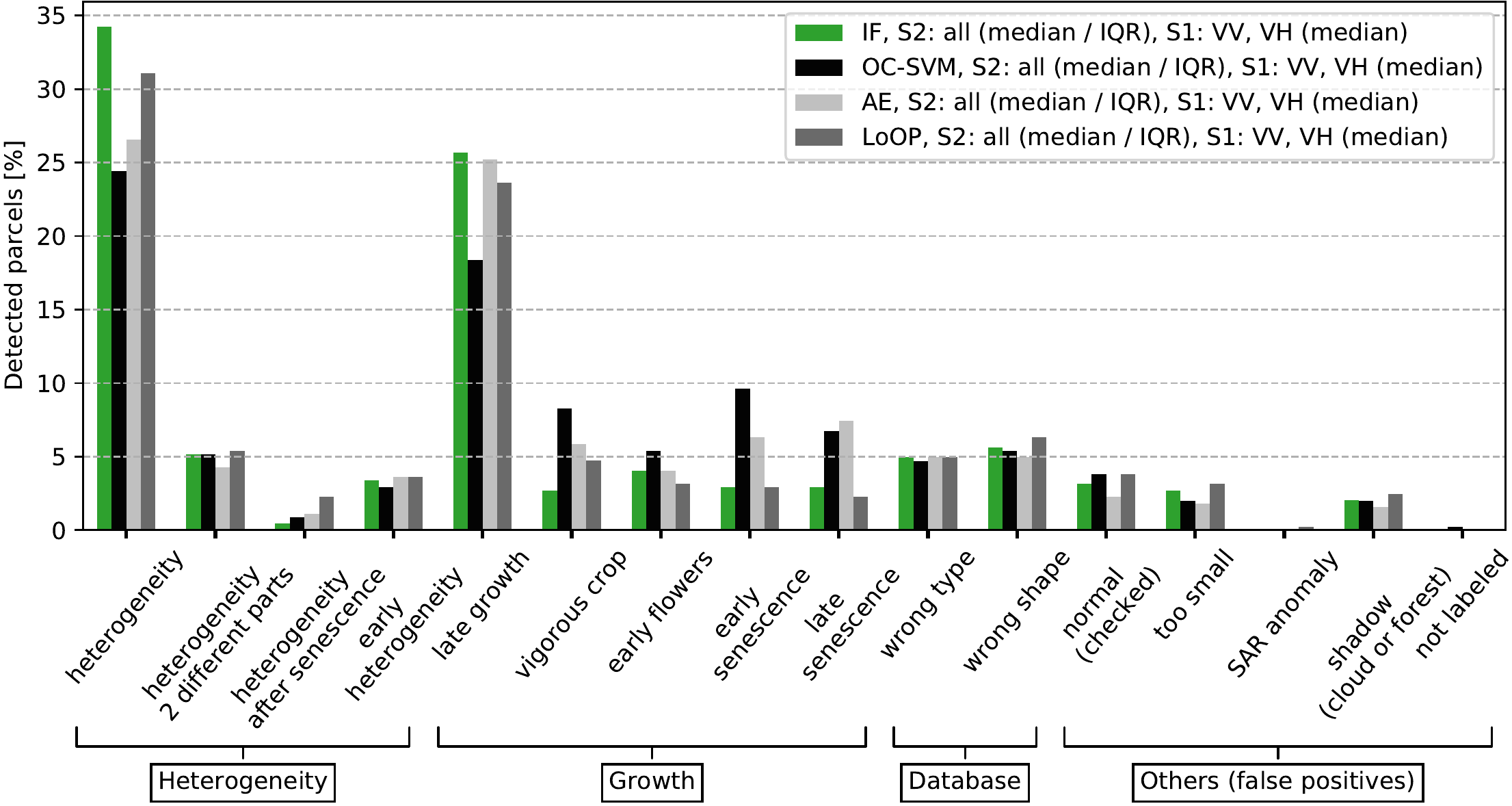}}
\caption{100$\times$(Number of detected parcels in each category / Number of detected parcels). Rapeseed parcels are analyzed with various outlier detection algorithm and with an outlier ratio equal to 20\%.}
\label{fig:histo_all_season_3algos}
\end{figure}

To analyze potential differences within the anomaly categories detected, the 4 algorithms were run with an outlier ratio fixed to 20\%. The percentages of the detected parcels within each category for LoOP, OC-SVM, AE and IF are displayed in \autoref{fig:histo_all_season_3algos}. Overall, all the algorithms detect a majority of heterogeneity (34\% for IF) and late growth (25\% for IF) anomalies, which can be understood as these anomalies generally affect the complete growing season. It also appears that almost all the ``wrong type'' and ``wrong shape'' of the dataset are detected, which seems also logical as these anomalies strongly affect the parcel time series. The histograms obtained with LoOP and IF are very similar, whereas OC-SVM and AE seems to detect less heterogeneity anomalies (44 heterogeneous parcels detected by IF are not detected by OC-SVM) in favor of anomalies related to delay in growth (late growth, vigorous crop, senescence anomalies). It is reasonable to detect heterogeneity issues having a global impact on the crop season before senescence problems, which occurred only at the end of the season. To that extent, results obtained with the LoOP and IF algorithms are more interesting.

\newpage
\subsection{Effect of the feature set used for crop anomaly detection}\label{sec:best_results}
\begin{figure}[ht!]
\centerline{\includegraphics[width=0.7\textwidth]{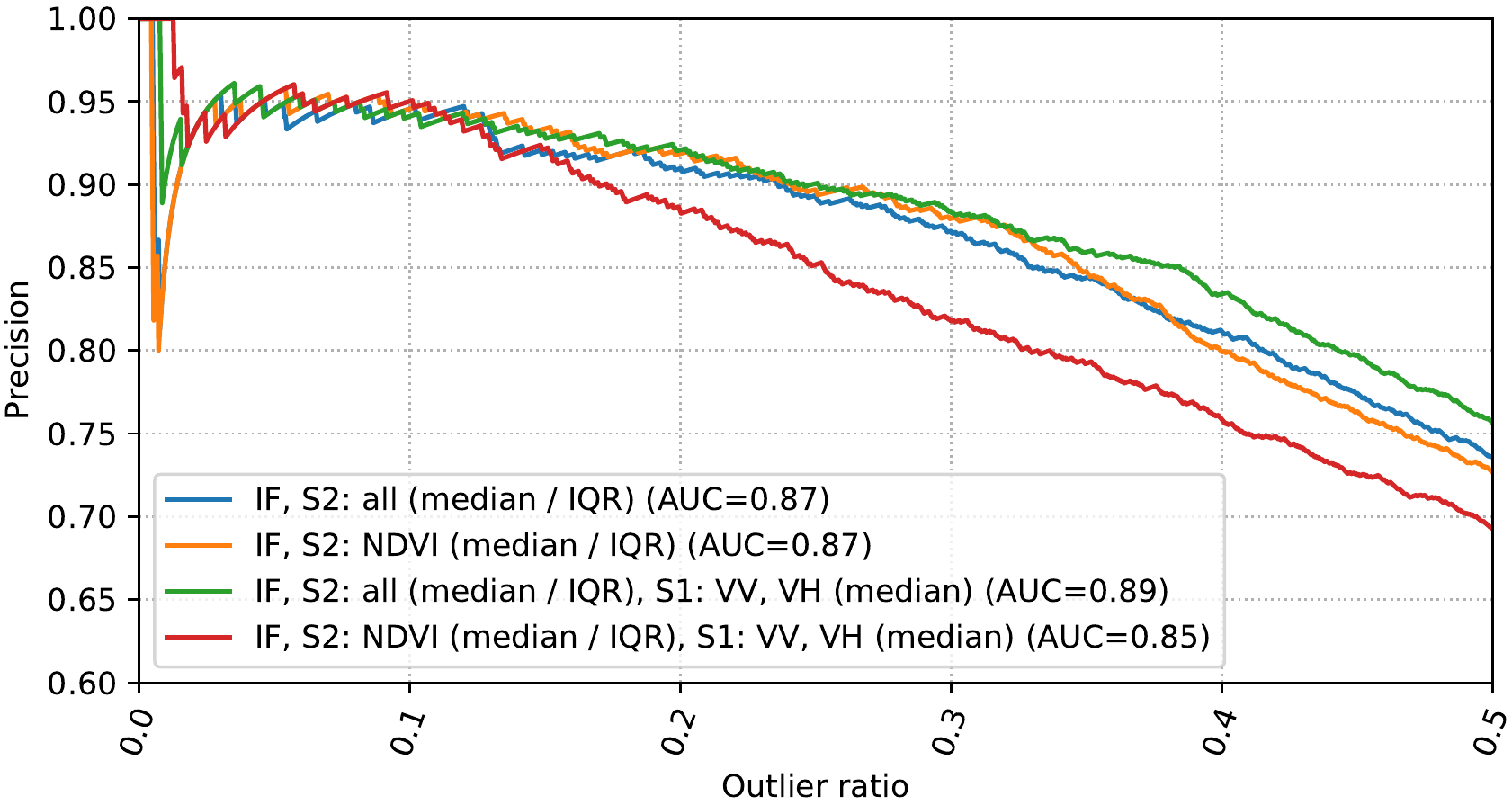}}
\caption{Precision vs. outlier ratio for a complete growing season analysis of the rapeseed parcels. Various sets of features using the IF algorithm are compared.}
\label{fig:PCR_all_season_all}
\end{figure}

\autoref{fig:PCR_all_season_all} shows the precision against outlier ratio for a selection of the best feature sets resulting from S1 and S2 data using the IF algorithm. The best AUC is obtained when using all S1 and S2 features jointly. Note that S1 data allow a larger amount of true positives to be detected accurately. However, when using S2 features only (or even only NDVI), a similar precision is reached for outlier ratios lower than 30\% (even if the percentage of detected parcels in each category is not identical, see details below). Using all S2 features instead of NDVI only increases slightly the precision of the results for outlier ratio higher than 40\%, confirming that NDVI is relevant to characterize efficiently a growing season but that some anomalies are better described using all 5 VIs. Finally, a lower AUC is obtained when using NDVI and S1 data when compared to the 5 VIs combined with VH and VV.

\begin{figure}[ht!]
\centerline{\includegraphics[width=1\textwidth]{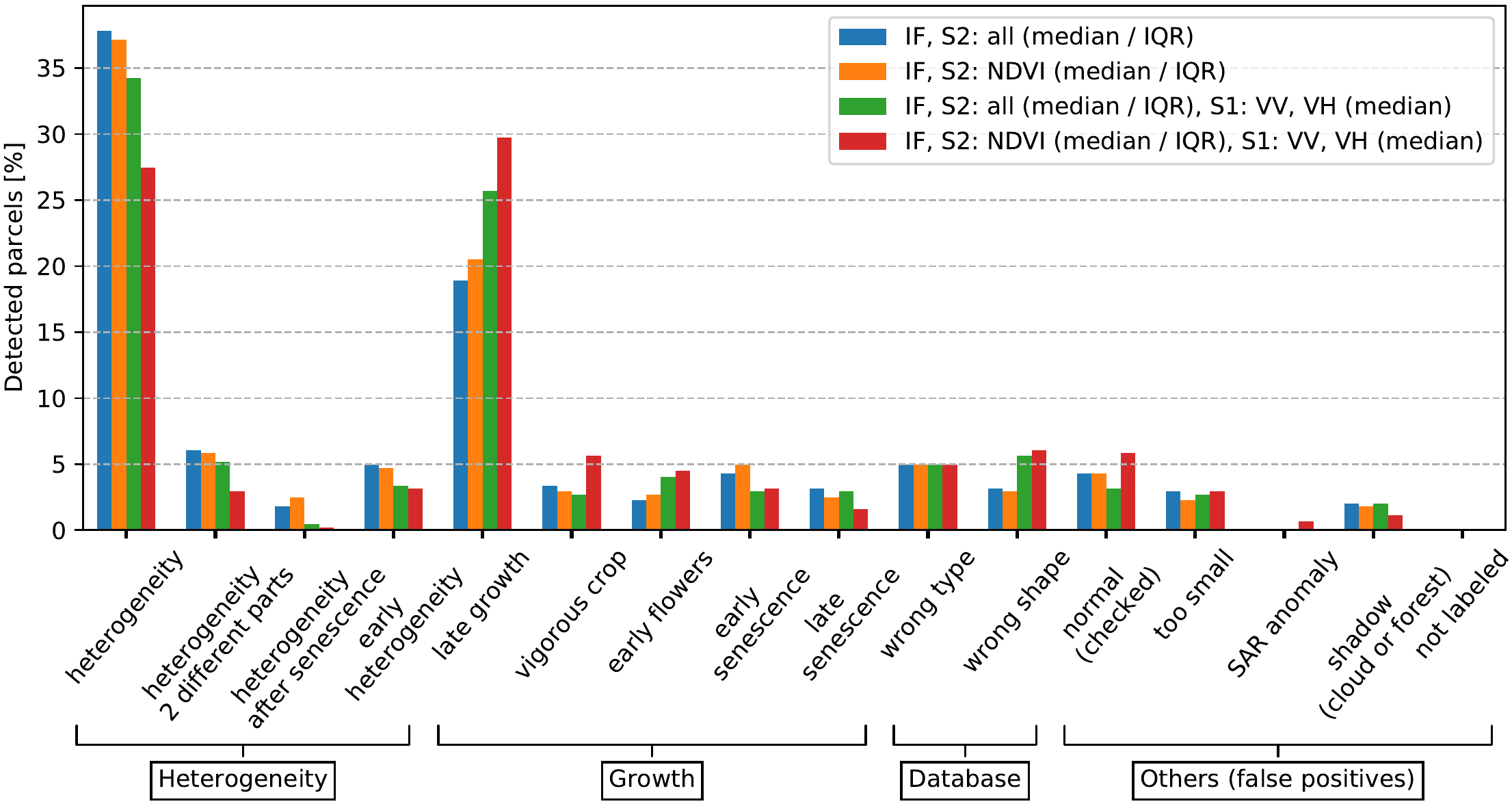}}
\caption{100$\times$(Number of detected parcels in each category / Number of detected parcels). Various sets of features are compared with the IF algorithm and an outlier ratio equal to 20\% for a complete growing season analysis (rapeseed crops).}
\label{fig:histo_all_season_all}
\end{figure}

The percentages of detected parcels within each category for an outlier ratio of 20\% are depicted in \autoref{fig:histo_all_season_all}. Again, the most frequent detected anomalies are due to heterogeneity and late growth. More growth anomalies (late/vigorous growth) and less heterogeneity are detected when S1 data is used. Results obtained when using all S2 features or NDVI only are close to each other in this example. However, the subsets of detected parcels by each configuration are not identical (55 parcels detected by one set of features are not detected by the other). Note that more false positives are detected when using NDVI only and S1 features, which leads to a precision of 89.4\% whereas it is close to 93\% for the other feature sets. The results of this section confirm the relevance of using median and IQR of all 5 S2 features jointly with median of VV and VH S1 features for detecting anomalous crop development.

\newpage
\subsection{Effect of the outlier ratio}\label{sec:outlier_ratio_effect}

\begin{figure}[ht!]
\centerline{\includegraphics[width=1\textwidth]{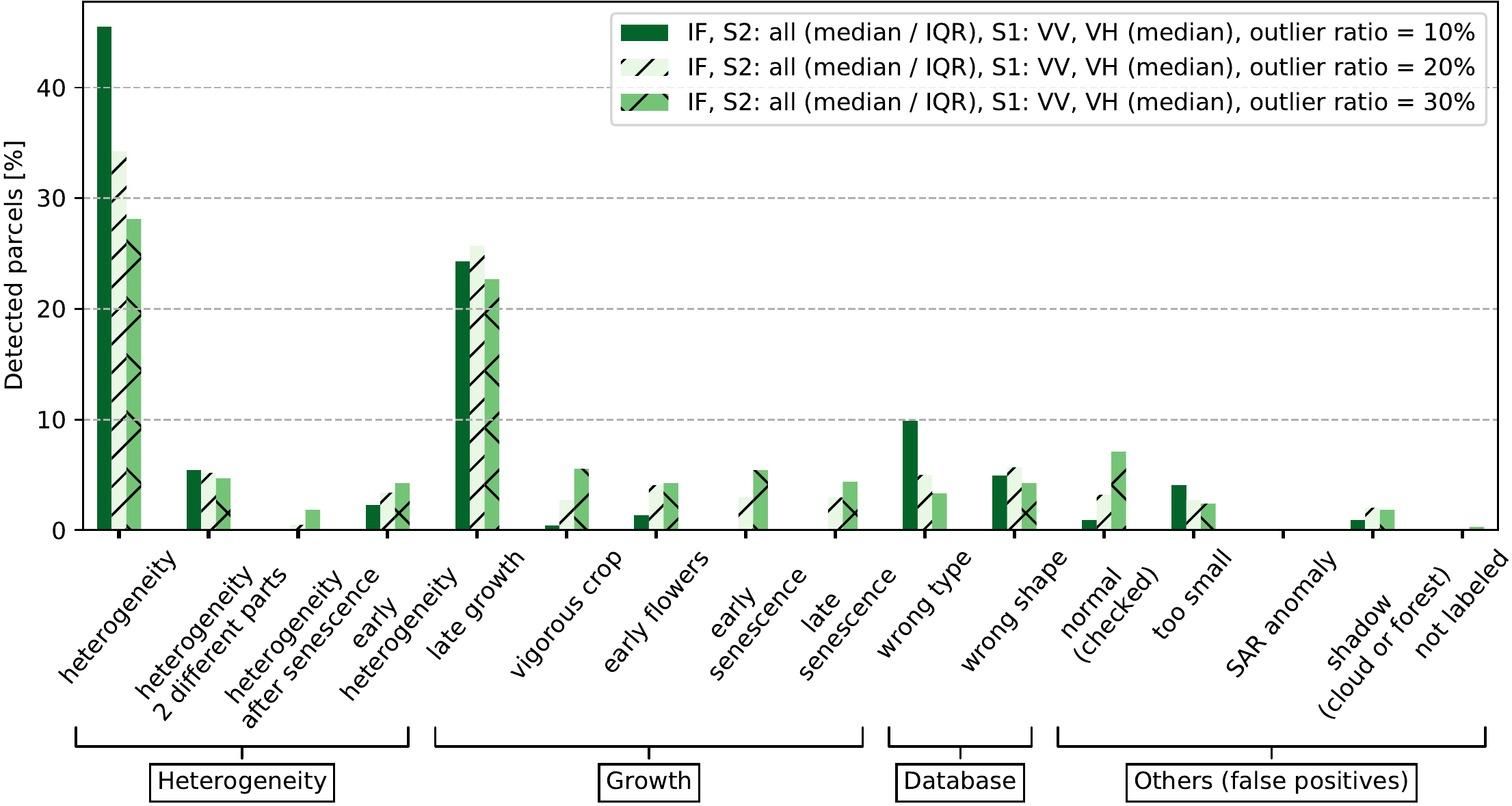}}
\caption{100$\times$(Number of detected parcels in each category / Number of detected parcels). Various outlier ratio are tested with the same set of features and the IF algorithm for a complete growing season analysis (rapeseed crops).}
\label{fig:histo_all_season_all_various_ratios}
\end{figure}

Three experiments were run using the median and IQR statistics derived from S2 images, the median statistics derived from S1 images and the IF algorithm, varying the outlier ratio in $\{0.1,0.2,0.3\}$. The percentages of detected parcels in the different anomaly categories for each of these experiments are depicted in \autoref{fig:histo_all_season_all_various_ratios}. For an outlier ratio of 10\%, the detected anomalies are mostly concentrated in wrong types, late growth and global heterogeneity which is relevant and confirms the observations made in the main document of this study. Moreover, for this outlier ratio, 45\% of the detected parcels belong to the category referred to as ``global heterogeneity'', which is coherent since this type of anomaly is (generally) strongly affecting the crop development of the parcels. Increasing the outlier ratio allows anomalies affecting smaller time periods of the season to be detected, such as early flowering and senescence problems in accordance to the observation made during labeling. For an outlier ratio of 30\%, much more false positives are detected (parcels labeled as normal). These results show that the IF algorithm provides a relevant anomaly score since more severe anomalies have higher anomaly scores. Moreover, because the score given by IF is computed only once, there is no need to run the algorithm several times when changing the outlier ratio and the amount of parcel to be detected can be easily adapted to the users' needs. Finally, for a generic analysis, choosing an outlier ratio of 20\% is a good balance between the precision of the detection results and the amount of parcel to be detected.

\subsection{Effect of adding new statistics for S2 data}\label{sec:statistics_used}

\begin{figure}[ht!]
\centerline{\includegraphics[width=0.7\textwidth]{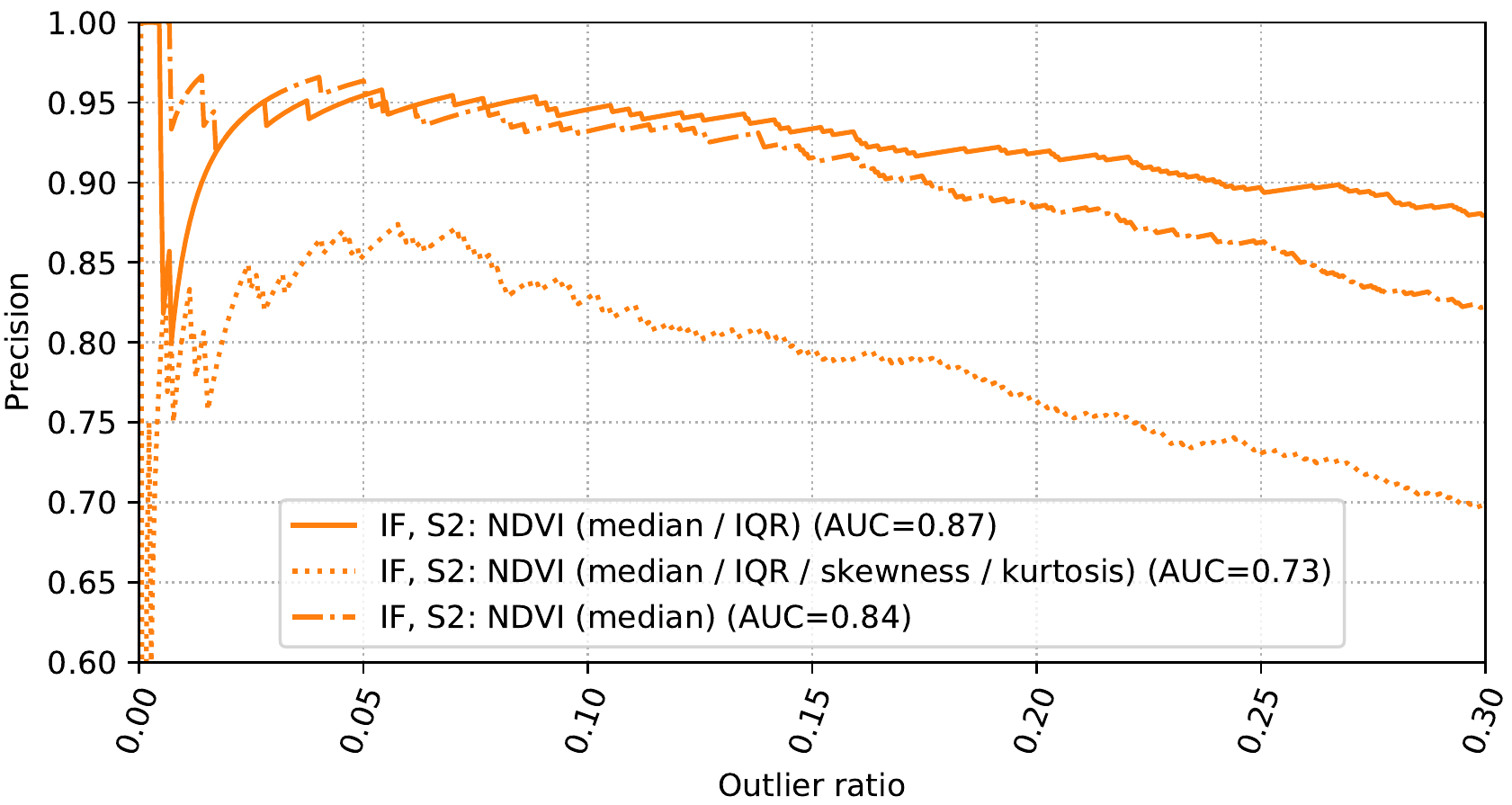}}
\caption{Precision vs. outlier ratio for a complete growing season analysis of the rapeseed parcels. Various statistics of the NDVI are compared using the IF algorithm.}
\label{fig:CR_NDVI_MIQR_vs_kurto_skew}
\end{figure}

All the previous experiments were conducted using the median and IQR of S2 data as statistics computed at the parcel-level. This section investigates two new statistics, namely the skewness and kurtosis (\textit{i.e.}, the normalized third and fourth order moments of the features). \autoref{fig:CR_NDVI_MIQR_vs_kurto_skew} shows the precision vs. outlier ratio when using the IF algorithm and these two additional statistics computed from S2 images to detect anomalies in rapeseed parcels. All the parcels are labeled for outlier ratios that are at least smaller than 10\% (less tests were made with skewness and kurtosis statistics as poor results were obtained). It can be observed in this figure that even for an outlier ratio lower than 5\%, using skewness and kurtosis statistics leads to a significant difference in the precision results. One issue encountered when using these new statistics is the detection of too subtle anomalies that are not always related to agronomic anomalies. Using the median only is also tested but provides a lower average precision score. This analysis confirms the importance of IQR statistics, which allows a larger number of relevant anomalies to be detected, and in particular heterogeneity problems. This section showed that using median and IQR statistics of S2 features computed at the parcel level is recommended for crop monitoring. 

\subsection{Effect of missing S2 images}\label{sec:missing_images}

Two scenarios were investigated to evaluate the effect of missing S2 images.

\begin{itemize}\setlength{\itemsep}{0pt}
    \item Scenario 1: the proposed approach was investigated using 6 S2 images instead of 13 to analyze the influence of a reduced amount of S2 images through the season. Only 1 image out of 2 was considered for the detection (the first S2 image was not used, the second S2 image was used and so on). Precision vs. outlier ratio curves are presented in \autoref{fig:PCR_all_season_missing_images}, where it can be observed that the proposed method is robust to missing S2 images. 

    \begin{figure}[ht!]
    \centerline{\includegraphics[width=0.7\textwidth]{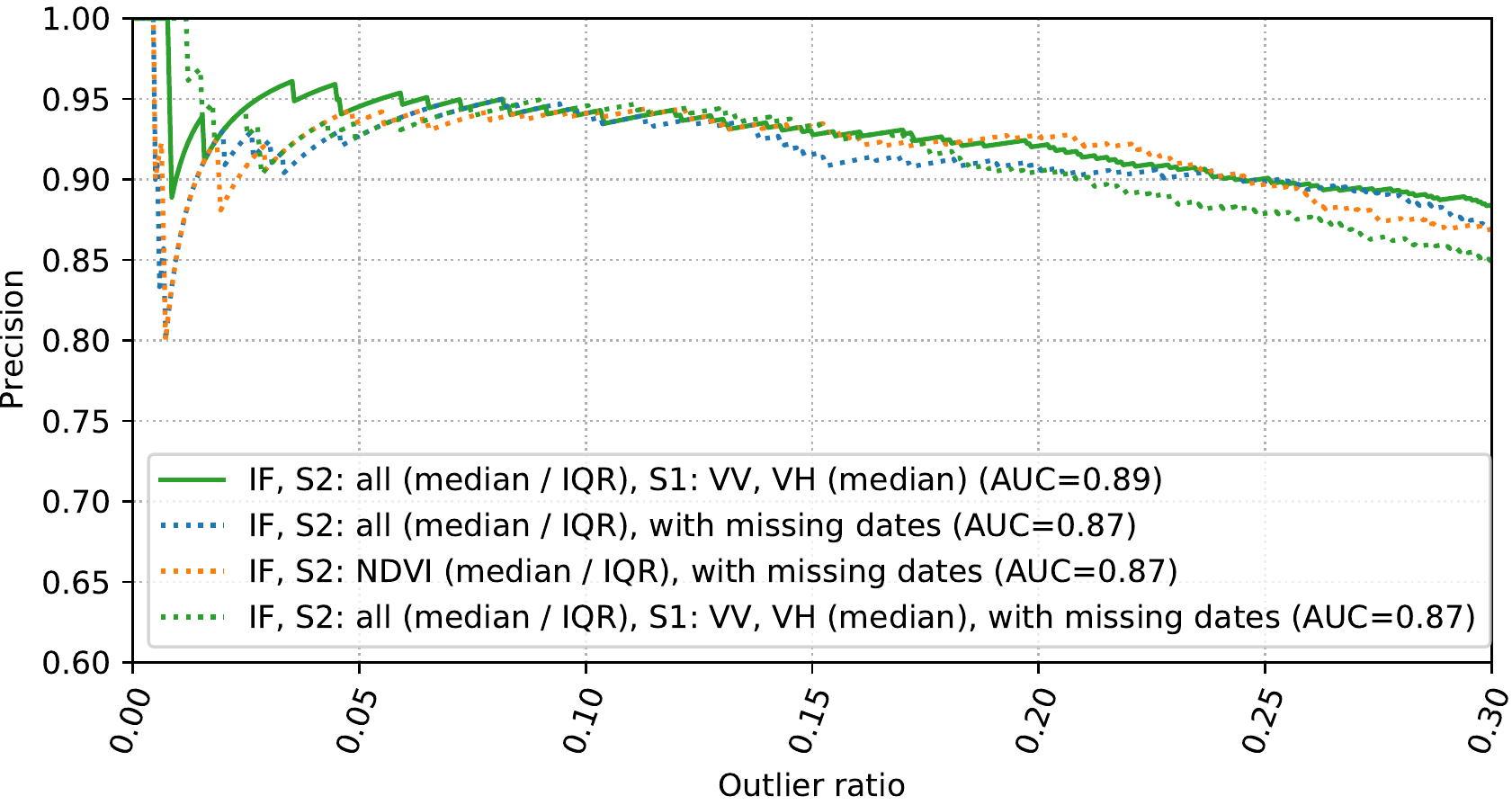}}
    \caption{Precision vs. outlier ratio for a complete season analysis of the rapeseed dataset. Missing dates means that only 1 S2 image out of 2 was taken (6 S2 images instead of 13).}
    \label{fig:PCR_all_season_missing_images}
    \end{figure}

    \item Scenario 2: another experiment was conducted to evaluate the effect of missing S2 images during the first part of the growing season (e.g, more clouds during winter). Precisely, we consider only 7 dates of S2 data between May and June that are used jointly with all S1 images. Precision vs. outlier ratio curves are presented in \autoref{fig:histo_end_season}. In that case, using S1 images improve significantly the precision of the results. The reason is that using S1 features allows the algorithm to detect almost the same amount of late growth crops when compared to using a complete season of S2 images which is understandable since S1 data are well suited to detect growth anomalies. These results confirm the interest of using S1 features as a complement to S2 sparse time series.

    \begin{figure}[ht!]
    \centerline{\includegraphics[width=0.7\textwidth]{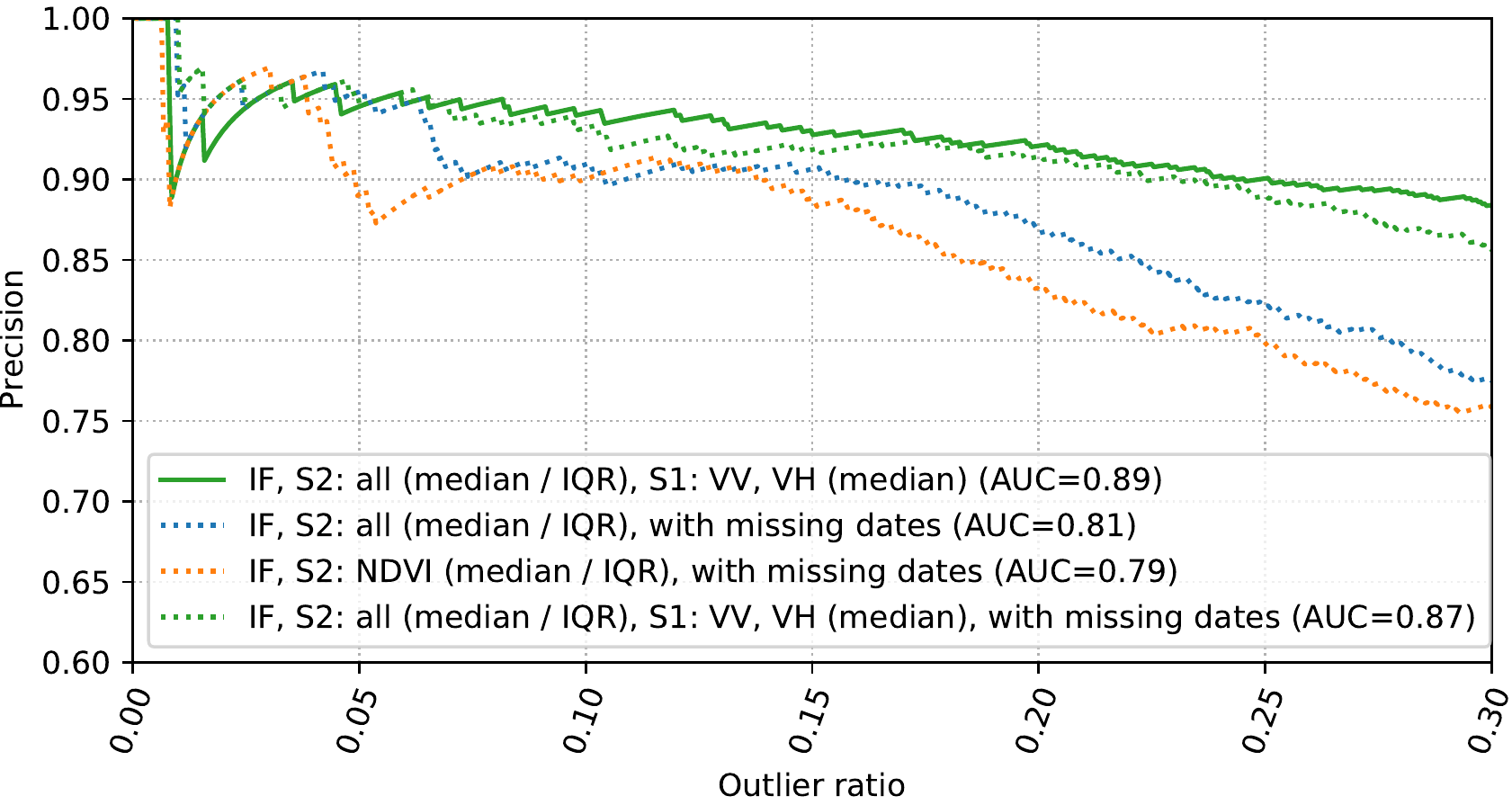}}
    \caption{Precision vs. outlier ratio for complete season analysis of the rapeseed dataset. Missing dates means that only the S2 images acquired after April were used (7 images).}
    \label{fig:histo_end_season}
    \end{figure}
    
\end{itemize}

\subsection{Mid-season analysis}\label{sec:mid_season}

\begin{figure}[ht!]
\centerline{\includegraphics[width=0.7\textwidth]{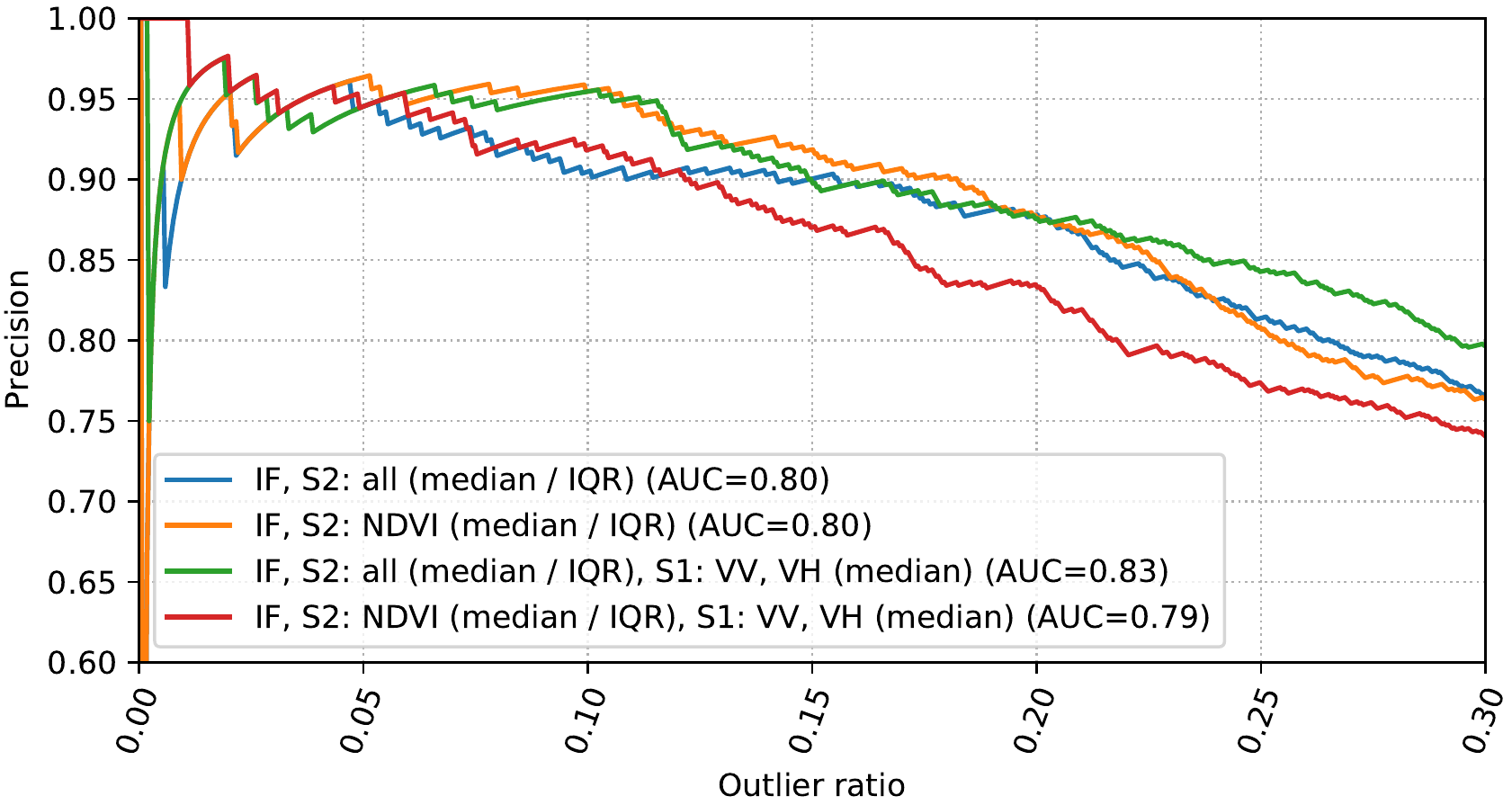}}
\caption{Precision vs. outlier ratio for a mid-season analysis of rapeseed parcels (all images available before February). Various sets of features are compared using the IF algorithm.}
\label{fig:PCR_mid_season_all}
\end{figure}

A mid-season analysis (using only dates before February) was conducted for multiple reasons detailed in the main manuscript of this study (see Section 3.5). A first experiment was made with the best sets of features selected in Section~\ref{sec:best_results} for a complete season analysis using rapeseed parcels. Results displayed in \autoref{fig:PCR_mid_season_all} show that even with a small number of images, many agronomic anomalies are detected (best precision=87.7\% for an outlier ratio equal to 20\%). This confirms the previous results found in the case of missing S2 images. \autoref{fig:PCR_mid_season_all} also shows that the best results are again obtained using all S1 and S2 features jointly with a higher average precision since more actual anomalies are detected for larger outlier ratios (\textit{e.g.}, the precision is 5\% better for an outlier ratio fixed to 30\%).

\begin{figure}[ht!]
\centerline{\includegraphics[width=1\textwidth]{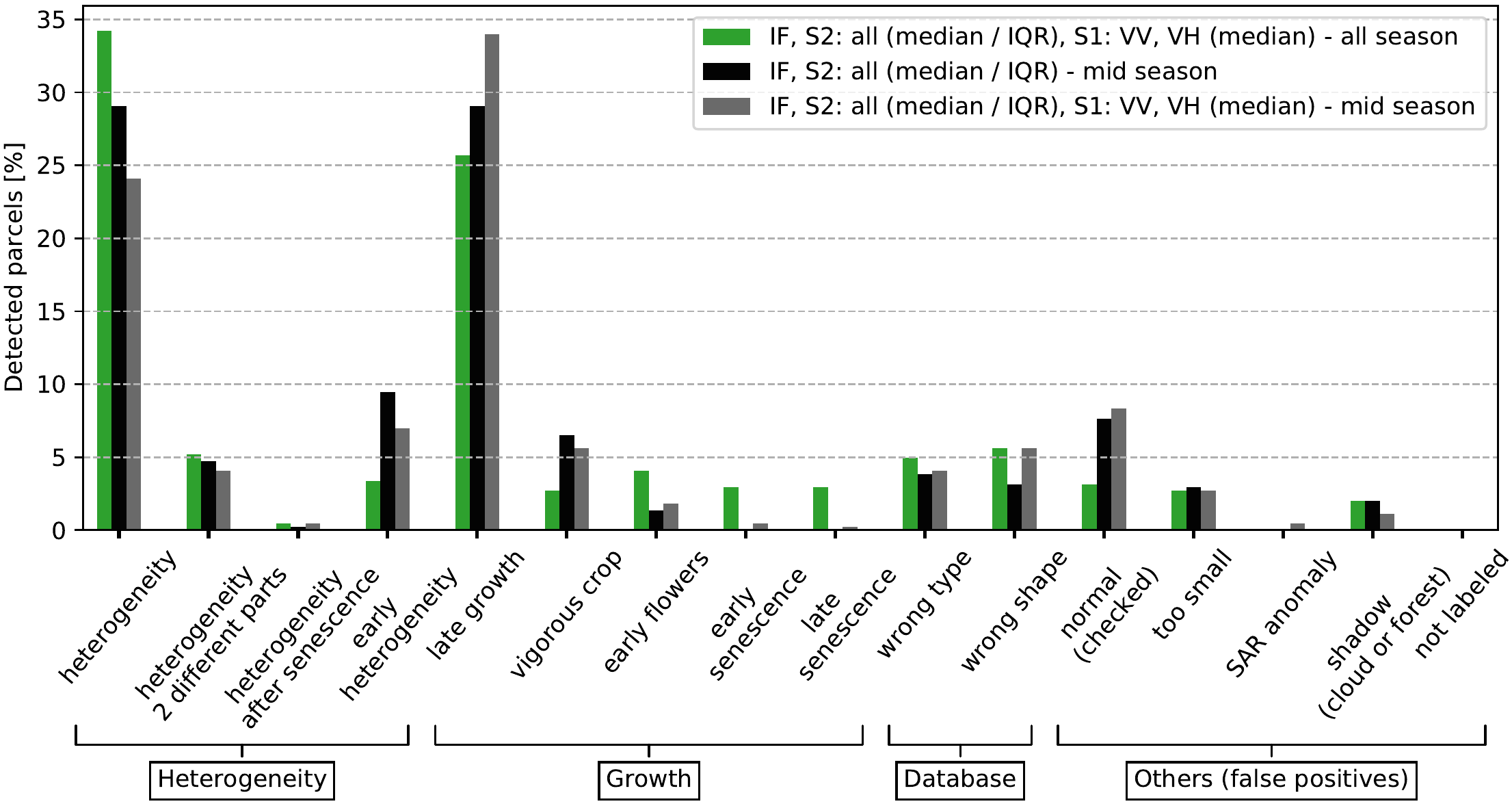}}
\caption{100$\times$(Number of detected parcels in each category / Number of detected parcels). Results obtained for a mid season analysis (before February) and a complete growing season analysis are compared for a outlier ratio equal to 10\% in the rapeseed dataset.}
\label{fig:histo_mid_season_vs_all_season}
\end{figure}

The impact of a mid-season analysis regarding the different categories of detected anomalies is depicted in \autoref{fig:histo_mid_season_vs_all_season}. In this case, almost no senescence problems are detected, which is easy to understand. Even with only 3 S2 images acquired between October and December, most other agronomic anomalies are detected by the algorithm. A mid-season analysis is able to detect more late growth anomalies and fewer heterogeneous parcels because late growth is impacting mostly the beginning of the season (especially for rapeseed crops). Finally, more false positives are detected with a mid-season analysis, which can be understood since the amount of potential anomalies to be detected is lower.

Complementary results for a mid season analysis are briefly presented in what follows since they confirm the observations made for a complete growing season analysis. The IF algorithm provides overall better results (AUC=0.83) and is more robust to changes. The AE performs slightly worse than IF (AUC=0.81), especially for outlier ratios greater than 20\%. OCSV (AUC=0.79) and LoOP (AUC=0.77) perform significantly worse in this case. These differences in performance can be explained by the fact that the parameters of OC-SVM, LoOP and AE algorithms are more difficult to tune compared to the IF algorithm. Regarding the influence of the outlier ratio, as for a complete season analysis increasing its value logically leads to detect more subtle anomalies (i.e., affecting a limited time interval) and more false positives, which confirms the relevance of the anomaly score given by IF. Almost no early heterogeneity and vigorous crop is detected with an outlier ratio of 10\%. Early heterogeneity is a more subtle anomaly than global heterogeneity, which confirms separation between these two categories. Finally, when using S1 data only, the detection results obtained for an outlier ratio of 10\% are still accurate with a precision equal to 89.6\%. These results confirm that S1 images are adapted to an early season analysis, especially thanks to an easier detection of late growth problems.

\newpage
\subsection{Robustness to changes in parcel boundaries}\label{sec:results_rpg}

The robustness of the proposed method to changes in the parcel boundaries was validated using another parcel delineation system for the rapeseed growing season. To that extent, 2118 parcel delineations resulting from the French Land Parcel Identification System (LPIS) was considered. The French LPIS is also known as \textit{Registre Parcellaire Graphique} (RPG). This database is available with an open license \footnote{\url{https://www.data.gouv.fr/fr/datasets/58d8d8a0c751df17537c66be/}, online accessed 8 July 2020} and is updated yearly (in general with a delay of 2 years) on the basis of the farmer's Common Agricultural Policy (CAP) \citep{BARBOTTIN2018281}. For comparison purposes, each parcel of database used in the main document was intersected with a corresponding LPIS parcel. Some parcels were not defined in the LPIS file, which explains why the number of parcels available for the LPIS analysis is slightly smaller than the number of parcels obtained when using the customer database.

\begin{figure}[ht!]
\centerline{\includegraphics[width=0.49\textwidth]{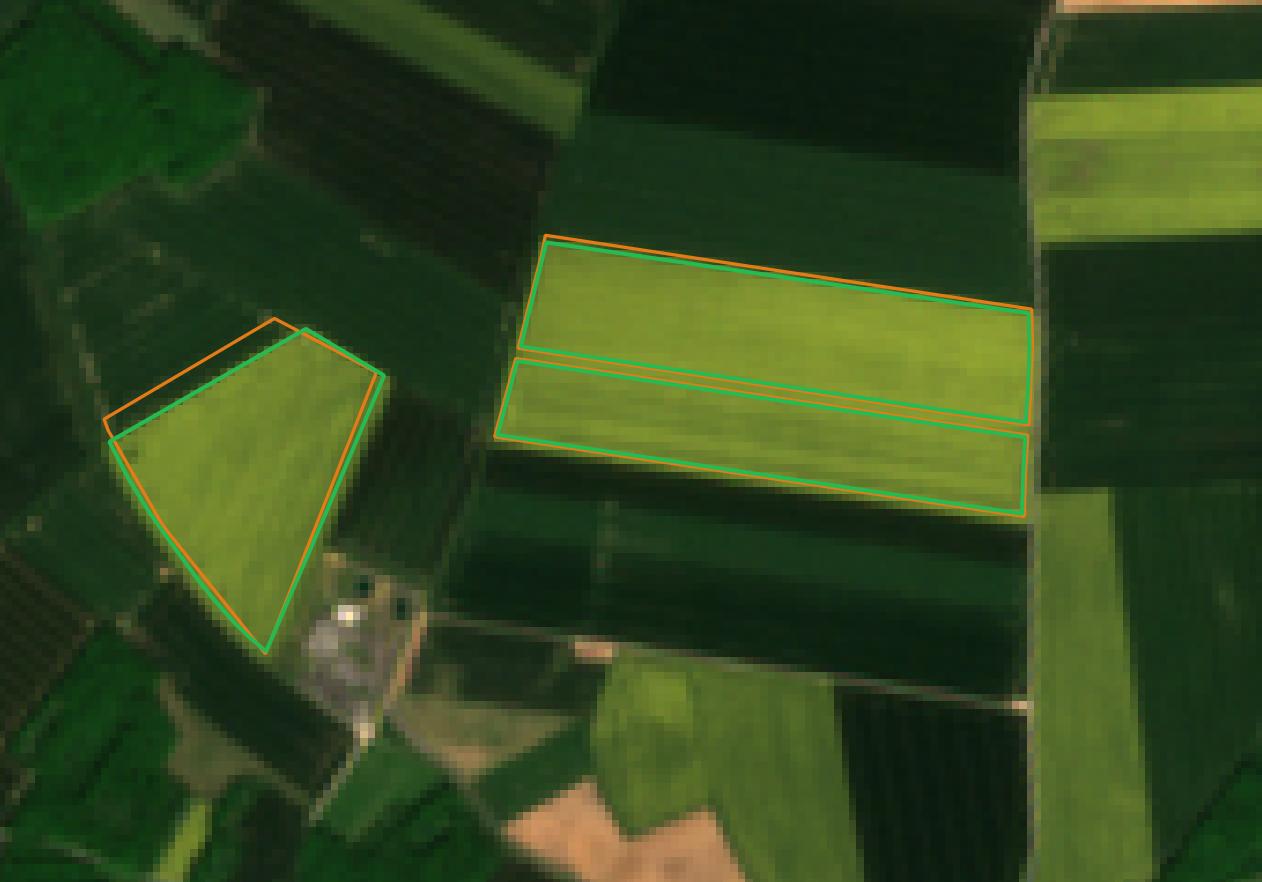}}
\caption{Example of parcel boundaries (rapeseed crop, growing season 2017/2018). In orange: customer database, in green: LPIS database.}
\label{fig:ex_LIPIS_custom}
\end{figure}

Examples of parcel delinations obtained with LPIS and the proprietary parcellation system are depicted in \autoref{fig:ex_LIPIS_custom}. The parcel frontiers obtained using LPIS are generally less accurate than those resulting from the proprietary system motivating the use of a buffer around the different parcels and robust zonal statistics.

\begin{figure}
\centerline{\includegraphics[width=1\textwidth]{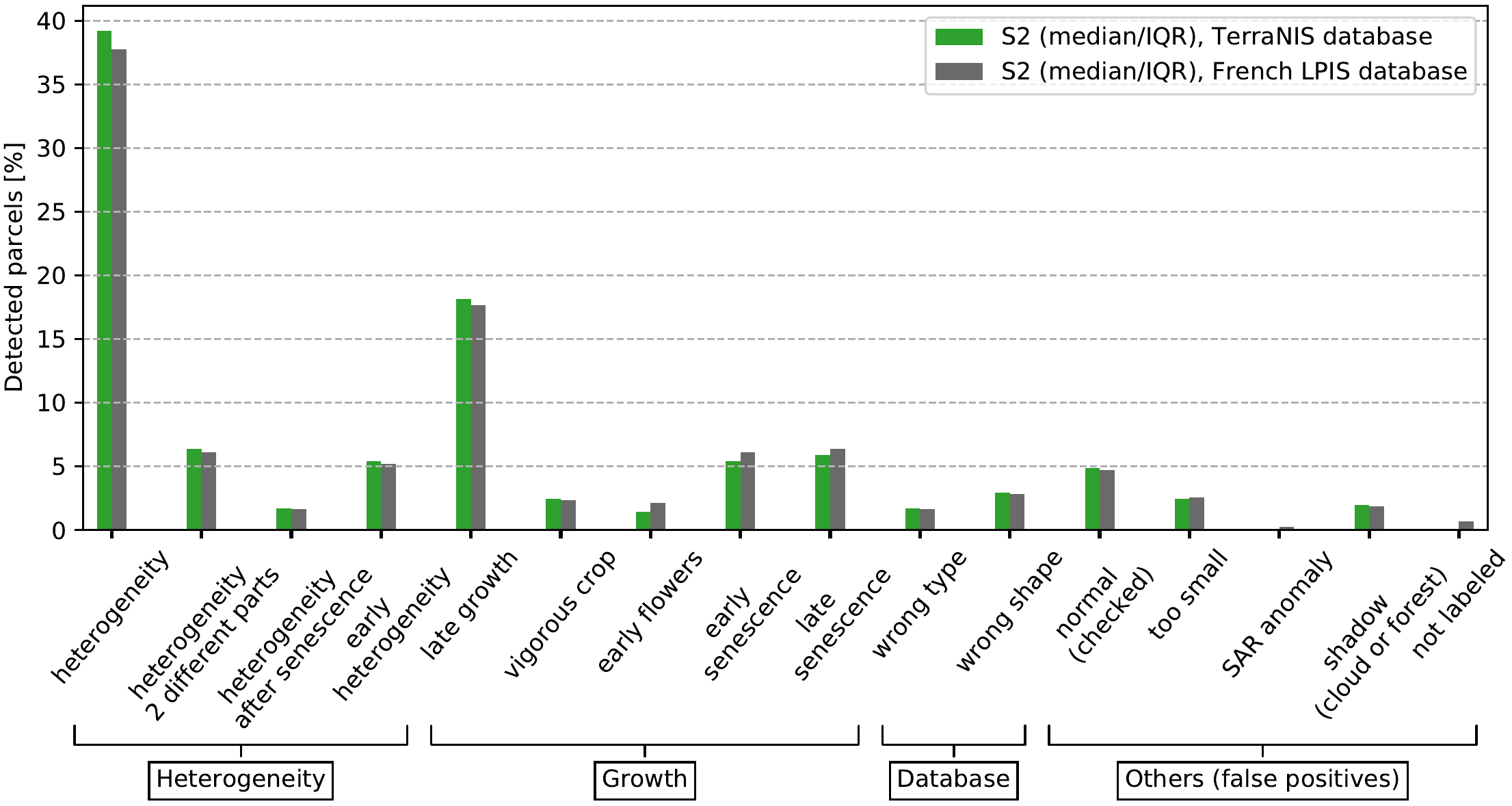}}
\caption{100$\times$(Number of detected parcels in each category / Number of detected parcels). LPIS and proprietary parcellation databases are compared with the IF algorithm and an outlier ratio equal to 20\%.}
\label{fig:histo_customer_vs_LPIS}
\end{figure}

Anomaly detection was run with an outlier ratio of 20\% using these two different databases. The numbers of detected anomalies for each category are depicted in \autoref{fig:histo_customer_vs_LPIS}. No significant difference can be observed when using the customer and LPIS parcels, showing that the proposed detection method is robust to this type of changes (probably because robust zonal statistics are used for anomaly detection).

\newpage
\bibliography{cas-refs}